\documentclass[]{aa} 
\usepackage{natbib} 
\usepackage{graphicx}

\def\rms{r{.}m{.}s{.}~}
\def\resp{resp{.}~}
\def\respn{resp{.}}
\def\ie{i{.}e{.}~}
\def\eq{Eq{.}~}                 
\def\eqs{Eqs{.}~}
\def\fg{Fig{.}~}
\def\fgs{Figs{.}~}
\def\sct{Sect{.}~}
\def\sctn{Sect{.}}
\def\scts{Sects{.}~}
\def\sctsn{Sects{.}}
\def\col{Col{.}~}

\def\colsn{Cols{.}}

\def\hkpc{$h^{-1}$~kpc~}
\def\hkpcp{$h^{-1}$~kpc}

\def\phiuni{~$h^3$~Mpc$^{-3}$~mag$^{-1}$}
\def\phiunit{~$h^3$~Mpc$^{-3}$~mag$^{-1}$~}

\begin{document}

\title{The ESO-Sculptor Survey: Luminosity functions of galaxies per
spectral type at redshifts $0.1-0.5$
\thanks{Based on observations collected at the European Southern 
Observatory (ESO), La Silla, Chile.}}

\author{Val\'erie de Lapparent\inst{1} 
   \and Gaspar Galaz\inst{2} 
   \and Sandro Bardelli\inst{3} 
   \and St\'ephane Arnouts\inst{4}}

\offprints{lapparen@iap.fr}

\institute{Inst. d'Astrophysique de Paris, CNRS, Univ. Pierre et
Marie Curie, 98 bis Boulevard Arago, 75014 Paris, France\\
\email{lapparen@iap.fr}
\and Depart. de Astronom\'{\i}a y Astrof\'{\i}sica, Pontificia Universidad
Cat\'olica de Chile, casilla 306, Santiago 22, Chile\\
\email{ggalaz@astro.puc.cl}
\and INAF-Osservatorio Astronomico di Bologna, via Ranzani 1, 40127 Bologna, Italy\\
\email{bardelli@excalibur.bo.astro.it}
\and European Southern Observatory, Karl-Schwarzschild-Strasse 2, D-85748, Garching, Germany\\
\email{sarnouts@eso.org}
}

\date{Received 14 January 2003 /  Accepted 26 March 2003}

\abstract{

We present the first statistical analysis of the \emph{complete}
ESO-Sculptor Survey (ESS) of faint galaxies. The flux-calibrated
sample of 617 galaxies with $R_\mathrm{c}\le20.5$ is separated into 3
spectral classes, based on a principal component analysis which
provides a continuous and template-independent spectral
classification.  We use an original method to estimate accurate
K-corrections: comparison of the ESS spectra with a spectral library
using the principal component analysis allows us to extrapolate the
missing parts of the observed spectra at blue wavelengths, then
providing a polynomial parameterization of K-corrections as a function
of spectral type and redshift. We also report on all sources of random
and systematic errors which affect the spectral classification, the
K-corrections, and the resulting absolute magnitudes.

\hspace{0.5cm} We use the absolute magnitudes to measure the
Johnson-Cousins $B$, $V$, $R_\mathrm{c}$ luminosity functions of the
ESS as a function of spectral class. The shape of the derived
luminosity functions show marked differences among the 3 spectral
classes, which are common to the $B$, $V$, $R_\mathrm{c}$ bands, and
therefore reflect a physical phenomenon: for galaxies of later
spectral type, the characteristic magnitude is fainter and the
faint-end is steeper. The ESS also provides the \emph{first} estimates
of luminosity functions per spectral type in the $V$ band.

\hspace{0.5cm} The salient results are obtained by fitting the ESS
luminosity functions with composite functions based on the
intrinsic luminosity functions per morphological type measured locally
by \citet{sandage85b} and \citet{jerjen97b}. The Gaussian luminosity
functions for the nearby Spiral galaxies can be reconciled with the
ESS intermediate and late-type luminosity functions if the
corresponding classes contain an additional Schechter contribution
from Spheroidal and Irregular dwarf galaxies, respectively. The
present analysis of the ESS luminosity functions offers a renewed
interpretation of the galaxy luminosity function from redshift surveys.
It also illustrates how luminosity functions per spectral type may be
affected by morphological type mixing, and emphasizes the need for a
quantitative morphological classification at $z\ga0.1$ which separates
the giant and dwarf galaxy populations.

 }

\authorrunning{de Lapparent et al.}
\titlerunning{The ESO-Sculptor Survey, Luminosity Functions at $z\simeq0.1-0.5$}
\maketitle

\section{Introduction \label{intro}} 

The galaxy luminosity function (LF hereafter) is a fundamental measure
for characterizing the large-scale galaxy distribution.  In the
current models of galaxy formation based on gravitational clustering,
the LF provides constraints on the mechanisms for the formation of
galaxies within the dark matter halos \citep{cole00,baugh02}.  The
bulge-dominated and disk-dominated galaxies can be traced separately
in the models and compared directly with the observations
\citep{baugh96,kauffmann97,cole00}. Nevertheless, due to the necessary
compromise between a large statistical volume and sufficient
resolution for simulating the individual galaxies, the N-body models
only describe a limited range of galaxy masses and morphological types
\citep{mathis02a}.  In contrast, observational studies of the local
galaxy distribution reveal a wealth of details. The galaxy LF spans
more than 12 magnitudes (that is 5 orders of magnitude in luminosity;
see for example \citealt{flint01a,trentham02b}). Moreover, each
morphological type has a distinct LF, denoted ``intrinsic'' LF, with
different parametric functions for the giant and dwarf galaxies
\citep[see the review by][]{binggeli88}.  The ``general'' galaxy LF,
averaged over all galaxy types, is then a composite of the intrinsic
LFs.

Specific studies of local galaxy concentrations have allowed detailed
insight into the intrinsic LFs per galaxy type. Co-addition of the
intrinsic LFs for the Virgo cluster \citep{sandage85b}, the Centaurus
cluster \citep{jerjen97b}, and the Fornax cluster \citep{ferguson91}
shows that the giant galaxies have Gaussian LFs, which are thus
bounded at bright \emph{and} faint magnitudes, with the Elliptical LF
skewed towards faint magnitudes. \citet{andreon98b} also shows that
the LFs for giant galaxies are invariant in shape among the Virgo,
Centaurus and Coma cluster; because these 3 clusters span a wide range
of cluster richness, the analysis suggests that these LFs may be
universal among galaxy concentrations. In contrast, the LFs for dwarf
galaxies may be ever increasing at faint magnitudes to the limit of
the existing surveys, with a steeper increase for the dwarf Elliptical
galaxies (dE), when compared with the dwarf Irregular galaxies
(dI). \citet{schaeffer88} have proposed an analytical description for
the bimodal behavior of the galaxy LF, which models the effect of the
galaxy binding energy onto the gas and the resulting efficiency in
star formation as a function of galaxy mass.

Because of the different intrinsic LFs for giant and dwarf galaxies,
the ``general'' LF in the local group and in nearby clusters and
groups has a varying faint-end behavior with the richness of the
concentration: this can be partly interpreted in terms of the varying
dwarf-to-giant galaxy ratio dE/E which increases with local density
(\citealp{ferguson91,trentham02a}; see also
\citealt{trentham02b}). The faint-end behavior of the dE and dI LFs is
however still controversial. Slopes as steep as $\alpha\sim-1.3$ are
measured for the Spheroidal/red dwarf galaxies in groups en clusters
\citep{ferguson91,andreon01c,conselice02}, whereas other less rich
environments yield $-1.2\la\alpha\la-1.1$
\citep{pritchet99,flint01a,trentham01,trentham02b}, with some
significant contribution from the dI galaxies in \citet{trentham01}.
It is unclear whether these differences are solely due to differences
in the detected dwarf populations (related to the ratio of dE to dI
galaxies), or to the different environments in terms of local density,
or to both.

In parallel, measurements of LF per galaxy type have been obtained
from systematic redshift surveys, with significant variations from
survey to survey. Estimates of intrinsic LFs using visual
morphological classification have been obtained from the ``nearby''
redshift surveys ($z\la 0.1$), based on photographic catalogues
\citep{efstathiou88a,loveday92,marzke94b,marzke98,marinoni99}.  At
$z\ga0.1$, visual morphological classification however becomes highly
uncertain and has been replaced by spectral classification
\citep{heyl97,bromley98,lin99,folkes99,fried01,madgwick02a,wolf03}. When
neither morphological nor spectral classification are available, the
intrinsic LFs are estimated using samples separated by color
\citep{lilly95,lin97,metcalfe98,brown01} or the strength of the
emission-lines \citep{lin96,small97b,zucca97,loveday99}.  However,
none of the existing redshift surveys separate the giant and dwarf
galaxy populations, despite the markedly different intrinsic LFs for
these 2 populations \citep{sandage85b,ferguson91,jerjen97b}.

In view of the discrepancy between the local measures of the intrinsic
LFs and the estimates from redshift surveys at larger distance, we
propose here a new approach for reconciling the various LFs. It is
based on the LFs per galaxy type measured from the ESO-Sculptor Survey
(ESS hereafter).  The ESS has the advantage to provide a nearly
complete redshift survey of galaxies at $z\la0.5$ over a contiguous
area of the sky \citep{bellanger95a}, supplemented by CCD-based
photometry \citep{arnouts97} and a detailed spectral classification
\citep{galaz98}.

\sct \ref{ess} gathers the analyses used to build the ESS database:
\sct \ref{sample} describes the spectroscopic sample selection; \sct
\ref{spclass} summarizes the results of the spectral classification
analysis, the classification technique itself being reported in details
elsewhere \citep{galaz98}; \sct \ref{kcor} describes the original
method used for deriving K-corrections for the ESS spectra; \sct
\ref{error} reports on all sources of random and systematic errors
which affect the spectral classification and the derived absolute
magnitudes in the ESS catalogue; \sct \ref{types} describes the choice
of the spectral classes on which are based the LF calculations.  

We then comment on the technique for deriving the ESS LFs in \sct
\ref{method}; the results are reported and discussed in \scts
\ref{LFess} and \ref{LFfilter}; in \sct \ref{LFother}, we compare the
ESS intrinsic LFs with those from the CNOC2 \citep{lin99}, the other
existing redshift survey to similar redshifts and with spectral
classification.  In \sct \ref{LFcomp}, we then propose a new approach
for interpreting the intrinsic LFs from redshift surveys. In \sct
\ref{LFlocal}, we first review the local measurements of intrinsic LFs
as a function of morphological type , and we derive the required
magnitude conversions for application to the ESS. In \sct
\ref{ESSlocal}, we propose composite fits of the ESS intrinsic LFs
which are based on the local LFs for giant and dwarf galaxies; we
discuss these composite fits for the ESS early, intermediate, and
late-type LFs in \scts \ref{LFearly}, \ref{LFintermediate}, and
\ref{LFlate} \resp \sct \ref{peak_sb} provides further evidence for
the presence of dwarf galaxy populations in the ESS, using the
distribution of peak surface brightness. Finally, we summarize the
results and discuss the prospects raised by the present analysis in
\sct \ref{concl}.

\section{The ESS spectroscopic survey \label{ess}}

The goal of the ESO-Sculptor Survey was to produce a complete
photometric and spectroscopic survey of galaxies with the following
scientific objectives: (i) to map the galaxy distribution of galaxies
at $z\simeq0.1-0.5$ and (ii) to provide a database for studying the
variations in the spectro-photometric properties of distant galaxies
as a function of redshift and local environment.  The ESO-Sculptor
Survey was successfully completed as an ESO key-programme, thanks to a
guaranteed allocation of $\sim 60$ clear nights of telescope time on
the ESO 3.6m and the NTT, performed over a period of 7 subsequent
years.

\subsection{Sample selection \label{sample} } 

The ESS photometric survey provides magnitudes in the Johnson $B$, $V$
and the Cousins $R_\mathrm{c}$ standard filters, for nearly 13000
galaxies to $V \simeq 24$ over a contiguous rectangular area of $\sim
0.37$ deg$^2$ [$1.53^\circ\mathrm{(R{.}A{.})}  \times
0.24^\circ\mathrm{(DEC{.})}$] \citep{arnouts97}. The survey region is
centered at $\sim 0^\mathrm{h}22^\mathrm{m}$ (R{.}A{.})  $\sim
-30^\circ06^\prime$ (DEC{.}) in J2000 coordinates, which is located
near the Southern Galactic Pole.  Multi-slit spectroscopy of the
galaxies with $R_\mathrm{c} \le 20.5$ \citep{bellanger95a} provided a
nearly complete redshift survey over a contiguous sub-area of $\sim
0.25$ deg$^2$ [$1.02^\circ\mathrm{(R{.}A{.})} \times
0.24^\circ\mathrm{(DEC{.})}$].  Selection of the galaxies to be
observed spectroscopically was solely based on their $R_\mathrm{c}$
magnitude.  Crowding on the mask left nearly 6\% of the galaxies with
$R_\mathrm{c}\le 20.5$ unobserved.  Instead, fainter galaxies could be
observed where there was remaining space on the multi-slit masks.  As a
result, the $R_\mathrm{c}$ completeness of the ESS spectroscopic
catalogue is not a pure step function.

\begin{figure}
  \resizebox{\hsize}{!}{\includegraphics{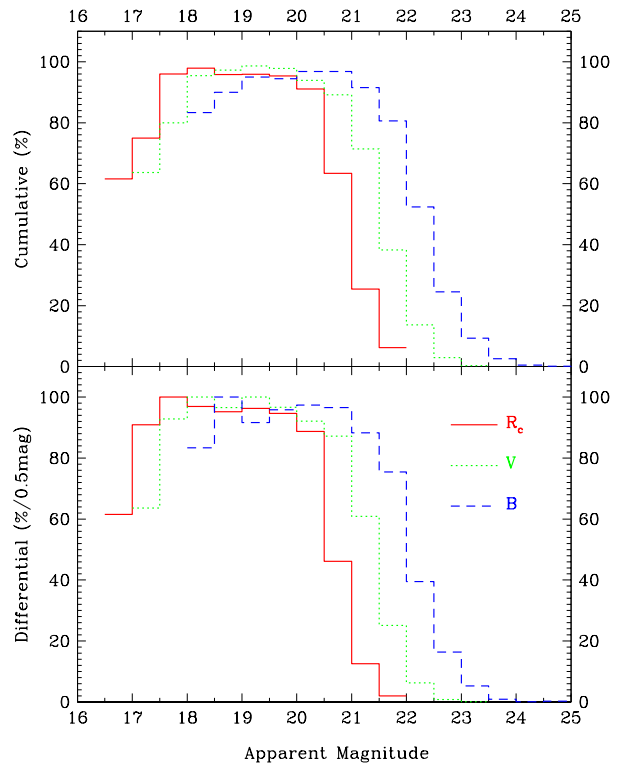}}
\caption{Fractional and cumulative completeness for the ESO-Sculptor
spectroscopic catalog, as a function of apparent magnitude, in the Johnson $B$, $V$ and Cousins $R_\mathrm{c}$ bands.}
\label{completeness}
\end{figure}

\begin{table*}
\caption{Differential completeness of the ESO-Sculptor redshift survey
in the Johnson $B$, $V$ and Cousins $R_\mathrm{c}$ bands.}
\label{comptab}
\begin{center}
\begin{tabular}{lcccccc}
\hline
\hline
\multicolumn{7}{c}{$R_\mathrm{c}$} \\ 
\hline
mag interval$^\mathrm{a}$      & $\le 20.0$ & $20.0-$\bf 20.5      & $20.5-21.0$      & $21.0-21.5$      & $21.5-22.0$      &                \\ 
completeness$^\mathrm{b}$       &   94.40 \% & \bf 92.23 (88.76) \% & 75.52 (46.19) \% & 52.28 (12.54) \% & 34.36 (1.96) \%  &                \\ 
galaxies with z $^\mathrm{c}$   &   388      & \bf  617 (229)       &  793 (176)       &  870 (77)        &  888  (18)       &                \\ 
\hline
\multicolumn{7}{c}{$V$} \\ 
\hline
mag interval          & $\le 20.0$ & $20.0-20.5$      & $20.5-$ \bf 21.0      & $21.0-21.5$      & $21.5-22.0$      & $22.0-22.5$    \\ 
completeness           &   95.41 \% & 94.03 (92.09) \% & \bf 91.45 (87.19) \%  & 80.40 (60.86) \% & 59.24 (25.10) \% & 39.53 (6.30) \% \\ 
galaxies with z      &   187      &    315 (128)     & \bf  492 (177)        &   677 (185)      &  808 (131)       &  859  (51)      \\ 
\hline
\multicolumn{7}{c}{$B$} \\ 
\hline
mag interval          & $\le 21.0$ & $21.0-21.5$      & $21.5-$ \bf 22.0      & $22.0-22.5$      & $22.5-23.0$      & $23.0-23.5$     \\
completeness           &   95.60 \% & 92.48 (88.32) \% & \bf 85.77 (75.48) \%  & 66.67 (39.46) \% & 45.10 (16.34) \% & 28.16 (5.25) \% \\ 
galaxies with z      &   174      &    295 (121)     & \bf    452 (157)      &    598 (146)     &    708 (110)     &   769 (61)      \\ 
\hline
\end{tabular}
\smallskip
\\
\end{center}
\begin{list}{}{}
\item[$^{\mathrm{a}}$] Apparent magnitude interval considered for the
completeness calculation.
\item[$^\mathrm{b}$] Cumulated completeness at the
faintest limit of the quoted apparent magnitude interval, calculated
as the ratio of the number of galaxies with redshift by the number of galaxies
in the photometric catalogue; in parentheses is indicated the
differential completeness in the quoted apparent magnitude interval. 
\item[$^\mathrm{c}$] Cumulated number of galaxies with a redshift measurement
brighter than the faintest limit of the quoted apparent magnitude
interval; in parentheses is indicated the differential number of
galaxies with a redshift measurement in the quoted apparent magnitude
interval. 
\end{list}
\end{table*}

Figure \ref{completeness} and Table \ref{comptab} show the
differential and cumulative redshift completeness in the $B$ $V$
$R_\mathrm{c}$ bands, in half-magnitude intervals. Table \ref{comptab}
shows that the differential completeness in $R_\mathrm{c}$ is nearly
flat from bright magnitudes to $R_\mathrm{c}=20.0$, with a
differential completeness larger than $94\%$, and decreases to 88.76\%
in the magnitude interval $20.0-20.5$, due to the increase in the
surface density of galaxies with magnitude; it then sharply drops to
46\%, 13\% and 2\% in the $R_\mathrm{c}$ intervals $20.5-21.0$,
$21.0-21.5$, $21.5-22.0$ respectively.  Despite the selection of the
spectroscopic sample in the $R_\mathrm{c}$ band, and the spread in $B
- R_\mathrm{c}$ and $V - R_\mathrm{c}$ colors (see right panels of \fg
\ref{abs_col} in \sct \ref{kcor}), the completeness functions in the
$V$ and $B$ bands have a similar behavior to that in $R_\mathrm{c}$.

For calculation of the LF in each band, we define a ``nominal
magnitude limit'' as the magnitude limit which provides the best
compromise between completeness, small color biases and sufficient
statistic.  In the $R_\mathrm{c}$ band, the choice is obvious and is
at $R_\mathrm{c}\le 20.5$, the spectroscopic selection limit (there is
no known color bias in the $R_\mathrm{c}$ sample at this limit).  Due
to the spectroscopic selection in the $R_\mathrm{c}$ band, the $V$ and
$B$ samples are deficient in objects with blue colors at faint
magnitudes.  We choose the nominal limits at $V\le 21.0$ and $B\le
22.0$ resp{.}, for the following reasons:
\begin{itemize}
\item the differential completeness is larger than $70\%$ in both the
  $B$ and $V$ samples at these limits (see Table \ref{comptab});
\item the $B$ and $V$ samples contain a sufficient number of galaxies
  for calculating intrinsic LFs based on 3 spectral classes;
\item the resulting combination of $B$, $V$, and $R_\mathrm{c}$
  magnitude limits is in agreement with the typical colors of the ESS
  galaxies at $R_\mathrm{c} \simeq 20.5$ ($B-R_\mathrm{c}\simeq1.5$
  and $V-R_\mathrm{c} \simeq0.5$, \citealt{arnouts97}).
\end{itemize}
We show in \sct \ref{LFess} that the LFs in the $B$ and $V$ bands vary
systematically when going to fainter limits than the nominal
magnitudes $V\le 21.0$ and $B\le 22.0$, due to the increasing color
biases at faint magnitudes in these samples.  Comparison with the LFs
for the $R_\mathrm{c}$ sample show that at the chosen $V$ and $B$
nominal limits, the color biases might nevertheless be comparable with
the random errors (see \sct \ref{LFess} and Table \ref{lf_BVR}). By
choosing brighter nominal magnitude limits in the $V$ and $B$ bands,
one would reduce the color biases in these samples; this would however
significantly reduce the number of galaxies (see Table
\ref{completeness}), and would not allow us to extract spectral-type
LFs in these filters.

As shown by \citet{disney83}, redshift surveys limited in apparent
magnitude also suffer selection effects in the central surface
brightness of galaxies. In the ESS photometric catalogue, the surface
brightness threshold in object detection used for the SExtractor image
analyses \citep{bertin96} is in the interval $\sim 25.5-26.5$ mag
arcsec$^{-2}$ in the $R_\mathrm{c}$ band, $\sim 25.5-27.0$ mag
arcsec$^{-2}$ in the $V$ band, and $\sim 26.0-27.5$ mag arcsec$^{-2}$
in the $B$ band (the $1$ to $1.5^\mathrm{mag}$ intervals are due to
variations in the depth of the individual images; most of it is caused
by the marked increase in depth when changing from the 3.6m telescope
to the NTT; a smaller part is due to the varying sky transparency with
time). Due to redshift dimming (see \sct \ref{peak_sb}), and to a
minor extent to K-corrections (see \sct \ref{kcor}), the resulting
rest-frame limiting peak surface brightness in the ESS redshift survey
is $~22.0$ mag arcsec$^{-2}$ in $R_\mathrm{c}$ for galaxies with
$R_\mathrm{c}\la21.5$ (see \fg \ref{sb}), $~22.5$ mag arcsec$^{-2}$ in
$V$ for galaxies with $V\la22.5$, and $~23.0$ mag arcsec$^{-2}$ in $B$
for galaxies with $B\la23.5$ (see \sct \ref{peak_sb} for definition of
ESS peak surface brightness).  The ESS distributions of rest-frame
peak surface brightness show no or weak correlation with apparent
magnitude, indicating that redshift effects have been appropriately
corrected for.

\citet{mcgaugh95} show that the low surface brightness population
sets in at a central surface brightness fainter than $\sim22.0$ mag
arcsec$^{-2}$ in $B$. The ESS spectroscopic sample reaches one
magnitude fainter in $B$, therefore detecting a fraction of this
population (see also \scts \ref{LFintermediate} and \ref{LFlate}). A
significant number of low surface brightness galaxies may nevertheless
have been missed in the ESS.  As shown by \citet{mcgaugh96} and
\citet[][see also \citealt{lobo99}]{dalcanton98}, the relatively
bright threshold in central surface brightness inherent to redshift
surveys may significantly affect the luminosity function at both the
bright and faint end.  Although low surface brightness galaxies may be
as numerous as the ``normal'' galaxies, they however contribute for
less than a factor 3 to the luminosity density
\citep{mcgaugh96,dalcanton97}.  We show in \sct \ref{peak_sb} that the
faintest $R_\mathrm{c}$ detected galaxies in the ESS also have a low
central surface brightness, with no evidence for intrinsically bright
though very extended galaxies above the sample limits.\footnote{We
find a tight correlation between the ESS rest-frame peak surface
brightnesses in the $R_\mathrm{c}$ and $B$ band, which implies that
the ESS surface brightness selection effects operate similarly in the
2 bands.}

\subsection{Spectral classification   \label{spclass} }  

Morphological types are not available for the ESS redshift survey. As
the survey describes the redshift range $0.1\la z \la0.6$, a large
fraction of the galaxies have diameters smaller than 10 arcseconds,
and identification of their morphology is severely limited by the
ground-based image quality \citep[see][]{arnouts97}.  We have
therefore chosen to perform the estimation of the intrinsic LFs based
on a spectral classification.  \citet{galaz98} show that using the ESS
data, a spectral classification method based on a Principal Component
Analysis (PCA hereafter) provides an objective spectral sequence,
which can be parameterized continuously using one or more parameters,
and is strongly related to the Hubble sequence of normal galaxies
\citep[see also][]{folkes96,bromley98,baldi01}.

The PCA allows us to describe each spectrum (in rest-wavelength) as a
linear combination of a reduced number of principal vectors, the
eigenvectors, also called principal components (PC hereafter), and
denoted $PC_i$. The PCs better discriminate the whole sample, and bear
decreasing variance with increasing index $i$. We denote $\alpha_i$
the projection of an observed spectrum onto vector $PC_i$. Galaz \& de
Lapparent (1998) show that in the ESS redshift survey, 3 PCs describe
$\sim 98\%$ of the flux of the spectra. The authors thus introduce the
coordinate change
\begin{equation}
\begin{array}{ll}
\delta &= \arctan(\alpha_2/\alpha_1)\\ 
\theta &= \arcsin \alpha_3,\\
\label{alpha_delta_theta}
\end{array}
\end{equation}
and show that $\delta$ and $\theta$ provide a robust 2-parameter
spectral sequence: the 2 parameters are continuous measures of the
relative fractions of old to young stellar populations, and the
relative strength of the emission lines, respectively.  Early-type
spectra, representative of red galaxies, without emission lines, lie
towards negative values along the $\delta$ direction.  Late-type
spectra, corresponding to blue galaxies, often have emission lines,
and lie at large values of $\delta$.  Note that by construction, the
$\delta-\theta$ classification is independent of absolute
normalization of the spectra (\ie luminosity).

\begin{figure}
  \resizebox{\hsize}{!}
    {\includegraphics{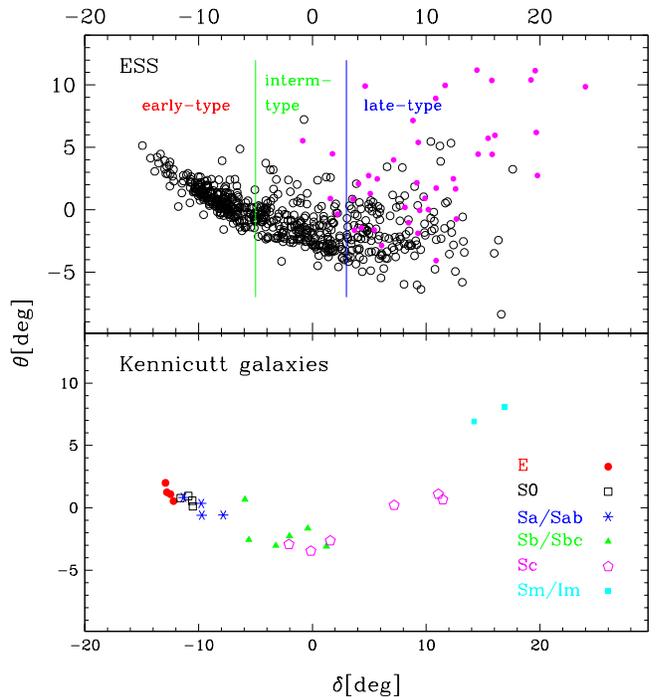}}
\caption{{\bf a)} Top panel: spectral classification parameters
$\delta$ and $\theta$ for the 603 ESO-Sculptor spectra with
$R_\mathrm{c}\le20.5$ and a PCA defined spectral class
$\delta_1-\theta_1$ (see \sct \ref{spclass} for details).  Spectra are
getting bluer toward larger values of $\delta$. Galaxies with EW[OII]
$<$ 30 \AA~ are shown as black open circles, those with EW[OII] $\ge$
30 \AA~as magenta filled circles.  The vertical lines at $\delta=-5$
and $\delta=3$ indicate the limits between the 3 spectral classes used
for the calculation of the luminosity functions (see \sct
\ref{types}). {\bf b)} Bottom panel: $\delta$ and $\theta$ parameters
obtained by projection of 26 Kennicutt spectra
\protect\citet{kennicutt92} onto the ESS PCs used in a). To correct
for the systematic color shift in the ESS spectra, a constant offset
of $-2.5^\circ$ is applied to the $\delta$ values for the Kennicutt
spectra (see \sct \ref{error}). The morphological types for the
Kennicutt galaxies provides indications on the type content of the ESS
spectral classes.}
\label{delta_theta}
\end{figure}

The top panel of \fg \ref{delta_theta}a shows the \emph{spectral
sequence} parameterized by $\delta$ and $\theta$ for 603 ESS spectra
with $R_\mathrm{c} \le 20.5$. This graph shows that spectra with
strong [OII]$\lambda$3727 emission line (EW[OII] $\ge$ 30 \AA, magenta
filled circles) tend to deviate from the $\delta-\theta$ sequence
defined by the no or low emission-line galaxies (black open circles),
in the direction of larger values of $\theta$. It also confirms that
there is an increasing frequency of high [OII]-emission for later
spectral types, and that early-type galaxies ($\delta \la -5^\circ$)
have no or weak emission lines. 

The \emph{classification plane} shown in \fg \ref{delta_theta}a is
obtained by restricting the spectra to the rest-wavelength interval
3700--5250 \AA~(a common wavelength interval must be used for
application of the PCA presented in \citealt{galaz98}), which is denoted
$\delta_1-\theta_1$. For the ESS spectra, this wavelength interval
provides the best compromise between having a large sample, and having
a large wavelength coverage which includes a sufficient number of
significant absorption and emission lines ([OII]$\lambda3727$,
[OIII]$\lambda$5007, Ca H\&K $\lambda 3934$, 3968 and Mgb $\lambda
5175$). Among the ESS spectra, 728 galaxies (511 with
$R_\mathrm{c}\le20.5$) have spectra which do cover the primary
wavelength interval 3700--5250 \AA. Most of the remaining galaxies can
be classified using 2 secondary wavelength ranges: 97 galaxies (50
with $R_\mathrm{c}\le20.5$) have spectra covering only the 3700--4500
\AA~interval, and 47 galaxies (42 with $R_\mathrm{c}\le20.5$), the
4500--6000 \AA~interval. We therefore perform 2 additional PCAs, each
using the spectra defined in each of the 2 secondary intervals; these PCAs
provide the $\delta_2-\theta_2$ and $\delta_3-\theta_3$ planes
respectively.

Comparison of the $\delta-\theta$ sequences for spectra covering both
the 3700--5250 \AA~primary interval and one of the 2 secondary
intervals then allows us to project all ESS spectra with a PCA type
onto the reference $\delta_1-\theta_1$ sequence. A total of 568
spectra (corresponding to 513 galaxies, as multiple spectra of
individual galaxies are included) can be projected onto both the
$\delta_1-\theta_1$ and the $\delta_2-\theta_2$ planes.  Note that
\emph{only} spectra observed in spectro-photometric conditions (see
\sct \ref{error}) are used in this projection analysis, with no
$R_\mathrm{c}$ limit. The derived conversion is a linear
transformation
\begin{equation}
\delta_1 = 1.0027 \delta_2 - 0.036,
\label{delta1_delta2}
\end{equation}
close to identity. From the 375 spectra (corresponding to 345
galaxies) which can be projected onto both the $\delta_1-\theta_1$ and the
$\delta_3-\theta_3$ planes, a third order polynomial transformation is
derived:
\begin{equation}
\delta_1 = 0.0013 \delta_3^3 -0.088 \delta_3^2 + 1.95 \delta_3 + 1.68.
\label{delta1_delta3}
\end{equation}
The residuals in the $\delta$ conversions resulting from the use of
\eqs\ref{delta1_delta2} or \ref{delta1_delta3} are comparable to the
random uncertainties in the measurement of $\delta$ (see
\eq\ref{sigma_delta_theta}).  The values of $\theta$ show no
systematic change from the $\delta_1-\theta_1$ plane to either of the
2 secondary planes. We therefore use
\begin{equation}
\begin{array}{ll}
\theta_1 &= \theta_2\\
\theta_1 &= \theta_3.\\
\label{theta1_theta23}
\end{array}
\end{equation}
\eq \ref{delta1_delta2} is then used to convert $\delta_2$ into
$\delta_1$ for the 97 galaxies which can only be projected onto the
restricted 3700--4500 \AA~interval, and \eq \ref{delta1_delta3} is
used to convert $\delta_3$ into $\delta_1$ for the 47 galaxies which
can only be projected onto the restricted 4500--6000 \AA~interval.

\begin{figure}
  \resizebox{\hsize}{!}
{\includegraphics{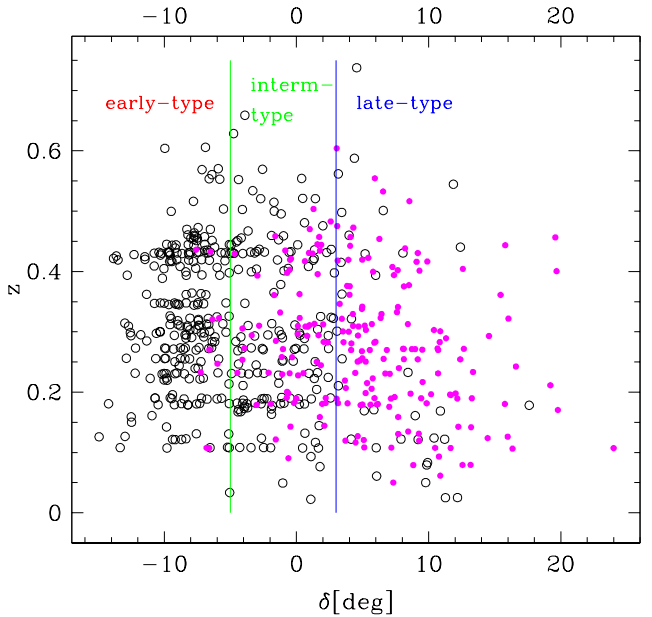}}
\caption{Spectral classification parameter $\delta$ as a function of
redshift for the ESO-Sculptor galaxies with
$R_\mathrm{c}\le20.5$. Galaxies with EW[OII] $<$ 10 \AA~ are shown as
black open circles, those with EW[OII] $\ge$ 10 \AA~as magenta filled
circles. The vertical lines at $\delta=-5^\circ$ and $\delta=3^\circ$
indicate the limits between the 3 spectral classes used for the
calculation of the luminosity functions (see \sct \ref{types}).}
\label{delta_z_OII}
\end{figure}

We emphasize that the rest-frame wavelength interval of each observed
spectrum is determined by (i) the position of the object on the
multi-object mask used for that specific observation, and (ii) the
redshift of the galaxy.  The first constraint affects the rest-frame
wavelength interval randomly, whereas the second causes a systematic
effect.  The 3 wavelength intervals used for application of the PCA
and derivation of the spectral type are therefore systematically
related to the redshift of the galaxies: high redshift galaxies tend
to be only defined in the restricted secondary interval 3700--4500
\AA, whereas low redshift galaxies tend to be preferentially defined
in the other secondary interval, 4500--6000 \AA. This effect can be
measured quantitatively using the mean redshift of the galaxies in
each sample: the galaxies defined in the primary wavelength interval
have $<z>=0.303\pm0.115$, those defined in the 2 secondary intervals
3700--4500 \AA~and 4500--6000 \AA~have $<z>=0.396\pm0.134$ and
$<z>=0.141\pm0.082$ \resp (the \rms dispersion among each considered
sample is indicated). We show below (see \fg \ref{delta_z_OII}) that
despite the relation between rest-wavelength and redshift, conversion
to a unique PCA sequence defined by $\delta_1$ is free from biases in
redshift.

To the remaining 17 galaxies (15 with $R_\mathrm{c}\le20.5$) which
have no PCA type, a spectral class in the $\delta_1-\theta_1$ plane is
assigned based on the relation between $\delta_1-\theta_1$ and the ESS
cross-correlation types. The cross-correlation types are determined by
cross-correlating each ESS spectrum with 6 templates representing an
E, S0, Sa, Sb, Sc, and Irr galaxy \respn; these were obtained by
averaging over Kennicutt spectra of the same morphological type
\citep{kennicutt92}, after discarding MK270, an untypical S0 galaxy
with strong emission lines (a total of 26 Kennicutt spectra, listed in
Table 2 of \citealt{galaz98}, are used).  Among the templates yielding
a cross-correlation peak at \emph{the} redshift of the object, the
cross-correlation type is defined as the morphological type of the
template yielding the highest correlation coefficient
\citep[see][]{bellanger95a}. Using the ESS galaxies with both a PCA
type in the $\delta_1-\theta_1$ plane and a cross-correlation type, we
calculate the median and dispersion of $\delta_1$ and $\theta_1$ for
each of the 6 cross-correlations types. Each of the 17 galaxies
without PCA type is then assigned (i) a randomly drawn value of
$\delta_1$ using a Gaussian probability distribution with the mean and
\rms dispersion measured for the corresponding cross-correlation type,
and (ii) the mean value of $\theta_1$ for that cross-correlation type.

Application of the various transformations described above provides
for each of the 889 galaxies with redshift (617 with
$R_\mathrm{c}\le20.5$) a PCA classification onto the common
$\delta_1-\theta_1$ plane.  Figure \ref{delta_z_OII} shows the type
parameter $\delta_1$ as a function of redshift for all ESS galaxies
with $R_\mathrm{c}\le20.5$. The full redshift range is represented at
all spectral types $\delta$, suggesting the absence of any
obvious bias related to redshift. Note that the major density
variations along the redshift axis are due to large-scale clustering
along the line-of-sight (some higher order variations with $\delta$,
interpreted as segregation effects, are described in
\citealt{lapparent03d}).  Figure \ref{delta_z_OII} thus confirms that
the conversion to a unique spectral sequence $\delta_1$ using the
transformations in \eqs \ref{delta1_delta2} and \ref{delta1_delta3}
above has been successful.

Figure \ref{delta_z_OII} also shows that the various spectral types
$\delta$ are represented at all redshifts.  The defined early-type,
intermediate-type and late-type spectral classes used for derivation
of the LFs below, can therefore be used for examining the variations
of the ESS galaxy populations with redshift
\citep[see][]{lapparent03b}.  Moreover, \fg \ref{delta_z_OII} shows
that galaxies with a significant equivalent width in the
[OII]$\lambda$3727 emission line, defined as EW[OII] $\ge$ 10 \AA,
have preferentially later spectral type $\delta$, and that this
relationship is homogeneous with redshift. This illustrates the
absence of another kind of possible bias: the preferential selection
of emission-line galaxies at the high redshift end of the ESS. This
demonstrates that the adjustment of the spectroscopic exposure times
for the ESS was successful in insuring that the absorption-line
galaxies at the high redshift-end of the survey have spectra with
sufficient signal-to-noise ratio for redshift measurement.

We also use \fg \ref{delta_z_OII} to justify that we do not report nor
discuss the ESS LFs which would be derived from sub-samples based on
the strength of the emission lines. As shown in \fg \ref{delta_z_OII},
the fraction ESS galaxies with EW[OII] $\ge$ 10 \AA~is $3.9$\% in the
early-type class, $32.3$\% in the intermediate-class, and $\sim
80.3$\% in the late-type class (for the $R_\mathrm{c}\le20.5$
sample).  The ESS LFs for the quiescent and star-forming galaxies are
therefore expected to closely resemble the LFs for the early-type and
late-type galaxies \respn, and therefore would not provide any
additional information over that based on the spectral-type LFs described
in the subsequent \sctsn

\subsection{K-corrections     \label{kcor} }  

Calculation of the absolute magnitudes necessary for derivation of the
galaxy LF requires knowledge of the K-corrections.  Historically,
K-corrections have been computed as a function of redshift and
\emph{morphological type} \citep{oke68,pence76,loveday92}, the latter
being based on visual classification.  However, it was shown that the
morphological type is strongly dependent on the expert who performs
the classification \citep{lahav95}. Galaxy classification is also
dependent on the central wavelength of the filter through which the
galaxy is observed \citep{kuchinski01}, and on the image quality
\citep{vandenbergh01}; both are in turn dependent on redshift, and the
latter also depends on seeing. Because K-corrections measure the
change in flux in a given filter caused by the redshifting of the
spectral energy distribution, a more direct and reliable approach for
computing K-corrections is the use of \emph{spectral types}, instead
of morphological types.

Here, we use the ESS PCA spectral classification to calculate
2-dimensional K-corrections as a continuous function $K(z,\delta)$ of
the spectral type $\delta$ and the redshift $z$.  These in turn
provide absolute $B$, $V$, and $R_\mathrm{c}$ magnitudes for the ESS
galaxies. Note that the absolute magnitudes cannot be calculated
directly from the observed spectra because: (i) their
spectro-photometric accuracy ($\sim 7-10\%$) is insufficient, and
$\sim 30\%$ of the spectra have a signal-to-noise ratio below 10; (ii)
the rest-wavelength intervals covered by the $B$, $V$, and
$R_\mathrm{c}$ filters are not always included in the observed
spectra, as it depends on the combination of redshift and position of
objects in the multi-object-spectroscopy mask. As a more robust and
precise alternative, we determine the K-corrections from the
spectrophotometric model of galaxy evolution PEGASE\footnote{``Projet
d'Etude des GAlaxies par Synth\`ese Evolutive.''}  \citep{fioc97}.
The model spectra extend from 2000 \AA~to 10000 \AA, thus allowing us
to derive K-corrections in the $B$, $V$, and $R_\mathrm{c}$ bands up
to $z\simeq 1.0$.

The PEGASE model allows one to generate a set of solar metallicity
spectra with different ages, stellar formation rates (SFR) and initial
mass functions (IMF). Although this feature is proposed in PEGASE, we
do not include in the model spectra any nebular emission line, because
line ratios depend on complex astrophysical conditions (gas densities,
temperatures, etc.) which are not intended to be explored in full
extent in the present analysis. Moreover, inclusion of the emission
lines only change the derived K-corrections by $\sim2\%$ in the most
extreme emission-line galaxies.  We have generated a large set of mock
spectra in the wavelength interval $2000-10000$~\AA~using a Scalo IMF
\citep{scalo86}, and a SFR of the form $\nu G_F$, where $\nu$ is a
constant and $G_F$ the fraction of stellar ejecta available for
further star formation. The adopted values of $\nu$ run from $\nu =
0.02 \times 10^{-3}$M$_{\sun}$Myr$^{-1}$ to $\nu = 10.0 \times
10^{-3}$M$_{\sun}$Myr$^{-1}$, with a typical step of $0.02 \times
10^{-3}$M$_{\sun}$Myr$^{-1}$, and the ages of the spectra vary from
0.01 Myr to 19.0 Gyr.  In order to simplify, we assume that $G_F =
1.0$ (other values do not change significantly the K-corrections). The
resulting set of templates amounts to 438 mock spectra.

For specific derivation of the K-corrections, a PCA of the ESS data is
performed using the observed spectra cleaned from their nebular
emission lines.  The $\delta-\theta$ sequence shown in top panel of
\fg \ref{delta_theta} flattens to a $\delta^\prime-\theta^\prime$
sequence in which $\theta^\prime\sim 0^\circ \pm 2^\circ$ as this
parameter measures the relative strength of the emission-lines; the
values of $\delta^\prime$, the classifying parameter, show no
systematic change: $<\delta-\delta^\prime> = 1.14^\circ\pm 1.3^\circ$
(in both cases, the quoted uncertainty is the \rms dispersion). This
analysis provides the observed PCs, onto which the PEGASE templates
described above are projected, after normalization by their scalar
norm \citep[see][]{galaz98}; a spectral type $\delta^\prime$ is thus
derived for all templates. Each template is then redshifted to all
redshifts between $z = 0$ to $z = 1.0$ using increments $\Delta
z=0.05$.  We finally compute for each of the Johnson $B$, $V$ and the
Cousins $R_\mathrm{c}$ bands, the K-corrections for the mock
spectrum $j$ with a spectral type $\delta^\prime$ and redshift $z$ as
$K_j = K_j(z,\delta^\prime)$ using the K-correction definition
\citep{oke68}:
\begin{eqnarray}
\label{K_corr}
K_j(z,\delta^\prime) = & 2.5\log(1+z) \nonumber \\
& + 2.5\log\left(\frac{\int_0^{\infty}f_j(\lambda)S_j(\lambda)d\lambda}
{\int_0^{\infty}f_j(\frac{\lambda}{1+z})S_j(\lambda)d\lambda}\right),
\end{eqnarray}
where $K_j(z,\delta^\prime)$ is expressed in magnitudes,
$f_j(\lambda)$ is the flux of spectrum $j$ at wavelength $\lambda$,
and $S_j(\lambda)$ is the response curve of the standard filter
\citep[see][]{arnouts97}. For each filter, the $K_j(z,\delta^\prime)$
are then fitted by a 2-D polynomial of degree 3 in $\delta^\prime$
and 4 in $z$, with the constraint that $K_j(z=0,\delta^\prime)=0$.
The derived analytical function $K(z,\delta^\prime)$ allows us to
compute for each observed spectrum its K-correction in each bandpass,
using only its $\delta^\prime$ value and its redshift.

Note that we have not included in the K-correction any evolutionary
correction, corresponding to the possible change of the spectrum
during the interval of time elapsed between the moment of light
emission and the present time. The evolutionary correction would
correct the absolute magnitude to what would be observed at the
present time.  This is however related to the formation age of the
objects, and is strongly model-dependent. The K-corrections derived
here only account for the redshift effect of the spectra, and provide
the absolute magnitudes of the objects at the time of emission (as is
used in most observational analyses).

\begin{figure}
  \resizebox{\hsize}{!}
    {\includegraphics{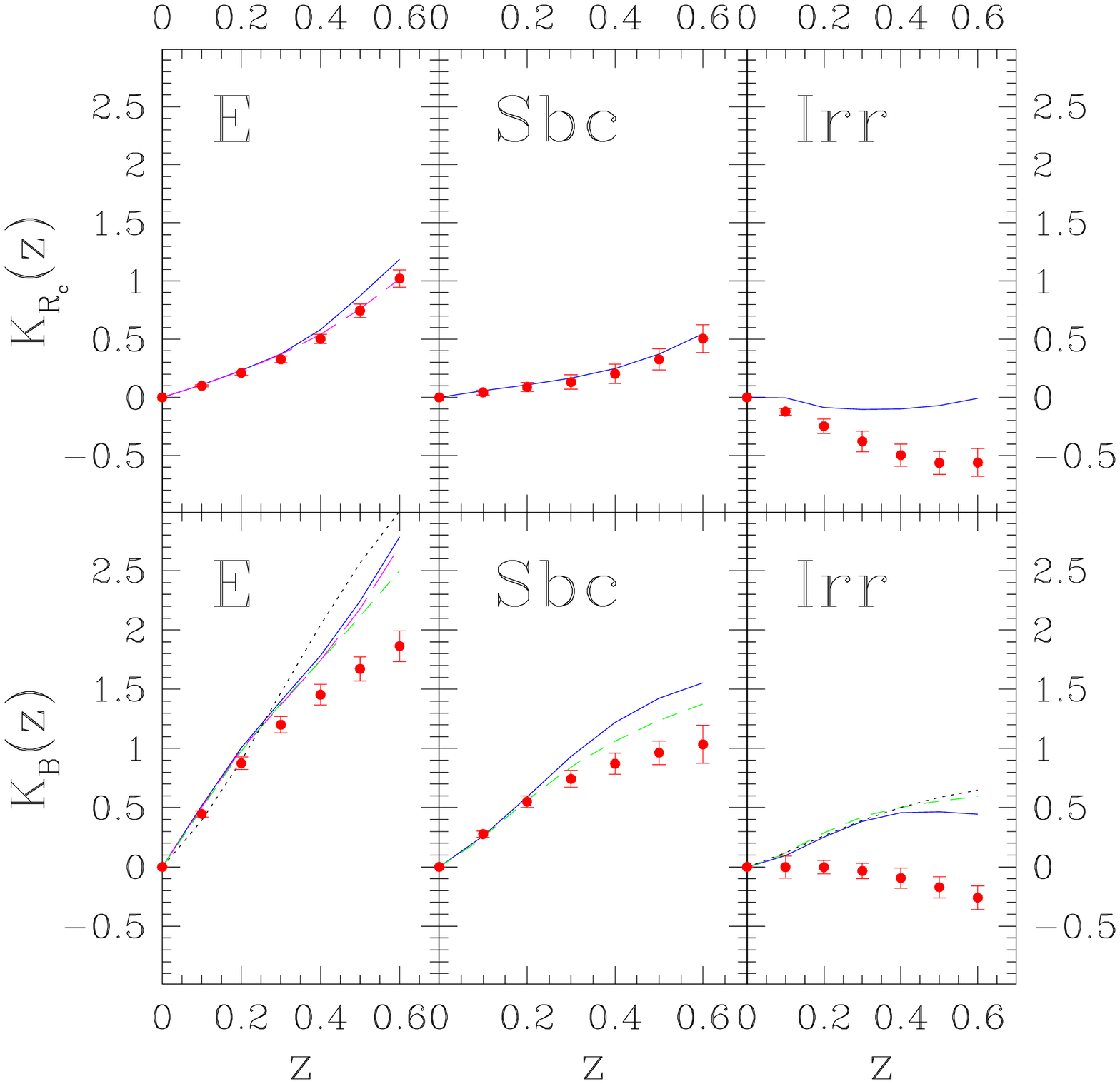}}
\caption{K-corrections obtained from the PEGASE templates
\protect\citep[][red filled circles with error-bars]{fioc97} in the
$R_\mathrm{c}$ and $B$ band, compared with the K-corrections from
\protect\citet[][blue solid lines]{coleman80}, \protect\citet[][black
dotted lines]{kinney96}, \protect\citet[][green short-dashed
lines]{pence76}, and \protect\citet[][magenta long-dashed
line]{poggianti97}; for the $B$ K-corrections by \protect\citet{kinney96}, the
K-corrections for type SB4 are used in panel ``Irr''.}
\label{K_compar}
\end{figure}

In \fg \ref{K_compar}, we show the $B$ and $R_\mathrm{c}$
K-corrections for the PEGASE templates, obtained as described above,
and we compare them with those obtained by other authors from observed
spectra \citep{pence76,coleman80,kinney96} and other
spectrophotometric models \citep{poggianti97}. The three ESS spectral
types included in \fg \ref{K_compar} (E, Sbc and Irr) are computed as
follows: E type is defined by $\delta^\prime \le -5.0^\circ$, Sbc by
$-1.0^\circ \le \delta^\prime \le 5.0^\circ$, and Irr by
$\delta^\prime \ge 10.0^\circ$ (see \fg \ref{delta_theta}a); each
point in \fg \ref{K_compar} represents an \emph{average} K-correction
at a given redshift, and the error-bars represent the \rms dispersion
in the given $\delta^\prime$ intervals. Figure \ref{K_compar} shows
that K-corrections for our templates agree well with the other
measures in the $R_\mathrm{c}$ band (and in the $V$ band, not included
in the graph), but they tend to be smaller in the $B$ band.  In other
words, our PEGASE templates are \emph{bluer} at short wavelengths
($\lambda \la 4200$\AA) than the spectra from which the K-corrections
of \citet{pence76}, \citet{coleman80}, \citet{kinney96}, and
\citet{poggianti97} are derived.  Moreover, the ESS K-corrections for
type Irr tend to be bluer than those from the other authors in all
bands; note that there exists few sources of K-corrections for Irr
types, and most of them are based on the results of
\citet{pence76}. We emphasize that in \fg \ref{K_compar}, we assume a
correspondence between the PCA spectral types for the PEGASE templates
and the Hubble morphological types used for the other measures
mentioned. This correspondence may however not be optimal, which could
explain part of the differences. For example, using $\delta^\prime \ge
3.0^\circ$ for defining Irr galaxies in the ESS (corresponding to the
``late-type'' class described in \sct \ref{types}) eliminates the
discrepancy with the Irr types of \citet{pence76} in the 3 bands.

We have also applied the above analysis to the GISSEL96 models
\citep{charlot96}, using solar metallicity and an instantaneous burst
of star formation. Because the GISSEL96 models have lower fluxes in
the wavelength interval $2000-4000$\AA\ compared with the PEGASE
models, the resulting K-corrections in the $B$ band for all 3 types
(E, Sbc, Irr), and for Irr type in the $V$ and $B$ bands are larger
than the K-corrections derived from PEGASE \citep{galaz97}, thus
providing intermediate values between the K-corrections derived from
PEGASE and those from \citet{pence76}, \citet{coleman80},
\citet{kinney96}, and \citet{poggianti97}. Our choice of using PEGASE
rather than GISSEL96 for estimating the ESS K-corrections is motivated
by the fact that PEGASE models provide a larger sample of templates,
which are not systematically based on an instantaneous burst.  Note
that using the GISSEL96 templates for deriving the ESS K-corrections
would only affect the $B$ and $V$ LFs. However, the major results
derived in this article are based on the LFs in the $R_\mathrm{c}$
band, which is the least affected by changes in the SFR via the
K-corrections (note that in the $V$ band, and to a greater extent, in
the $B$ band, the LFs are also biased by color incompleteness, see
\sct \ref{LFess}).

The K-corrections for the ESS spectra are then calculated according to
the redshift $z$ and spectral type $\delta^\prime$ of each
galaxy. Here, we do \emph{not} need to use a single spectral type
scale for the whole sample, as designed in \sct \ref{spclass} (see
\eqs \ref{delta1_delta2}--\ref{delta1_delta3}), which would introduce
additional dispersion.  The PEGASE templates are projected onto the 3
sets of PCs obtained with the spectra defined in the 3 wavelength
ranges: the primary interval 3700--5250 \AA, and the 2 secondary
intervals 3700--4500 \AA~ and 4500--6000 \AA; the corresponding
spectral classification parameters $\delta_1^\prime$,
$\delta_2^\prime$, and $\delta_3^\prime$ are derived. The polynomial
fits $K_i(z,\delta_i^\prime)$ are calculated for the 3 sets of PCs and
spectral type sequences $\delta_i^\prime$ ($i=1,2,3$).  Then, for each
ESS spectrum, we use its spectral type $\delta_i^\prime$ and the
corresponding polynomial function $K_i(z,\delta_i^\prime)$ to
calculate its K-correction (with $i$ defined by the wavelength range
of the rest-frame spectrum). The absolute magnitude $M$ can
subsequently be derived from the apparent magnitude $m$ and the
redshift $z$ using
\begin{equation}
\label{abs_mag}
M = m - 5\log d_\mathrm{L}(z) - K(z,\delta^\prime) - 25,
\end{equation}
where 
\begin{equation}
\label{d_lum}
d_\mathrm{L}(z) = \frac{c}{H_0\;q_0^2}
\left[zq_0+(q_0-1)(\sqrt{1+2q_0z}-1)\right]
\end{equation}
is the luminosity distance in Mpc \citep{weinberg76}. Throughout the
article, we use $H_0 = 100 h$ km s$^{-1}$ Mpc$^{-1}$ for the Hubble
constant, and $q_0 = 0.5$ (for $\Omega_m=1.0$ and
$\Omega_\Lambda=0.0$).

\begin{figure}
  \resizebox{\hsize}{!}
    {\includegraphics{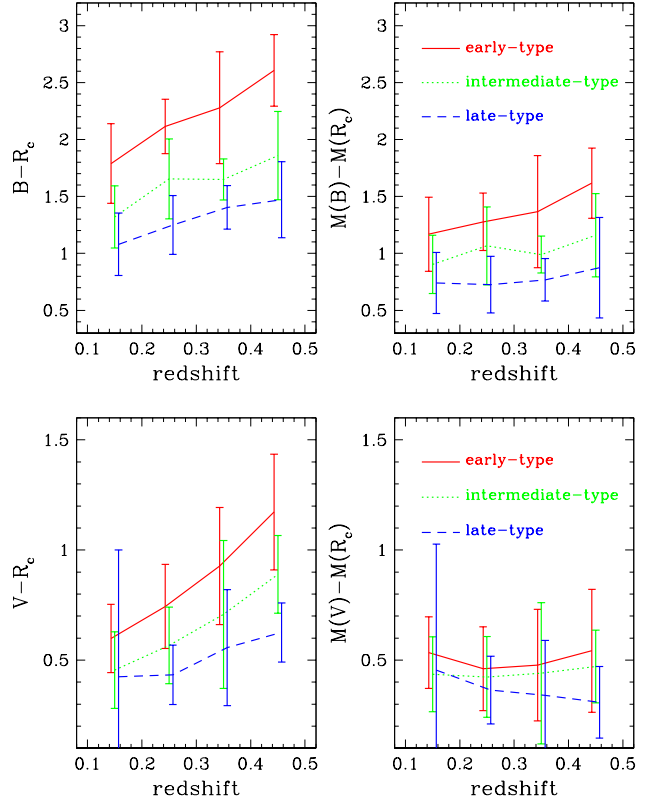}}
\caption{The mean absolute and apparent colors for galaxies with
$R_\mathrm{c}\le20.5$ in each of the 3 ESO-Sculptor spectral classes
used for calculation of the LFs, and in the 4 redshifts intervals
bounded by 0.1, 0.2, 0.3, 0.4, and 0.5; red solid, green dotted, blue
dashed lines connect the points for the early, intermediate, and
late-type classes respectively. The left panels show the apparent
colors and the right panels the absolute colors.  The upper panels
show the $B-R_\mathrm{c}$ colors, and the lower panels, the
$V-R_\mathrm{c}$ colors.  The error bars indicate the \rms dispersion
measured in the corresponding redshift interval. For clarity, the
points for early and late-type galaxies are offset along the
redshift axis by $-0.007$ and $0.007$, respectively.}
\label{abs_col}
\end{figure} 

Figure \ref{abs_col} provides indirect evidence that the
PEGASE/PCA-based K-corrections yield adequate corrections of the ESS
apparent magnitudes into absolute magnitudes.  The left panels of \fg
\ref{abs_col} show the ESS $B-R_\mathrm{c}$ and $V-R_\mathrm{c}$
\emph{apparent} colors. These show significant variations with
redshift, as a result of the redshifting of the spectra. For the
early-type galaxies, for which the effect is the largest, there is a
$0.7 ^\mathrm{mag}$ reddening from $z=0.15$ to $z=0.45$.  In contrast,
the ESS $M(B)-M(R_\mathrm{c})$ and $M(V)-M(R_\mathrm{c})$
\emph{absolute} colors, shown in the right panels of \fg
\ref{abs_col}, display only small variations with redshift.  The $0.3
^\mathrm{mag}$ increase of $M(B)-M(R_\mathrm{c})$ for early-type
galaxies between $z=0.15$ and $z=0.45$ might not be an intrinsic color
effect, as the models of galaxy spectral evolution
\citep{bruzual93,fioc97} indicate little evolution in the interval
$0\la z\la0.5$. This increase could be caused by insufficient (\ie too
low) K-correction in the $B$ band, due to the relatively high flux of
the PEGASE templates at wavelengths in the interval $2000-4000$\AA~(as
discussed above; see also \fg \ref{K_compar}).  The bluing of
$M(V)-M(R_\mathrm{c})$ for the late-type galaxies by $0.15
^\mathrm{mag}$ between $z=0.15$ and $z=0.45$ might be related to the
strong evolution detected in this population
\citep[see][]{lapparent03b}. Overall, the residual variations in
absolute colors with redshift are small, and confirm the reliability
of the ESS K-corrections.

\subsection{Random and systematic uncertainties in the ESS spectral sample \label{error} } 

We now estimate the uncertainties in the ESS parameters
used in this article for the calculation of the LFs: spectral type
$\delta$, K-corrections, absolute magnitudes.  The main source of
error in the absolute magnitudes originate from the
K-corrections. Once the spectral library is chosen (see \sct
\ref{kcor}), the K-corrections are essentially determined by the
spectral classification, which in turn results from the errors in the
flux calibration. Therefore, all mentioned parameters are dependent on
the flux-calibration of the spectra, which we first examine.

The ESS spectra were flux-calibrated using spectro-photometric
standards observed several times per observing night (see
\citealp{galaz98}).  Among the 889 galaxies with a redshift
measurement in the ESS spectroscopic sample (617 with
$R_\mathrm{c}\le20.5$), 606 galaxies have at least 1 spectrum obtained
in spectro-photometric conditions (402 with $R_\mathrm{c}\le20.5$);
for the remaining 283 galaxies (215 with $R_\mathrm{c}\le20.5$), the
single, 2 or 3 spectra of them were observed in either obvious
non-spectro-photometric conditions or suspected as such. Among the 889
galaxies in the ESS spectroscopic sample, 204 of them have double
spectroscopic measurements, and 35 have triple spectroscopic
measurements. These multiple measurements provide 228 pairs of spectra
with each a $\delta_1-\theta_1$ defined spectral type, which we use to
assess our internal random errors. Among them, 102 pairs have both
spectra taken in spectro-photometric observing conditions, and 126
pairs have at least one spectrum taken during a
non-spectro-photometric night.  The resulting \rms dispersion in the
spectral classification, and in the resulting K-corrections and
absolute magnitudes is:
\begin{equation}
\begin{array}{ll}
\sigma(\delta)&\simeq 2.5^\circ \; (2.8^\circ) \\
\sigma(\theta)&\simeq 1.8^\circ \; (2.1^\circ) ;\\ 
\label{sigma_delta_theta} 
\end{array}
\end{equation}
\begin{equation}
\begin{array}{ll}
\sigma(K_{B})&\simeq 0.14 \;\; {\rm mag} \;\; (0.21 \;\; {\rm mag}) \\
\sigma(K_{V})&\simeq 0.11 \;\; {\rm mag} \;\; (0.14 \;\; {\rm mag}) \\
\sigma(K_{R_\mathrm{c}})&\simeq 0.07 \;\; {\rm mag} \;\; (0.11 \;\; {\rm mag}) ; \\
\label{sigma_K} 
\end{array}
\end{equation}
\begin{equation}
\begin{array}{ll}
\sigma(M_{B})&\simeq 0.16 \;\; {\rm mag} \;\; (0.24 \;\; {\rm mag}) \\
\sigma(M_{V})&\simeq 0.13 \;\; {\rm mag} \;\; (0.15 \;\; {\rm mag}) \\
\sigma(M_{R_\mathrm{c}})&\simeq 0.09 \;\; {\rm mag} \;\; (0.12 \;\; {\rm mag}) \\
\label{sigma_M} 
\end{array}
\end{equation}
Note that in the present \sctn, $\delta$ and $\theta$ stand for
$\delta_1$ and $\theta_1$ \respn, as described in \sct \ref{spclass}.
In \eqs \ref{sigma_delta_theta}--\ref{sigma_M}, the first quoted
dispersion is calculated from the 102 pairs of spectra observed in
spectro-photometric conditions, whereas the value in parentheses
indicates the dispersion calculated from the 126 pairs in which at
least one spectrum was taken during a non-spectro-photometric night.
The dispersion is calculated using a $2.5$-$\sigma$ rejection of the
outliers.

We first note that adding in quadrature the $0.05 ^\mathrm{mag}$
uncertainties in the $B$ $V$ and $R_\mathrm{c}$ magnitudes (for $R_\mathrm{c}\la
21.0$; see \citealp{arnouts97}) to the values in \eq \ref{sigma_K}
yield values close to those in \eqs \ref{sigma_M}.  Second, as
expected, the random errors are systematically larger for spectra
which where taken in \emph{non} spectro-photometric conditions. This
sensitivity to the spectro-photometric observing conditions after the
full sequence of data treatment performed to obtain absolute magnitude
testifies on the quality of the ESS spectroscopic data-reduction,
including the flux-calibration stage.  A crude measure of the
uncertainties in the flux calibration is obtained by calculating the
\rms deviation in the ratios of the spectra for each pair; the ratio
of two spectra is measured as the ratio which most deviates from 1 in
the wavelength interval $\sim 4000-9000$~\AA. For the 102 pairs of
spectro-photometric spectra, and for the 126 pairs with at least one
non-spectrophotometric spectrum, the \rms deviation in the ratios is
$\sim 7-10$\% and $\ga$ 10\% respectively.

We also evaluate the contribution to the uncertainties in the absolute
magnitudes caused by the errors in the redshifts. From the 228 pairs
of independent spectra mentioned above, we measure an ``external''
\rms uncertainty of $\sigma\sim0.00055$ in the redshifts, which would
correspond to an uncertainty of $\sim165$ km/s in the recession
velocity at small distances. From \eqs \ref{abs_mag} and \ref{d_lum},
we measure that the contribution from the uncertainty in the redshift
to the absolute magnitude is caused by the luminosity distance term
$d_L$, with a contribution $\sigma(M|d_L)\simeq2.5\sigma(z) f(z)$,
where $f(z)$ varies from $0.99$ at $z=0.1$ to $0.67$ at $z=0.5$.
Therefore, the contribution to the total $\sigma(M)$ from the
uncertainties in the redshifts is $\sigma(M|d_L)\simeq0.0009-0.0014$
for $0.1\le z\le0.5$, which is negligeable compared with the values in
\eq \ref{sigma_M}.

A robust way to evaluate both the random and systematic uncertainties
in the flux calibration for the ESS spectroscopic sample is to
calculate ``spectroscopic colors'' by ``observing'' the spectra
through the standard $B$, $V$, and $R_\mathrm{c}$ filters and compare
them with the photometric colors. This procedure is only possible for
a fraction of the spectra for which the appropriate wavelength range
is available: $\sim300$ spectra for which a $V-R_\mathrm{c}$ color can
be calculated from the redshifted spectra (covering the $4800-8500$
\AA~interval), and another $\sim 300$ spectra for which a rest-frame
$M_{B}-M_{V}$ color can be calculated from the rest-wavelength spectra
(covering the $3600-6500$~\AA~interval).  Because the spectroscopic
colors are a function of the relative normalization of the filter
transmission curves, these colors must be calibrated onto a sequence
of standard stars.  We use the spectra of the CTIO spectro-photometric
standard stars which were originally obtained by \citet{stone83} and
\citet{baldwin84}, and were subsequently re-observed by
\citet{hamuy92} and \citet{hamuy94}. We also use the $B$ $V$ and
$R_\mathrm{c}$ photometry provided by \citet{landolt92b} for these
standard stars. The resulting calibrations are adjusted by linear
regression and the dispersion in the $B-V$ and $V-R_\mathrm{c}$ color
residual is in the range $0.005-0.015 ^\mathrm{mag}$ (which is
negligeable compared with the 0.05 uncertainties in the ESS apparent
magnitudes and to those in \eqs \ref{sigma_M}).

``Spectroscopic colors'' are then calculated from the ESS spectra, and
the resulting mean offset between the photometric and spectroscopic
colors and the dispersion around the mean are:
\begin{eqnarray}
\label{spec_cola}   (V-R_\mathrm{c})_\mathrm{spec} & -\, (V-R_\mathrm{c})_\mathrm{phot} & =\, 0.06 \pm 0.23 \\ 
\label{spec_colb} (M_B-M_V)_\mathrm{spec} & -\, (M_B-M_V)_\mathrm{phot} & =\, 0.10 \pm 0.31 
\end{eqnarray}
When the response curves for the standard filters are taken from other
sources, they result in insignificant changes in \eqs
\ref{spec_cola}--\ref{spec_colb}, thanks to the prior calibration of
the spectroscopic colors with the CTIO standards. Removal of the
atmospheric O$_2$ absorption bands from the spectra, near 6900 \AA~and
7600 \AA, by linear interpolation from the surrounding continuum also
yields insignificant changes in \eqs \ref{spec_cola}--\ref{spec_colb}.

We first consider the dispersion in the color offsets in \eqs
\ref{spec_cola}--\ref{spec_colb}: 0.23 for $V-R_\mathrm{c}$ and $0.31$
for $M_B-M_V$.  The \rms uncertainties of $0.05 ^\mathrm{mag}$ in the
$B$ $V$ and $R_\mathrm{c}$ magnitudes for $R_\mathrm{c}\la 21.0$
represent a negligeable contribution to these values. Part of
dispersion in the color offsets calculated from apparent magnitudes
(\eq \ref{spec_cola}) originates from the random errors in the flux
calibration. As mentioned above, these can contribute by $\ga 0.10$ to
the dispersion in the spectroscopic magnitude, thus by $\ga 0.10
\sqrt{2}\simeq0.14$ to dispersion in the spectroscopic color
$V-R_\mathrm{c}$.  The $0.31 ^\mathrm{mag}$ dispersion in the color
offset for absolute colors (\eq \ref{spec_colb}) is larger than in \eq
\ref{spec_cola} because it includes the dispersion in the
K-corrections (\eq \ref{sigma_K}).

We then examine the systematic offsets between the photometric and
spectroscopic colors themselves, which can be interpreted as a
magnitude scale offset. Because the \rms dispersion in the color
offsets given in \eqs \ref{spec_cola} and \ref{spec_colb}~is measured
over the $\sim 300$ spectra considered in each case, the uncertainties
in the scale offsets are obtained by dividing the dispersion values by
$\sim\sqrt{300}$, which yields $0.013 ^\mathrm{mag}$ and $0.018
^\mathrm{mag}$ respectively.  These are negligeable compared with the
$0.06$ and $0.10 ^\mathrm{mag}$ offsets in \eqs \ref{spec_cola} and
\ref{spec_colb}, making these offsets highly significant.  If we now
assume that the mean scale offsets in \eqs
\ref{spec_cola}--\ref{spec_colb}~originate from a systematic error in
the flux-calibration, \emph{both} offsets are consistent with the
\emph{single} interpretation that the ESS spectra have a 9\% redder
continuum every 1000~\AA~in the wavelength range
$\sim4000-8000$~\AA. Because the effect is present in both the
observed colors (\eq \ref{spec_cola}) and the rest-frame colors (\eq
\ref{spec_colb}), the contribution from the ESS K-corrections to the
color offset must be small -- as these would only affect \eq
\ref{spec_colb}. We suggest that the systematic color offset is
related to the shape of the transmission curves of the various CCDs
used for the multi-object spectroscopic observations: the
spectro-photometric calibrations may have under-corrected the lower
sensitivity in the blue parts of the spectra, a common feature of CCD
detectors.

Note that there may be a contribution to \eqs
\ref{spec_cola}--\ref{spec_colb} from aperture effects: the ESS
spectra were obtained using long slits centered on the galaxies, which
sample a larger fraction of the nuclei of galaxies as compared with
their outer parts. Because color gradients are present in
galaxies of varying types
\citep{segalovitz75,boroson87,vigroux88,balcells94}, and in most cases
correspond to several tenths of a magnitude bluer colors when going
from the central to the outer regions of a galaxy, the spectroscopic
colors may be biased towards redder colors. This effect is likely to
contribute to both the systematic offset and the dispersion in the
difference between the photometric and spectroscopic colors in \eqs
\ref{spec_cola}--\ref{spec_colb}. Here, we cannot however separate the
relative contributions of the intrinsic galaxy color gradients and of
the instrumental response curve; this would require detailed
simulations based on galaxy surface photometry.

Measurement of the (steep) slopes of the PCA classification parameter
$\delta$ as a function of $(V-R_\mathrm{c})_\mathrm{phot}$ and
$(M_{B}-M_{V})_\mathrm{phot}$ for the ESS spectra, allows us to
convert the systematic offsets in \eqs
\ref{spec_cola}--\ref{spec_colb}~into a systematic offset in the
spectral type $\delta$. Both \eqs \ref{spec_cola} and \ref{spec_colb}
yield $\Delta\delta\simeq-2.5^\circ$, which contributes to validating
our interpretation of the systematic color offsets in terms of a
general flux-calibration error affecting all spectra over a wide
wavelength range.  Note that the derived systematic offset in $\delta$
is comparable in absolute value to the random error given in \eq
\ref{sigma_delta_theta}, and it is small compared with the wide range
of $\delta$ covered by the galaxy types in the ESS,
$-15^\circ\la\delta\la20$ (see \fg \ref{delta_theta}a). This offset
has the net effect of shifting the ESS spectral sequence towards
earlier-type spectra.  It has the advantage of explaining the apparent
systematic offset between the ESS spectra and the Kennicutt spectra in
\fg 8 of \citet{galaz98}, the latter appearing shifted towards
later-type spectra when projected onto the ESS PCA plane.

The above analysis of the systematic errors in the flux-calibration
therefore indicates that when comparing the ESS $\delta$ spectral
sequence with that for other samples, the values of $\delta$ for the
comparison sample obtained by projection onto the ESS PCs should be
offset by $-2.5^\circ$. If not, ESS galaxies would appear of
earlier-type (too red) compared with other databases. This is used in
the next \sct where we compare the ESS spectral sequence with the
Kennicutt spectra \citeyearpar{kennicutt92}, with the goal to make a
correspondence between the ESS spectral type LFs and the intrinsic LFs
per morphological class.

\subsection{Sub-samples in spectral type \label{types} } 

Although the full sequence of galaxy spectral types are present in the
ESS (see \fg \ref{delta_theta}a), the moderate number of objects in the
survey limits the number of spectral classes which can be analyzed.
We choose to separate the sample into 3 classes defined by
$\delta\le-5.0^\circ$, $-5.0^\circ\le \delta\le3.0^\circ$
$3.0^\circ\le \delta$; the corresponding galaxies are labeled
``early-type'', ``intermediate-type'', and ``late-type''
respectively. These values separate the ESS sample into 3 sub-samples
with comparable numbers of objects in the $R_\mathrm{c}\le20.5$ sample
($\sim 200$ galaxies, see Table \ref{lf_BVR} below), and therefore
allow us to measure the 3 LFs with comparable signal. The 3 samples are
indicated in \fg \ref{delta_theta}a by vertical lines.

Because the PCA spectral classification is continuous, the
$\delta=-5.0^\circ$ and $\delta=3.0^\circ$ boundaries are
arbitrary. A correspondence can nevertheless be made with the Hubble
morphological classification by projecting Kennicutt spectra
\citep{kennicutt92} onto the ESS $\delta-\theta$ sequence: we use the
26 Kennicutt spectra listed in Table 2 of \citet{galaz98}, discarding
MK270, an untypical S0 galaxy with strong emission lines. As discussed
in the previous \sctn, this comparison requires that we offset the
projections of the Kennicutt spectra onto the ESS PCs by
$\Delta\delta=-2.5^\circ$. The resulting Kennicutt spectral sequence
is plotted in \fg \ref{delta_theta}b above, and confirms that the
morphological types vary continuously along the Hubble sequence as
$\delta$ increases, as already shown by \citet{galaz98}.  

Comparison of \fgs \ref{delta_theta}a and \ref{delta_theta}b suggest
that the ESS early-type class contains predominantly E, S0 and Sa
galaxies, the intermediate-type class, Sb and Sc galaxies, and
the late-type class, Sc and Sm/Im galaxies.  The chosen $\delta$
boundaries at $-5^\circ$ and $3^\circ$ therefore make physical sense
as far as differentiating between intrinsically different LFs: they
may help in separating the contributions from the bounded LFs for the
Elliptical, Lenticular and Spiral galaxies, and the unbounded LF for
the Irregular galaxies.

\begin{figure}
  \resizebox{\hsize}{!}
{\includegraphics{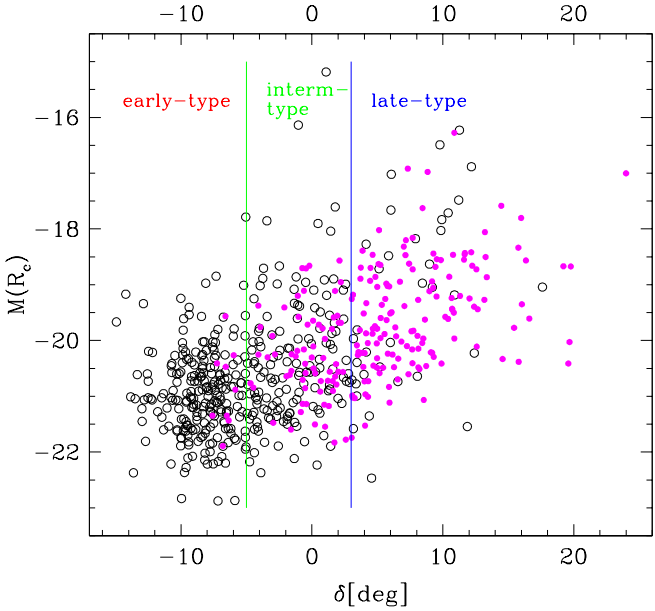}}
\caption{Spectral classification parameter $\delta$ as a function of
absolute magnitude $M(R_\mathrm{c})$ for the ESO-Sculptor galaxies
with $R_\mathrm{c}\le20.5$ sample. Galaxies with EW[OII] $<$ 10 \AA~
are shown as black open circles, those with EW[OII] $\ge$ 10 \AA~as
magenta filled circles. The vertical lines at $\delta=-5^\circ$ and
$\delta=3^\circ$ indicate the limits between the 3 spectral classes
used for the calculation of the luminosity functions (see \sct
\ref{types}).}
\label{delta_mr_OII}
\end{figure}

Figure \ref{delta_mr_OII} shows the ESS absolute magnitude
$M(R_\mathrm{c})$ as a function of the spectral classification
parameter $\delta_1$ for all galaxies with $R_\mathrm{c}\le20.5$. Here,
there is a systematic correlation between spectral-type and luminosity
of the galaxies, with a dimming by nearly $2.0 ^\mathrm{mag}$ from
$\delta\simeq-10^\circ$ to $\delta\simeq10^\circ$: this effect is a
real property of the galaxies which causes the shift of $M^*$ towards
fainter magnitudes for galaxies of later spectral type
(\citealt{bromley98,madgwick02a}; see also \sct \ref{LFess} below).

\section{The shape of the ESS luminosity functions \label{shape}}

\subsection{Method \label{method}}

The ESS shows remarkable clustering in the galaxy distribution
\citep{bellanger95b}.  As far as the determination of the shape of the
LF is concerned, simple methods such as the $1/V_{max}$ method
\citep{schmidt68} are strongly biased by the large-scale structures in
the survey \citep{willmer97}. Instead, one must use statistical
estimators based on ratios of number of galaxies, thus cancelling out
the variations in density with distance. We also use maximum
likelihood estimators which involve the probability that each galaxy
in the survey is observed with its redshift and absolute magnitude.
Two variants are used here: the step-wise maximum likelihood method
(SWML hereafter) developed by \citet{efstathiou88a}, which does not
assume any specific parameterization but requires to bin the data in
steps of absolute magnitude; and the STY method \citep{sandage79},
which does not require to bin in magnitude intervals, but assumes a
specific form for the LF. The SWML and STY solutions both account for
the incompleteness per apparent magnitude interval according to the
prescription by \citet{zucca94}.

Because the ESS spectral-type LFs can be fitted by an exponential
fall-off at bright magnitude and a power-law behavior at faint
magnitudes, we use a \citet{schechter76} parameterization for the STY
fit (but see \sct \ref{LFcomp}).  This function is defined by 3
parameters, $\phi^*$ the amplitude, $L^*$ the ``characteristic
luminosity'', and $\alpha$ which determines the behavior at faint
luminosities:
\begin{equation}
\label{schechter_lum}
\phi(L) dL = \phi^* \left({L\over L^*}\right)^\alpha e^{-{L\over L^*}} d\left({L\over L^*}\right)
\end{equation}
Rewritten in terms of absolute magnitude, \eq\ref{schechter_lum} becomes:
\begin{equation}
\begin{array}{ll}
\label{schechter_mag}
\phi(M) dM & = 0.4 \ln 10\; \phi^* e^{-X} X^{\alpha+1}\; dM \\
{\rm with}& \\
X & \equiv {L\over L^*} = 10^{\;0.4\,(M^* - M)} \\
\end{array}
\end{equation}
where $M^*$ is the ``characteristic magnitude''.

The performances of the SWML and STY techniques, and various other
methods for deriving the LF have been tested on simulated samples by
several authors \citep{willmer97,takeuchi00}. We refer the reader to
these articles for discussion of the strengths and weaknesses of the
SWML and STY methods. We did verified by application to various
simulations matching the ESS configuration that these estimators are
able to measure the input LF for an ESS-type survey, despite the
large-scale spatial inhomogeneities (with the accuracy allowed by the
number of galaxies in the sample).  These simulations are mock ESS
surveys with $\sim 240$, $\sim 2400$ and $\sim 24\,000$ points, and
various types of large-scale inhomogeneities characterized by a
modulation of the density in the redshift distribution (variations in
density with position on the sky at constant redshift have no impact
on the luminosity function). In all cases, the measured values of
$M^*$, $\alpha$ and $\phi^*$ differ from the input values by the
expected \rms accuracy from the number of galaxies in the sample.  We
are therefore confident that the LFs measured here are unbiased by the
ESS large-scale structure and other possible numerical effects.

Note that we have not incorporated into our STY fits the uncertainties
in the absolute magnitudes: these can be accounted for by replacing
the Schechter function by its convolved analog under the assumption of
Gaussian errors in the magnitudes (with an \rms dispersion denoted
$\sigma_\mathrm{M}$ hereafter). Several analyses have been performed
for evaluating the effect of the magnitude errors onto the Schechter
parameters \citep{lin96,zucca97,ratcliffe98a}.  For
$\sigma_\mathrm{M}=0.1 ^\mathrm{mag}$, \citet{lin96} find systematic
offsets in the STY Schechter parameters of $\Delta M^*=+0.03
^\mathrm{mag}$ and $\Delta\alpha=+0.03$, for nearly flat LFs with
$\alpha\simeq-1$. \citet{lin97} then show that neglecting photometric
errors with $\sigma_\mathrm{M}\le 0.1 ^\mathrm{mag}$ only biases $M^*$
and $\alpha$ by at most $\Delta M^*=-0.02 ^\mathrm{mag}$ and
$\Delta\alpha=-0.01$.  For $\sigma_\mathrm{M}=0.2 ^\mathrm{mag}$,
\citet{zucca97} measure $\Delta M^*=+0.10 ^\mathrm{mag}$ and
$\Delta\alpha=+0.05$ for $\alpha$ in the range $-0.9$ to $-1.4$.  For
$\sigma_\mathrm{M}=0.22 ^\mathrm{mag}$, \citet{ratcliffe98a} measure
$\Delta M^*=+0.04 ^\mathrm{mag}$ and $\Delta\alpha=+0.10$ for an LF
with $\alpha\simeq-1$. Based on these results, we expect that the
random errors in the ESS absolute magnitudes, which are in the range
$0.09 ^\mathrm{mag}\la\sigma_\mathrm{M}\la0.24 ^\mathrm{mag}$ (\eq
\ref{sigma_M}), would yield systematic offsets $0.03 ^\mathrm{mag}\la
\Delta M^*\la0.10 ^\mathrm{mag}$ and $0.03\la \Delta\alpha\la0.10$.
The random errors in the Schechter parameters for the ESS LFs are in
the range $0.15-0.30$ (see Table \ref{lf_BVR} below) and are thus
larger than these systematic errors.  We therefore neglect the
uncertainties in the absolute magnitudes in the calculation of the STY
solution.

\subsection{The ESS luminosity functions per spectral type \label{LFess} } 

\begin{figure*}
\centering
\includegraphics[width=17cm]{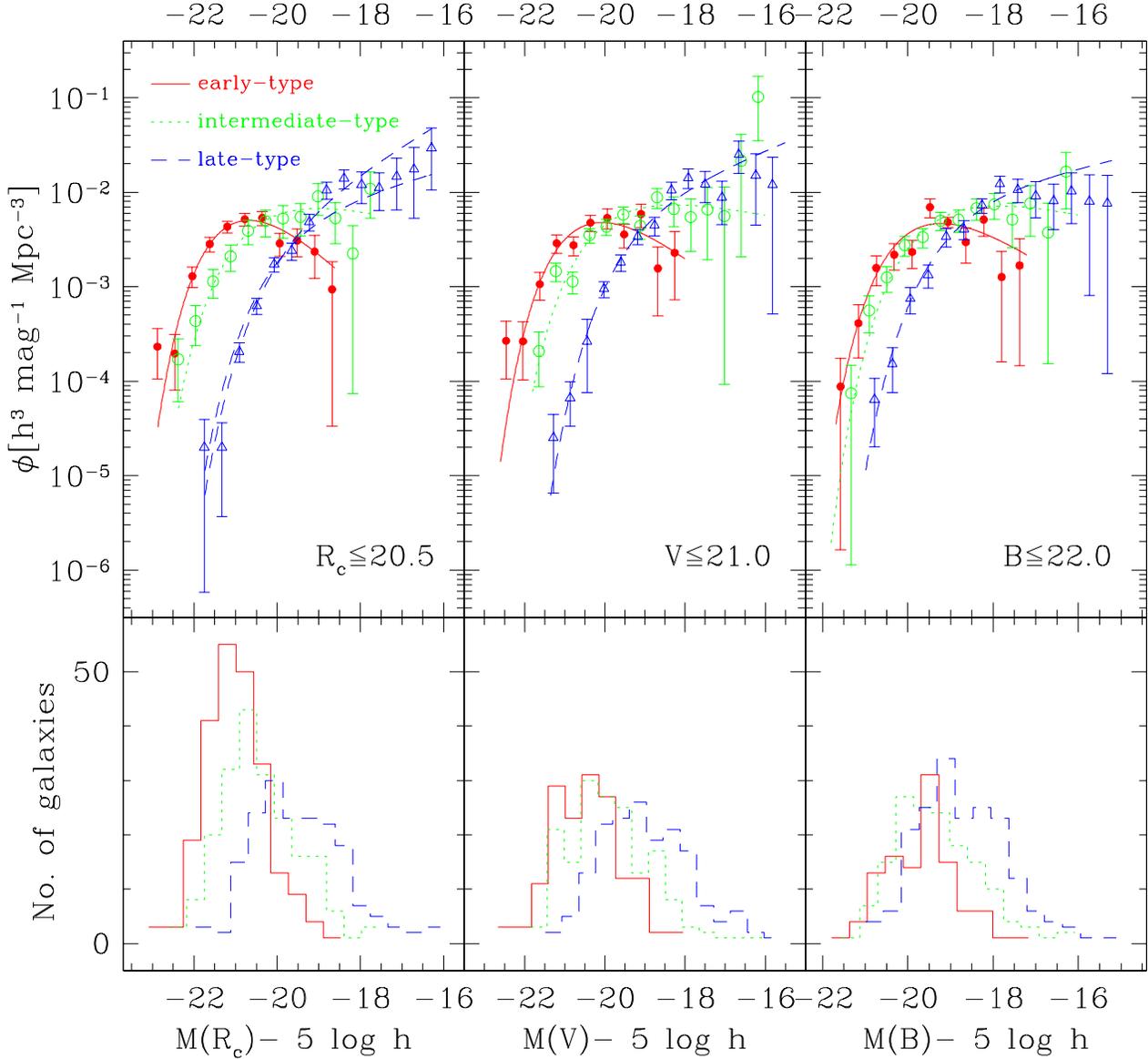}
\caption{The ESO-Sculptor luminosity functions for the early-type,
intermediate-type, and late-type galaxies at the nominal limits in the
3 filters: $R_\mathrm{c}\le20.5$, $V\le21.0$ and $B\le22.0$. Symbols
indicate the SWML solution, and lines the STY solution: early-type are
shown as red filled circles and red solid lines; intermediate-type as
green open circles and green dotted lines; late-type as blue open
triangles and blue dashed lines.  The flatter curve for the late-type
galaxies in the $R_\mathrm{c}\le20.5$ sample has $\alpha=-1.3$ (see
text for details).  The Schechter parameters of the STY solutions are
listed in Tables \ref{lf_BVR} and \ref{phistar_BVR}. The amplitudes of
the SWML points are adjusted by least-square fits to the STY
solutions. The histograms show for each filter the number of galaxies
used in the calculation of the corresponding LFs as a function of
absolute magnitude (with the same line coding as for the STY
solutions).}
\label{lf_filters}
\end{figure*}

Figure \ref{lf_filters} plots the measured LFs for the 3 galaxy types
in each filter, restricted to the nominal limits given in bold face in
Table \ref{comptab}.  The points represent the SWML solution, and the
curves show the STY fit using a Schechter parameterization whose
parameters $M^*$ and $\alpha$ are listed in Table \ref{lf_BVR}.  Figure
\ref{lf_filters} also shows the histograms of absolute magnitude,
which allow one to evaluate how the ESS samples populate the measured
LFs.  Contrary to clusters of galaxies, where all galaxies occupy
approximately the same volume, these histograms cannot be used as
such, as galaxies with fainter magnitudes are detected in shallower
samples.

\begin{table}
\caption{Schechter parameters for the ESO-Sculptor luminosity
functions, in the Cousins $R_\mathrm{c}$ and Johnson $V$ and $B$
filters.}
\label{lf_BVR}
\begin{center}
\begin{tabular}{lcccc}
\hline
\hline
\multicolumn{5}{c}{early-type galaxies} \\
\hline
Sample                       & $N$ & $<\delta>$ &  $M^* - 5 \log h$ & $\alpha$ \\
(1)                          & (2) & (3)        &  (4)              & (5)      \\
\hline
$R_\mathrm{c}\le20.5$ & 232 & -8.469 & $-20.56\pm 0.14$ & $+0.11\pm 0.23$ \\
$R_\mathrm{c}\le21.0$       & 278 & -8.385 & $-20.61\pm 0.14$ & $-0.02\pm 0.20$ \\ 
$R_\mathrm{c}\le21.5$       & 291 & -8.376 & $-20.69\pm 0.14$ & $-0.13\pm 0.17$ \\
                                                              
\hline
$V\le21.0$ & 156 & -8.576 & $-20.26\pm 0.18$ & $-0.16\pm 0.24$ \\
$V\le21.5$       & 210 & -8.497 & $-20.08\pm 0.14$ & $+0.03\pm 0.22$ \\
$V\le22.0$       & 266 & -8.420 & $-20.08\pm 0.15$ & $-0.37\pm 0.21$ \\
$V\le22.5$       & 285 & -8.379 & $-20.05\pm 0.15$ & $-0.40\pm 0.20$ \\

\hline
$B\le22.0$ & 108 & -8.511 & $-19.52\pm 0.24$ & $-0.24\pm 0.33$ \\
$B\le22.5$       & 150 & -8.448 & $-19.50\pm 0.22$ & $-0.55\pm 0.29$ \\
$B\le23.0$       & 204 & -8.397 & $-19.18\pm 0.21$ & $-0.74\pm 0.31$ \\
$B\le23.5$       & 240 & -8.400 & $-19.07\pm 0.22$ & $-0.97\pm 0.30$ \\
\hline

\multicolumn{5}{c}{intermediate-type galaxies} \\
\hline
Sample                      & $N$ & $<\delta>$ &  $M^* - 5 \log h$ & $\alpha$ \\
\hline
{\bf $R_\mathrm{c}\le20.5$} & 204 & -1.082 & $-20.43\pm 0.17$ & $-0.73\pm 0.19$ \\
$R_\mathrm{c}\le21.0$       & 247 & -0.995 & $-20.63\pm 0.17$ & $-0.80\pm 0.15$ \\
$R_\mathrm{c}\le21.5$       & 270 & -1.006 & $-20.85\pm 0.19$ & $-1.07\pm 0.13$ \\

\hline
{\bf $V\le21.0$} & 169 & -0.848 & $-19.96\pm 0.18$ & $-0.79\pm 0.19$ \\
$V\le21.5$       & 216 & -0.931 & $-19.94\pm 0.14$ & $-0.58\pm 0.17$ \\
$V\le22.0$       & 249 & -0.987 & $-19.86\pm 0.12$ & $-0.52\pm 0.15$ \\
$V\le22.5$       & 266 & -0.979 & $-19.93\pm 0.12$ & $-0.76\pm 0.14$ \\

\hline
{\bf $B\le22.0$} & 154 & -0.681 & $-19.37\pm 0.20$ & $-0.75\pm 0.21$ \\
$B\le22.5$       & 193 & -0.795 & $-19.37\pm 0.17$ & $-0.69\pm 0.18$ \\
$B\le23.0$       & 225 & -0.850 & $-19.02\pm 0.15$ & $-0.44\pm 0.19$ \\
$B\le23.5$       & 242 & -0.920 & $-19.07\pm 0.16$ & $-0.61\pm 0.17$ \\
\hline

\multicolumn{5}{c}{late-type galaxies}\\ 
\hline

Sample                      & $N$ & $<\delta>$ &  $M^* - 5 \log h$ & $\alpha$ \\
\hline
{\bf $R_\mathrm{c}\le20.5$} & 181 & 8.215 & $-19.84\pm 0.24$ & $-1.64\pm 0.23$ \\
$R_\mathrm{c}\le21.0$       & 268 & 8.549 & $-19.92\pm 0.19$ & $-1.46\pm 0.18$ \\
$R_\mathrm{c}\le21.5$       & 309 & 8.787 & $-20.08\pm 0.21$ & $-1.48\pm 0.16$ \\
\hline

{\bf $V\le21.0$} & 168 & 8.393 & $-19.34\pm 0.23$ & $-1.42\pm 0.22$ \\
$V\le21.5$       & 251 & 8.626 & $-19.44\pm 0.17$ & $-1.22\pm 0.17$ \\
$V\le22.0$       & 293 & 8.653 & $-19.41\pm 0.15$ & $-0.93\pm 0.15$ \\
$V\le22.5$       & 308 & 8.738 & $-19.49\pm 0.14$ & $-0.86\pm 0.14$ \\
\hline

{\bf $B\le22.0$} & 190 & 8.670 & $-19.00\pm 0.20$ & $-1.25\pm 0.20$ \\
$B\le22.5$       & 255 & 8.808 & $-18.95\pm 0.17$ & $-1.06\pm 0.17$ \\
$B\le23.0$       & 279 & 8.825 & $-19.00\pm 0.16$ & $-0.80\pm 0.16$ \\
$B\le23.5$       & 287 & 8.765 & $-18.96\pm 0.15$ & $-0.62\pm 0.14$ \\
\hline

\end{tabular}
\smallskip
\\
\end{center}
\begin{list}{}{}
\item[\underline{Definition of \colsn:}]
\item[1] Limiting magnitude.
\item[2] Number of galaxies in the sub-sample used for computation of the derived LF.
\item[3] Average spectral type $\delta$ for the sub-sample.
\item[4] Characteristic magnitude of the LF obtained by an STY Schechter fit (see \eq \protect\ref{schechter_mag}).
\item[5] Slope at faint magnitudes of the LF obtained by an STY Schechter fit (see \eq \protect\ref{schechter_mag}).
\end{list}
\end{table}

\begin{table}
\caption{Amplitude $\phi^*$ of the LFs in the Johnson $B$, $V$ and Cousins $R_\mathrm{c}$ bands
for the 3 spectral classes in the ESO-Sculptor redshift survey.}
\label{phistar_BVR}
\begin{center}
\begin{tabular}{lccc}
\hline 
\hline 
Sample                 &  early-type        & intermediate-type  & late-type \\ 
                       &  $z\le 0.55$       & $z\le 0.55$        & $0.1\le z \le 0.2$ \\
\hline
$R_\mathrm{c}\le20.5$  &   0.01477          &     0.01361        & 0.00652           \\
$V\le21.0$             &   0.01392          &     0.01366        & 0.00848           \\
$B\le22.0$             &   0.01336          &     0.01416        & 0.01013           \\
\hline
\end{tabular}
\smallskip
\end{center}

\begin{list}{}{}
\item[\underline{Note:}] This table is extracted from Table 2 of \protect\citet{lapparent03b}, to be consulted for details.
$\phi^*$ is in \phiunit.
\end{list}
\end{table}

Table \ref{lf_BVR} also lists the number of galaxies and average
spectral type $<\delta>$ for each sub-sample for which we calculate a
LF: the 3 spectral classes, in the 3 filters $B$ $V$ $R_\mathrm{c}$,
to the nominal magnitude limits (see Table \ref{comptab}) and to
fainter limits.  Note that in the calculation of the LF, a
K-correction is calculated for each galaxy using the individual values
of $\delta^\prime$ and the calculated transformation
$K(z,\delta^\prime)$ described in \sct \ref{kcor} (\eq \ref{K_corr});
the average spectral types $<\delta>$ listed in Table \ref{lf_BVR} are
thus only shown as indicative.

For the SWML points in \fg \ref{lf_filters}, a bin size of $\Delta M =
0.48 ^\mathrm{mag}$ is used in all 3 filters. Note that the SWML
solution is weakly dependent on $\Delta M$ \citep{efstathiou88a},
which we have checked using varying values of $\Delta M$ for the ESS
LFs: smaller or larger bin sizes within a factor of 2 yield similar
curves. For the STY solutions, we set the brightest and faintest
limits to $-23.0$ and $-16.0$ \resp in $R_\mathrm{c}$, $-22.7$ and
$-16.0$ \resp in $V$, $-21.6$ and $-15.0$ \resp in $B$; these bounds
only exclude a couple of galaxies with anomalously bright or faint
absolute magnitude.  Because the amplitudes of both the STY and SWML
solutions are undetermined, we adopt the following: we use for all STY
curves in \fg \ref{lf_filters} the $\phi^*$ values listed in Table
\ref{phistar_BVR} (see \citealt{lapparent03b} for details); then, for
each sample, the SWML points are adjusted by least-square fit to the
STY solutions.  Because the amplitude $\phi^*$ strongly evolves with
redshift for the late-type galaxies \citep[see][]{lapparent03b}, Table
\ref{phistar_BVR} lists for that sample the average amplitude in the
interval $0.1\le z\le 0.2$; in contrast, the integrated estimate of
$\phi^*$ for $z\le0.55$ is used for the early-type and late-type
samples \citep[see][]{lapparent03b}.

\begin{figure}
\resizebox{\hsize}{!}{\includegraphics{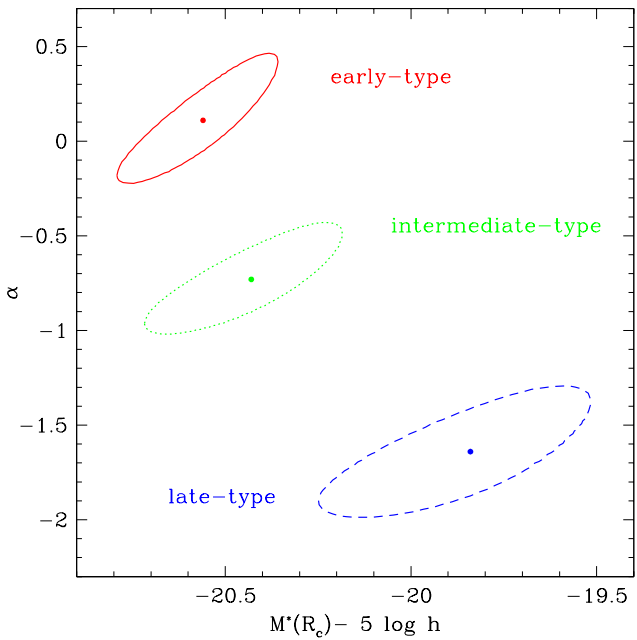}}
\caption{The best-fit parameters (filled circles) and the 1-$\sigma$
error ellipses for $M^*$ and $\alpha$ as derived from the STY fit of
the LFs to $R_\mathrm{c}\le20.5$ for the 3 galaxy classes in the
ESO-Sculptor Survey: early-type (red solid line); intermediate-type
(green dotted line); late-type (blue dashed line).}
\label{ellip}
\end{figure}

Figure \ref{lf_filters} shows that the ESS ``general'' LF is a
composite function of at least 3 different galaxy populations: at
bright magnitudes ($M[R_\mathrm{c}]\la -21$), early-type and
intermediate-type galaxies dominate the population, whereas at the
faint-end, they are outnumbered by the late-type galaxies, which show
a steep increase in number density. The fact that these trends are
observed in all 3 filters $B$ $V$ $R_\mathrm{c}$, suggests that
differences in the LFs between the 3 spectral classes are not due to a
color-dependent effect (such as star formation, for example), but
rather reveal truly different mass distributions for the various
galaxy types.  Figure \ref{ellip} shows the 1-$\sigma$ error ellipses
for the LFs measured at $R_\mathrm{c}\le20.5$ in each of the 3
spectral classes: the error ellipses are well separated, and the slope
$\alpha$ is significantly steeper at more than the 3-$\sigma$ level
from one class to the next, when going from the early-type to the
late-type galaxies.

\begin{figure}
\resizebox{\hsize}{!}
{\includegraphics{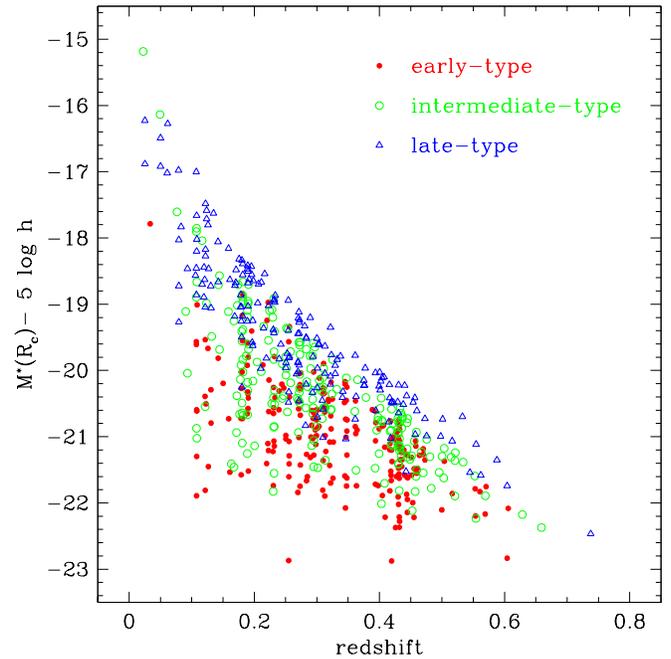}}
\caption{Absolute $R_\mathrm{c}$ magnitudes as a function of redshift
$z$ for the ESO-Sculptor Survey galaxies with $R_\mathrm{c}\le20.5$. This graph shows
how the limit in apparent magnitude biases the range of absolute
magnitudes detected at increasing redshift, and how the varying
K-corrections per spectral type affect the faintest absolute magnitude
reached at a given redshift.}
\label{abs_z}
\end{figure}

We also show in \fg \ref{abs_z}, the distribution of absolute
magnitude $M(R_\mathrm{c})$ versus redshift for the 3 ESS spectral
classes. Although all spectral classes are detected at all redshifts
in the ESS, as shown in \fg \ref{delta_z_OII}, there is a strong
correlation between absolute magnitude and redshift, due to the limit
in apparent magnitude. Figure \ref{abs_z}, shows that at the high
redshift end of the ESS ($z \ga 0.4$), only galaxies brighter than
$M_{R_\mathrm{c}}\simeq-20.0$ can be detected whereas faint galaxies
(with $M(R_\mathrm{c})\ga-18.0$) can only be detected the low redshift
end of the ESS ($z \la 0.15$).  Only galaxies in the magnitude
interval $-22.0 \la M_{R_\mathrm{c}} \la -20.5$ can be observed in the
full ESS redshift range $z\sim 0.1$ to $z\sim 0.5$.  Figure
\ref{abs_z} also shows that the small volume probed at $z \la 0.1$
tends to under-sample the number of galaxies at low levels of the LF:
at $M(R_\mathrm{c})\la-20.0$ and $z \la 0.1$, no ESS galaxies of any
class is detected, as the amplitude of all 3 LFs are below the minimal
threshold for detecting at least one galaxy in the small sampled
volume.

Note that the fainter absolute magnitudes probed by the ESS LF at a
given redshift when going from early-type to late-type in \fg
\ref{lf_filters} are also partly due to the decrease of K-corrections
for later galaxy spectral types (see \fg \ref{K_compar} and \eq
\ref{abs_mag} in \sct \ref{kcor}): the faint bound of the absolute
magnitude distribution is a function of redshift and K-correction and
is defined by replacing $m$ in \eq \ref{abs_mag} with the
$R_\mathrm{c}=20.5$ apparent magnitude limit.  

Table \ref{lf_BVR} shows that for the early-type galaxies, the slope
$\alpha$ at the nominal magnitudes is in the range $-0.24$ to $0.11$
for the 3 filters, which results in a decrease in the number density
of galaxies a faint magnitudes, whereas for the intermediate-type
galaxies, $\alpha$ is close to the value $\alpha=-1$ for a flat
slope,\footnote{$\alpha=-1$ is called a ``flat slope'' because it results
in a constant $\phi(M)$ at faint $M$, see \eq \ref{schechter_mag}.}
and remains nearly constant in all filters at the nominal magnitudes:
$-0.76 \le \alpha \le -0.75$.  In contrast, the faint-end slope for
the late-type galaxies is significantly steeper than for the
early-type and intermediate-type galaxies, and varies at the nominal
magnitudes from $-1.64$ in the $R_\mathrm{c}$ filter, to $-1.25$ in
the $B$ filter.  This corresponds to a steep increase in the number
density of Sc+Sm/Im galaxies at faint magnitudes.

To estimate quantitatively whether the Schechter parameterization is a
good description of each LF, we compare the SWML solution with the STY
fits using the likelihood ratio defined by \citet{efstathiou88a},
which is distributed asymptotically like a $\chi^2$ probability
distribution $P_\phi(\nu)$ with $\nu$ the number of degrees of
freedom in the STY fit. To the nominal magnitude limits in the
$R_\mathrm{c}$, $V$ and $B$ samples, the likelihood ratios are $0.81$,
$0.83$ and $0.73$ \resp for the early-type LFs, $0.75$, $0.71$ and
$0.52$ \resp for the intermediate-type LFs, and $0.46$, $0.44$ and
$0.31$ \resp for the late-type LFs. The high values of the likelihood
ratios for the early-type and intermediate-type classes indicate that
the corresponding Schechter parameterizations are good representations
of these LFs in the 3 photometric bands.

For the late-type galaxies, although the likelihood ratios of the STY
solution remain within the range corresponding to an acceptable fit,
they are systematically smaller than for the early-type and
intermediate-type galaxies in each band. We interpret this effect as
symptomatic of the difficulty to match both the intermediate magnitude
range of the late-type LF ($-20\la M\la-18$ in $R_\mathrm{c}$ and $V$;
$-19\la M\la-17$ in $B$) and the faint end ($M\ga-18.0$ in
$R_\mathrm{c}$ and $V$; $M\ga-17$ in $B$) when using a Schechter
parameterization.  Figure \ref{lf_filters} shows that the faintest 4
points of the SWML solution with $-18\la M\la-16$ in the
$R_\mathrm{c}\le20.5$ late-type LF lie systematically below the STY
fits. The same effect is observed in the $B$ band, but could be partly
due to incompleteness (see \sct \ref{LFfilter} below); we then
restrict the discussion to the $R_\mathrm{c}$ late-type LF.  Because
of the inherent under-sampling of the faint-end of the LF (see above),
the faintest 4 magnitude bins in the late-type SWML solution contain 5
or less galaxies each, and thus poorly constrain the STY fit.  The
steep faint-end slope $\alpha=-1.64\pm0.23$ is therefore determined by
the 93 galaxies in the interval $-20.0\la M(R_\mathrm{c})\la-18.0$.
Ideally, the faint end slope $\alpha$ should be determined by the
faint end points of the SWML solution. We also plot in \fg
\ref{lf_filters} the late-type STY solution with $\alpha=-1.3$, which
corresponds to the flattest slope allowed by the STY fit at the
1-$\sigma$ level (see \fg \ref{ellip}).  Whereas $\alpha=-1.3$ better
matches the 4 faint-end points of the late-type LF, it lies
systematically below the SWML points in the brighter interval $-19\la
M(R_\mathrm{c})\la-18$.  A similar effect is observed for the
late-type LF obtained from the fainter sample $R_\mathrm{c}\le21.5$:
this sample contains 128 additional galaxies, and yields a steep slope
for the STY fit $\alpha=-1.48\pm0.16$ (see Table \ref{lf_BVR}) which
is determined by 169 galaxies with $-20.0\la M(R_\mathrm{c})\la-18.0$
and provides a good visual match to the SWML points in this interval;
the faintest 3 points of the SWML solution (with
$M(R_\mathrm{c})\ga-17.5$) however lie systematically below the STY
solution.  This illustrates the difficulty to fit the ESS late-type
LFs using a single Schechter function.  In \sct \ref{LFlate} we show
that a two-component function (Gaussian + Schechter) provides a better
adjustment.

Figure \ref{lf_filters} also indicates that the bright magnitude
fall-off of the $V$ and $R_\mathrm{c}$ LFs for the late-type galaxies
is fainter than for the early-type and intermediate-type galaxies by
more than $1 ^\mathrm{mag}$. The smaller offset of the LF bright-end
fall-off in the $B$ band can be interpreted as follows.  At the median
redshift $z\sim0.3$ of the ESS, the portions of the galaxy spectra
shifted into the $R_\mathrm{c}$ and $V$ filter correspond
approximately to the $V$ and $B$ region \resp in rest-wavelength. The
measured LFs thus detect the optical parts of the rest-wavelength
spectral energy distribution. In contrast, at $z\sim0.3$, the observed
$B$ band probes the rest-frame spectral energy distribution in the
near UV, which is highly sensitive to star formation; because the
late-type galaxies have higher star formation than the earlier types,
they appear relatively brighter in the $B$ band as compared with the
$R_\mathrm{c}$ and $V$ bands.

Note that in a Schechter parameterization, offsets in the bright-end
fall-off of the LF are poorly measured by the differences in $M^*$.
In \fg \ref{lf_filters}, the magnitude shift between the bright-ends
for the early and late-type LFs is $\sim1.7 ^\mathrm{mag}$ for the
$R_\mathrm{c}\le20.5$ sample, $\sim1.5 ^\mathrm{mag}$ for the
$V\le21.0$ sample, and $\sim1.0 ^\mathrm{mag}$ for the $B\le22.0$
sample (we measure it at $\phi\simeq10^{-3}$\phiuni). In contrast, the
difference in $M^*$ between the early and late-type LF is $0.72$ mag,
$0.92$ mag, and $0.52 ^\mathrm{mag}$ for the $R_\mathrm{c}\le20.5$,
$V\le21.0$, and $B\le22.0$ LFs respectively (see Table
\ref{lf_BVR}). This effect is due to the strong correlation between
the $M^*$ and $\alpha$ Schechter parameters \citep{schechter76}: for
differing values of the slope $\alpha$, $M^*$ shifts to different
parts of the LF and marks differently the fall-off of the
bright-end. This indicates that in a comparison of Schechter LFs, the
difference in $M^*$ must be increased by $0.5$ to $1.0 ^\mathrm{mag}$
to derive the shift in the bright-end between a LF with $\alpha\sim0$
and a LF with $-1.6\la\alpha\la-1.2$. This effect is conveniently
overcome by using Gaussian LFs for the giant galaxies, which have a
well defined peak and \rms dispersion (see \sct \ref{LFcomp}).

\subsection{Variations with filter and magnitude limit \label{LFfilter}}

\begin{table*}
\caption{Comparison of the differences in the Schechter characteristic
magnitudes and the mean absolute colors of the galaxies in the Johnson
$B$, $V$ and Cousins $R_\mathrm{c}$ bands for the 3 spectral classes
in the ESO-Sculptor redshift survey to $R_\mathrm{c}\le20.5$.}
\label{diff_mag}
\begin{center}
\begin{tabular}{lcccc}
\hline 
\hline 
Spectral type          & $M^*(B) - M^*(R_\mathrm{c})$ & $<M(B)-M(R_\mathrm{c})>$ 
                       & $M^*(V) - M^*(R_\mathrm{c})$ & $<M(V)-M(R_\mathrm{c})>$ \\
\hline
early-type             & $1.04\pm0.28$                & $1.4\pm0.4$
                       & $0.30\pm0.23$                & $0.5\pm0.2$   \\
intermediate-type      & $1.06\pm0.25$                & $1.0\pm0.3$
                       & $0.47\pm0.25$                & $0.4\pm0.2$   \\
late-type              & $0.84\pm0.31$                & $0.8\pm0.3$
                       & $0.50\pm0.33$                & $0.4\pm0.4$   \\
\hline
\end{tabular}
\smallskip
\end{center}
\end{table*}

We now discuss how the ESS LFs per spectral-type vary among the
$R_\mathrm{c}$, $V$ and $B$ bands, and with magnitude limit.  
Table \ref{diff_mag} lists the differences $M^*(B)-M^*(R_\mathrm{c})$
and $M^*(V)-M^*(R_\mathrm{c})$ obtained from the LFs parameters measured
at the nominal magnitudes as a function of galaxy spectral type, and
compares them with the mean absolute colors per spectral class for the
galaxies with $R_\mathrm{c}\le20.5$, calculated as the mean
difference between the absolute magnitudes in the 2 considered
filters (see also left panels of \fg \ref{abs_col}, showing the
variations in the absolute colors with redshift). Table \ref{diff_mag}
shows that for a given spectral type, the differences in the
characteristic magnitudes $M^*$ from one filter to another simply
reflect the mean {\rm absolute} colors for the corresponding galaxy
types.

As shown in Table \ref{lf_BVR}, going to deeper magnitude limits than
the nominal values increases the 3 spectral classes by a significant
number of galaxies ($\sim$ 50--100 objects). For the $R_\mathrm{c}$
LF, when going to the fainter limits listed in Table \ref{lf_BVR}, the
STY solution remains remarkably stable, despite the increasing
incompleteness of the spectroscopic samples: the STY fits have
consistent $M^*$ and $\alpha$ values within less than 2-$\sigma$. This
is evidence for robustness of the $R_\mathrm{c}$ LFs, as the number of
early-type, intermediate-type and late-type galaxies increases by
25\%, 32\% and 71\% respectively from the nominal limit to the
faintest limit $R_\mathrm{c}\le21.5$ (the large increase in the number
of late-type galaxies is caused by a strong evolution in this
population, see \citealt{lapparent03b}). Note that the variations of
the LFs with the $R_\mathrm{c}$ magnitude limit provides a good
illustration of the correlation between the 2 shape coefficients of
the Schechter parameterization: when going from $R_\mathrm{c}\le21.0$
to $R_\mathrm{c}\le21.5$, the extreme bright-end bin of the SWML
solution shifts from 1 to 2 galaxies; despite the large error bars,
this causes a brightening of $M^*$ by 0.2 magnitudes; to compensate
and match the SWML points at other magnitudes, $\alpha$ becomes
steeper by $\sim0.35$.

In contrast, the $V$ and $B$ faint spectroscopic samples suffer color
biases which affect the corresponding LFs. Because the completeness of
the spectroscopic catalogue sharply drops to nearly $50$\% at
$R_\mathrm{c}\sim21.5$, the $V$ and $B$ catalogue are biased in favor
of red objects for galaxies at or fainter than the nominal limiting
magnitudes $V\le21.0$ and $B\le22.0$: near these limits, the $V$ and
$B$ spectroscopic catalogues are be deficient in galaxies with bluer
colors than $B-R_\mathrm{c}\simeq1.5$ and $V-R_\mathrm{c} \simeq0.5$
respectively. We measure that the resulting reddening in the observed
$B-R_\mathrm{c}$ and $V-R_\mathrm{c}$ colors beyond the nominal $V$
and $B$ limits varies from $\sim0.15$ to $\sim0.40 ^\mathrm{mag}$
depending on the color and class considered, with, as expected, a
larger value for earlier-type galaxies and in the $B$ band. Because at
fainter limiting magnitudes, one probes more distant objects which are
therefore redder (due to the K-correction), betters estimates of the
color biases are given by the {\it absolute} colors.  Whereas the
average $M(B)-M(R_\mathrm{c})$ colors change by at most $\sim\pm 0.13
^\mathrm{mag}$ when going from $R_\mathrm{c}\le20.5$ to $20.5\le
R_\mathrm{c}\le21.5$ sample, for the 3 spectral classes, the colors
become redder by $0.21-0.24 ^\mathrm{mag}$ for the early-type and
intermediate-type galaxies, when going to fainter limiting magnitudes
in $V$ and $B$ respectively. The effect is smaller for the late-type
galaxies, with a reddening in $M(B)-M(R_\mathrm{c})$ of $0.05
^\mathrm{mag}$ and $0.13 ^\mathrm{mag}$ in the fainter $V$ and $B$
samples respectively. The change in the $M(V)-M(R_\mathrm{c})$ color
when going to fainter magnitudes than the nominal limits are in the
range $-0.06$ to $0.06$ for the 3 filters and 3 spectral types.

Overall, these colors biases are likely to be responsible for the
$0.45 ^\mathrm{mag}$ dimming of the $M^*(B)$ magnitude from the
$B\le22.0$ to the $B\le23.5$ sample for the early-type galaxies; and
for the flattening of $\alpha$ by $\sim0.6$ with nearly constant $M^*$
at fainter $V$ and $B$ magnitudes for the late-type galaxies (the
other variations, for intermediate-type galaxies in the $B$ filter,
and for early-type and intermediate-type galaxies in the $V$ filter,
are smaller and correspond to less that 1-$\sigma$
deviations). Moreover, it is likely that the color biases affecting
the $V$ and $B$ samples cause the flatter slope $\alpha$ for the
late-type $B$ and $V$ LFs as compared with that in $R_\mathrm{c}$:
even at the nominal magnitudes in the $B$ and $V$, these samples are
deficient in the blue galaxies which populate the fainter magnitudes
for late-type galaxies.

\subsection{Comparison with the CNOC2 survey \label{LFother}}

The only comparable survey to the ESS is the CNOC2 (for
``Canadian Network for Observational Cosmology'') redshift survey
\citep{lin99}: as the ESS, the CNOC2 survey is based on medium
resolution spectroscopy from which redshifts and spectral types are
measured.  The ESS and the CNOC2 also are the \emph{only} redshift
surveys providing spectral-type LFs in the $R_\mathrm{c}$ band at
$z\sim0.5$.  The CNOC2 covers $0.692$ deg$^2$ and is limited to
$R_\mathrm{c}\le21.5$.  At variance with the ESS, the CNOC2 spectral
classification is obtained by least-square fit of the
$UBVR_\mathrm{c}I_\mathrm{c}$ colors to those calculated from the
galaxy spectral energy distributions linearly interpolated between the
4 templates of E, Sbc, Scd and Im galaxy types defined by
\citet{coleman80}; the ``early'', ``intermediate'', and ``late''
spectral classes are then defined as corresponding to the E, Sbc, and
Scd+Im templates \citep[see][]{lin99}.  The CNOC2 intrinsic LFs are
measured from 611 early-type, 517 intermediate-type, and 1012
late-type galaxies.

Both the CNOC2 and ESS detect evolutionary effects in their
$R_\mathrm{c}$ LFs \citep{lin99,lapparent03b}.  Here we only consider
the following LFs: for the ESS, the ``average'' LFs for each spectral
type obtained in \sct \ref{LFess}, by calculating the LFs over the
full redshift range of the survey (see Table \ref{lf_BVR}); for the
CNOC2, we use the listed values of $\alpha$, for which no evolution is
detected, and the listed values of $M^*$ at $z=0.3$ by \citet{lin99},
as it nearly corresponds to both the median redshift of the survey and
the peak of the redshift distribution (see \fg 6 of \citealp{lin99};
$z=0.3$ is also close the peak redshift for the ESS).

\begin{figure}
  \resizebox{\hsize}{!}
    {\includegraphics{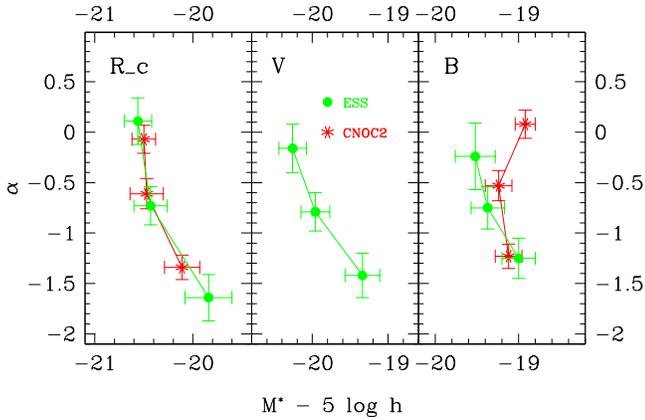}}
\caption{Comparison of the Schechter parameters $M^*$ and $\alpha$ for
the ESO-Sculptor Survey in the $R_\mathrm{c}$, $V$, and $B$ bands with
those from the CNOC2 \protect\citep{lin99}. Galaxies of later spectral
type are in the direction of steeper slopes $\alpha$.}
\label{mstar_alpha_RVB}
\end{figure} 

Figure \ref{mstar_alpha_RVB} plots the $M^*$ and $\alpha$ parameters
for the ESS and the CNOC2 in the $R_\mathrm{c}$, $V$, and $B$
bands. The points for each survey are connected from one class to
the next (red stars for the CNOC2, green filled circles for the ESS).
Left panel of \fg \ref{mstar_alpha_RVB} shows that the values of $M^*$
and $\alpha$ for the CNOC2 $R_\mathrm{c}\le21.5$ sample are in close
agreement with those for the ESS $R_\mathrm{c}\le20.5$ sample at the
1-$\sigma$ level. As in the ESS, the CNOC2 intrinsic LFs show a
steepening in $\alpha$ and a dimming in $M^*(R_\mathrm{c})$ when going
from early to late spectral types, with most of the dimming occurring
between intermediate and late types.  In the next \sctn, we show that
for the ESS, this dimming is a signature of the fainter magnitude
late-type Spiral galaxies (Sc, Sm) detected in the late-type class,
compared with the earlier Spiral types Sa and Sb included in the
early and intermediate-type classes, respectively.

The agreement of the ESS and CNOC2 intrinsic LFs in the $R_\mathrm{c}$
band is a result of the similar morphological content of the spectral
classes: the early, intermediate, and late-type classes contain
predominantly E/S0, Sbc, and Scd/Im \resp in the CNOC2; in the ESS,
they contain E/S0/Sa, Sb/Sc, and Sc/Sm/Im \resp (see \sct
\ref{types}). We further check the similar content of the ESS and
CNOC2 by comparing the relative number of galaxies in each class. At
$R_\mathrm{c}\le20.5$, the ESS early, intermediate and late-type class
contain 38\%, 33\% and 29\% of the galaxies, respectively.  At
$R_\mathrm{c}\le21.5$, the CNOC2 early, intermediate, and late-type
classes contain 29\%, 24\%, and 47\% of the galaxies, respectively.
The 1-mag fainter limiting magnitude in the $R_\mathrm{c}$ band for
the CNOC2, and the detected evolution in the amplitude of the
late-type LFs in both the CNOC2 \citep{lin99} and the ESS
\citep{lapparent03b}, is likely to be responsible for the increase in
the fraction of late-type galaxies in the CNOC2 compared with the
ESS. For direct comparison with the CNOC2, we estimate the expected
fraction of ESS galaxies per spectral class at $R_\mathrm{c}\le21.5$
as follows: in each of the 3 spectral classes lying in the 2 magnitude
intervals $20.5< R_\mathrm{c}\le21.5$ and $20.5< R_\mathrm{c}\le21.5$,
we correct the number of galaxies with a redshift measurement by the
incompleteness in that magnitude interval (given in parenthesis in
Table \ref{comptab}). This assumes that the incompleteness is
independent of spectral class beyond the $R_\mathrm{c}\le20.5$ nominal
limit, which is plausible as the observed galaxies beyond the nominal
limit where chosen on the basis of total luminosity and crowding on
the multi-object masks. The lower success rate in measuring redshifts
for low signal-to-noise absorption-line spectra compared with
emission-line spectra of similar signal-to-noise ratio might bias the
galaxies with measured redshifts toward later spectral type; this is
however a small effect, which we ignore here.  The resulting estimated
fractions of ESS galaxies per spectral class at $R_\mathrm{c}\le21.5$
are: 27\%, 30\%, and 43\% for early, intermediate, and late-type
respectively. The $\sqrt{N}$ uncertainties in the ESS and CNOC2
fractions are $1-2$\% (taking into account 2-point clustering would
slightly increase these uncertainties).  The CNOC2 and ESS early-type
classes therefore contain a consistent fraction of galaxies.  In
contrast, the CNOC2 intermediate-type class contains fewer galaxies
than in the ESS, whereas the opposite is true for the late-type
class. This suggests that the CNOC2 late-type class includes galaxies
of earlier type than in the ESS late-type class. This might explain
why the late-type LF for the CNOC2 has a flatter $\alpha$ and brighter
$M^*$ than in the ESS (see \fg \ref{mstar_alpha_RVB}).

There are 2 other surveys providing estimates of intrinsic LFs at
$z\sim0.5$ in a red filter: the sample of field galaxies extracted
from the CNOC1 cluster survey \citep{lin97}, based on $gr$ photometry
in the \citet{thuan76} system, in which the intrinsic LFs are derived
from 2 color sub-samples; and the COMBO-17 survey \citep{wolf03},
based on the $r^*$ band \citep{fukugita96}, in which LFs are measured
for 4 spectral classes.  The results from these 2 surveys, and those
from 3 other surveys at smaller redshifts
\citep{lin96,brown01,nakamura03} are analyzed in \citet{lapparent-lc},
which provides an exhaustive comparison of all estimates of intrinsic
LFs in the optical bands derived from surveys ranging from
$z\simeq0.03$ to $z\simeq0.6$. The analysis of \citet{lapparent-lc}
includes surveys in which the intrinsic LFs are based on either
spectral classification, morphological type, rest-frame color, or
strength of the emission-lines.

In the Johnson $V$ band, the ESS provides the \emph{first} estimates
of intrinsic LFs at $z\sim0.3$. The corresponding Schechter parameters
are plotted in the middle panel of \fg \ref{mstar_alpha_RVB}, and show
the similar dimming in $M^*$ and steepening in $\alpha$ for later
types as detected in the $R_\mathrm{c}$ band. The only other existing
measurements in the $V$ band are those provided by the Century Survey
\citep{brown01} based on 2 intervals of $V-R_\mathrm{c}$ rest-frame
color; these are compared to the ESS in \citet{lapparent-lc}.

Right panel of \fg \ref{mstar_alpha_RVB} shows the Schechter
parameters for the ESS and CNOC2 LFs in the $B$ band. For the CNOC2,
we have converted the listed values of $M^*(B_\mathrm{AB})$ for
$z=0.3$ and $q_0=0.5$ into the Johnson $B$ band using
$B-B_\mathrm{AB}=0.14$ \citep[see][]{fukugita95}.  The CNOC2 $B$ LFs
are based on samples with nearly identical numbers of galaxies as in
the $R_\mathrm{c}$ filter.  The $B$ band intrinsic LFs for the 2
surveys also show the steepening in $\alpha$ from the early to the
late-type classes. The agreement between the CNOC2 and ESS $B$ LFs is
however not as good as in the $R_\mathrm{c}$ band, with a
$\sim2$-$\sigma$ difference between the $M^*$ values for the
early-type LFs.  This could be caused by the incompleteness of the ESS
$B$ samples due to the $R_\mathrm{c}$ selection of the spectroscopic
sample (see \scts \ref{sample} and \ref{LFfilter}).

Several other redshift surveys provide estimates of $B$ LFs to
$z\sim0.5$: the Canada-France Redshift Survey \citep[CFRS][]{lilly95};
the CNOC1 \citep{lin97}; the Norris survey \citep{small97b}; the
Autofib survey \citep{heyl97}; the CADIS \citep{fried01}; and the
COMBO-17 survey \citep{wolf03}.  We refer the reader to
\citet{lapparent-lc}, for comparison of the $B$ LFs among these
surveys and with those measured at lower redshifts.

\section{Composite adjustments of the ESS luminosity functions \label{LFcomp}}

In this section, we derive composite fits of the ESS luminosity
functions per spectral-type by comparison with the LFs per
morphological type measured from local groups and clusters (see
\sct\ref{intro}). This analysis has the advantage of providing clues
on the underlying morphological mix in the ESS spectral classes.

\subsection{The local luminosity functions per morphological type \label{LFlocal}}

Comparing the local LFs to the ESS measurements requires to relate the
extrapolated $B_\mathrm{T}$ magnitudes from the \emph{Third Reference
Catalogue of Bright Galaxies} \citep{rc3} to the Johnson-Cousins
system. To this end, we use the apparent photo-electric magnitudes in
the Johnson $B$ band measured for Virgo cluster galaxies\footnote{This
catalogue was obtained from the VizieR database provided by the ``Centre de
Donn\'ees de Strasbourg'' (CDS; \citealp{vizier}).} by \citet{schroeder96a}.
The resulting $B_\mathrm{T}-B$ distribution as a function of
morphological type has a bell shape with a large dispersion of
$\sim0.2 ^\mathrm{mag}$.  We empirically adopt the values listed in
column $M(B_\mathrm{T})-M(B)$ of Table \ref{conv_BT}, which lies
within the $B_\mathrm{T}-B$ distribution and vary smoothly with
morphological type (between $0.0$ and $0.3$, with a peak for type
Sa). Note that although \citet{schroeder96a} provide apparent colors,
these are close to colors in absolute magnitudes at the small redshift
of the Virgo cluster, hence the notation of absolute color in Table
\ref{conv_BT}.  We also list the Johnson-Cousins $B-V$ and
$B-R_\mathrm{c}$ colors calculated by \citet{fukugita95} at redshift
$z=0$, and deduce by combination with the $B_\mathrm{T}-B$ values the
color transformation from $B_\mathrm{T}$ to the other ESS bands; these
are also listed in Table \ref{conv_BT} as absolute colors. For the ESS
galaxies, derivation of the corresponding apparent colors would
require use of the K-corrections described in \sct \ref{kcor}.

\begin{table*}
\caption{Color terms for converting absolute magnitudes from the $B_\mathrm{T}$ system into the Johnson-Cousins system used in the ESO-Sculptor survey.}
\label{conv_BT}
\begin{center}
\begin{tabular}{lccccc}
\hline
\hline
Type$^{\mathrm{a}}$ & $M(B_\mathrm{T})-M(B)$ & $M(B)-M(V)$ $^{\mathrm{b}}$     & $M(B_\mathrm{T})-M(V)$ & $M(B)-M(R_\mathrm{c})$ $^{\mathrm{b}}$ & $M(B_\mathrm{T})-M(R_\mathrm{c})$ \\
\hline
E                   &  0.10         & 0.96       & 1.06             & 1.57             & 1.67             \\
S0                  &  0.20         & 0.85       & 1.05             & 1.39             & 1.59             \\
Sab                 &  0.30         & 0.78       & 1.08             & 1.34             & 1.64             \\
Sbc                 &  0.20         & 0.57       & 0.77             & 1.09             & 1.29             \\
Scd                 &  0.10         & 0.50       & 0.60             & 1.00             & 1.10             \\
Sm/Im               &  0.00         & 0.27       & 0.27             & 0.58             & 0.58             \\
\hline
Spiral $^{\mathrm{c}}$  & \ 0.20    & 0.57      & 0.77             & 1.09    & 1.29             \\
\hline
\end{tabular}
\smallskip
\\
\end{center}
\begin{list}{}{}
\item[$^{\mathrm{a}}$] Hubble morphological type.
\item[$^{\mathrm{b}}$] From \citet{fukugita95}.
\item[$^{\mathrm{c}}$] The intermediate colors for type Sbc are used.
\end{list}
\end{table*}

\begin{table*}
\caption{Parameters of the Gaussian and Schechter LFs for the different morphological types, derived from local galaxy concentrations.}
\label{Vir_Cen_lf}
\begin{center}
\begin{tabular}{lcccccl}
\hline 
\hline 
Morphological type     & Type for   &\multicolumn{4}{c}{Gaussian $M_0 - 5 \log h$} & Gaussian $\Sigma$ \\  
                       & color term $^{\mathrm{a}}$ & $B_\mathrm{T}$  & $B$                 & $V$                 & $R_\mathrm{c}$      &                         \\
\hline
E                      & E    & $18.33$ $^{\mathrm{b}}$         & $-18.43$  & $-19.39$ & $-20.00$ & $2.15\pm0.36/1.32\pm0.21$ $^{\mathrm{b}}$ \\
S0                     & S0   & $-18.90\pm0.12$ $^{\mathrm{b}}$ & $-19.10$  & $-19.95$ & $-20.49$ & $1.13\pm0.10$ $^{\mathrm{b}}$   \\
Spiral                 & Sbc  & $-18.20\pm0.09$ $^{\mathrm{b}}$ & $-18.40$  & $-18.97$ & $-19.49$ & $1.37\pm0.07$ $^{\mathrm{b}}$           \\
\hline
Sa/Sb                  & Sab  & $-19.6\pm0.2$ $^{\mathrm{c}}$  & $-19.9$ & $-20.7$  & $-21.2$  & $0.9\pm0.1$ $^{\mathrm{c}}$ \\
Sc                     & Sbc  & $-18.5\pm0.2$ $^{\mathrm{c}}$  & $-18.7$ & $-19.3$  & $-19.8$  & $1.2\pm0.1$ $^{\mathrm{c}}$ \\
Sd/Sm                  & Scd  & $-17.1\pm0.2$ $^{\mathrm{c}}$  & $-17.1$ & $-17.4$  & $-17.7$  & $0.8\pm0.1$ $^{\mathrm{c}}$ \\
\hline
                       & Type for   &\multicolumn{4}{c}{Schechter $M^* - 5 \log h$} & Schechter $\alpha$\\  
                       & color term $^{\mathrm{a}}$ & $B_\mathrm{T}$  & $B$                 & $V$                 & $R_\mathrm{c}$      &                         \\
\hline
dE+dS0 (Virgo)         & Sab  & $-17.79\pm0.32$ $^{\mathrm{b}}$  & $-18.09$ & $-18.87$ & $-19.43$ & $-1.33\pm0.06$ $^{\mathrm{b}}$\\
dE+dS0 (Centaurus)     & Sab  & $-18.67\pm4.06$ $^{\mathrm{b}}$  & $-18.97$ & $-19.75$ & $-20.31$ & $-1.68\pm0.56$ $^{\mathrm{b}}$\\
\hline
Im+BCD  (Virgo)        & Sm/Im  & $-16.16\pm0.24$ $^{\mathrm{b}}$  & $-16.16$ & $-16.43$ & $-16.74$ & $-0.31\pm0.18$ $^{\mathrm{b}}$\\
Im+BCD  (Centaurus)    & Sm/Im  & $-17.55\pm3.42$ $^{\mathrm{b}}$  & $-17.55$ & $-17.82$ & $-18.13$ & $-1.35\pm0.79$ $^{\mathrm{b}}$\\
\hline
\end{tabular}
\smallskip
\\
\end{center}
\begin{list}{}{}
\item[$^{\mathrm{a}}$] Galaxy type from which colors from Table \ref{conv_BT} are assigned to the considered class of galaxies,
thus providing the conversion of $M(B_\mathrm{T})$ into the $BVR_\mathrm{c}$ bands.
\item[$^{\mathrm{b}}$] From \protect\citet{jerjen97b}.
\item[$^{\mathrm{c}}$] LF parameters for individual Spiral types are estimated visually from \fg 18 of \protect\citet{sandage85b}.
\end{list}
\end{table*}

Table \ref{Vir_Cen_lf} shows the parameters of the local intrinsic LFs
reported by \citet{jerjen97b} in the $B_\mathrm{T}$ system, along with
the conversion of the LF characteristic magnitudes (Gaussian peak or
Schechter $M^*$) from the $B_\mathrm{T}$ band into the Johnson-Cousins
system using the transformations in Table \ref{conv_BT}.
\citet{sandage85b} were the first to demonstrate that in the Virgo
cluster, the LFs of Elliptical, Lenticular and Spiral galaxies are
bounded at both bright and faint magnitudes. Here, we use the more
recent analysis of \citet{jerjen97b}, which has the advantage of
averaging the LFs for giant galaxies over 3 clusters (Virgo, Fornax,
Centaurus), and thus yields a robust determination of the parametric
forms for these LFs: the S0 and Spiral LFs have Gaussian shapes; the E
LF has a Gaussian shape which is skewed towards fainter magnitudes,
and can be fitted by a Gaussian with a different dispersion at the
bright and faint end \citep{jerjen97b}. Interpretation of the ESS
spectral-type LFs requires to split the Spiral LF into the LFs for
individual Spiral types.  In their \fg 18, \citet{sandage85b} sketch
the LFs for types Sa/Sb, Sc, and Sd/Sm respectively. Because the
authors do not provide the functional forms nor the parameters for
these curves, we have estimated them visually, assuming Gaussian
profiles. The resulting parameters are listed in Table
\ref{Vir_Cen_lf}, and the corresponding curves appear in reasonable
agreement with the histograms for each Spiral type in the Virgo and
Centaurus clusters \citep[see \fg 3 of][]{jerjen97b}.

In contrast, the LFs for dwarf Spheroidal galaxies (dE and dS0) have
an ever increasing LF at the faint end, which is well fitted by a
Schechter function with a steep slope $-1.6\la \alpha \la -1.3$,
depending on the local density
\citep{sandage85b,ferguson91,pritchet99,jerjen00,flint01a,flint01b,conselice02}.
The LF for late-type dwarf galaxies (Im+BCD, where BDC stands for
``blue compact galaxy'') also has a varying behavior depending on the
environment: at magnitudes brighter than $M(B_\mathrm{T})\la-14$, it
may be fitted by Schechter functions with a widely varying slope
$-1.35\la \alpha \la -0.35$. Nevertheless, in all cases considered,
the LF for late-type dwarf galaxies appears to decrease at the
faintest magnitudes with a poorly determined shape
\citep{ferguson89a,jerjen97b,jerjen00}, and to be flatter than the LF
for early-type dwarf galaxies \citep{pritchet99}.
\citet{drinkwater96} confirmed by obtaining redshift measurements in
the Virgo cluster, that the decrease of the late-type dwarf LF at
faint magnitudes is not due to incompleteness (as would be caused by
misidentification of some of the dwarf cluster members with background
galaxies).  Because the measured LFs for early-type and late-type
dwarf galaxies in the Virgo and Centaurus clusters
\citep{sandage85b,jerjen97b} are representative of the range of
results obtained from concentrations of galaxies of varying richness
(see above mentioned references), we only list the results for these 2
clusters in Table \ref{Vir_Cen_lf}. Note that the dE and Im galaxies
largely dominate in numbers over the dS0 and BCD galaxies \respn, in
both the Virgo and Centaurus clusters. The LFs for dE+dS0 and Im+BCD
galaxies therefore essentially describe the LFs for types dE and Im
respectively. In the following, we denote these 2 populations dSph and
dI respectively.

\subsection{Applicability to the ESS luminosity functions \label{ESSlocal}}

Most analyses of the local LFs were performed on galaxy concentrations
of varying richness. A non-exhaustive list, excluding rich clusters
like Coma, contains: the Virgo cluster
\citep{sandage85b,ferguson91,trentham02a}; the Fornax cluster
\citep{ferguson89a,ferguson91}; the Centaurus cluster
\citep{jerjen97b}: the Ursa Major cluster \citep{trentham01}; the
Perseus cluster \citep{conselice02}; the Leo group
\citep{ferguson91,flint01a,flint01b,trentham02b}; the Dorado, NGC
1400, NGC 5044, Antlia groups \citep{ferguson91}; the Coma I, NGC
1407, and NGC 1023 groups \citep{trentham02b}. By studying the
relationship between the measured LF and the richness of a
concentration, \citet{ferguson91}, \citet{trentham02a} and
\citet{trentham02b} have shown that the dwarf-to-giant galaxy ratio is
a increasing function of richness.  Moreover, \citet{binggeli90}
showed from a local wide-angle survey of low surface brightness
galaxies with $M(B)\la-16$, that although dwarf galaxies delineate the
same large-scale structures as the giant galaxies, there is a strong
segregation among dwarf galaxies: dE lie preferentially in
concentrations of galaxies, whereas dI are more dispersed; outside
clusters, dE also tend to be satellites of giant galaxies. Visual
detection in the ESS of numerous ``fingers-of-god'' with densities
corresponding to groups of galaxies suggests that the survey does
contain a large number of groups
\citep{bellanger95b,lapparent03e}. Nearby redshift surveys indicate
that a fraction as large as $\sim30-40$\% of the total number of
galaxies in a redshift survey is expected to lie in groups
\citep{ramella02}. Group and field galaxies in the ESS should
therefore provide significant samples of early-type and late-type
dwarf galaxies \respn, which should in turn produce non-negligeable
contributions to the ESS spectral-type LFs. 

Following the idea that both the early-type and late-type dwarf
galaxies may contribute to the ESS LF, we adjust the ESS spectral-type
LFs in the $R_\mathrm{c}$ band with composite functions suggested
by the local LFs listed in Table \ref{Vir_Cen_lf}: a two-wing
Gaussian for the early-type galaxies, and the sum of a Gaussian and a
Schechter function for the intermediate-type and late-type galaxies.
The parameters of the composite functions adjusted to the ESS are
listed in Table \ref{ess_local}, and are plotted in \fgs
\ref{lf_local} and \ref{lf_local_2}, together with the observed ESS
LFs (SWML points) for early-type, intermediate-type, and late-type
galaxies with $R_\mathrm{c}\le20.5$ (top panels) and
$R_\mathrm{c}\le21.5$ (bottom panels).  The ESS LFs for
$R_\mathrm{c}\le20.5$ are already shown in \fg \ref{lf_filters} (\sct
\ref{LFess}), with ``pure'' Schechter functions fitted to each
curve. Here, we also consider the ESS LFs at $R_\mathrm{c}\le21.5$, as
the fainter limiting magnitude of that sample provides tighter
constraint on the LF component for dwarf galaxies (see \scts
\ref{LFintermediate} and \ref{LFlate}).  

\begin{table*}
\caption{Parameters of the Gaussian and Schechter components of the composite LFs 
fitted to the ESO-Sculptor $R_\mathrm{c}$ LFs.}
\label{ess_local}
\begin{center}
\begin{tabular}{lcccccccc}
\hline
\hline
Sample                  & Morphol. & \multicolumn{3}{c}{Gaussian component} & \multicolumn{3}{c}{Schechter component} & lik.        \\
Type of LF       & type     & $M_0-5\log h$   & $\Sigma$ $^{\mathrm{a}}$  & $\phi_0$ $^{\mathrm{b}}$  & $M^*-5\log h$   & $\alpha$      & $\phi^*$ $^{\mathrm{b}}$   &  ratio      \\
\hline
\multicolumn{9}{c}{\bf early-type galaxies}  \\
\hline
$R_\mathrm{c}\le20.5$:  & \\
Pure Schechter          &      &                 &               &  & $-20.56\pm0.14$ & $\;\;0.11\pm0.23$ & $0.01477$ & $0.81$\\
{\bf 2-wing Gaussian} & E+S0+Sa     & $-20.68\pm0.24$ & $0.76\pm0.12$ & $0.00538$ & & & & $0.89$\\
                        &             &                 & $1.15\pm0.32$ &           & & & &     \\
\hline
$R_\mathrm{c}\le21.5$:  & \\
Pure Schechter          &      &                 &               &  & $-20.69\pm0.14$ & $-0.13\pm0.17$    & $0.01477$ & $0.84$\\
{\bf 2-wing Gaussian} & E+S0+Sa     & $-20.57\pm0.23$ & $0.84\pm0.24$ & $0.00533$ & & & & $0.92$\\
                        &             &                 & $1.37\pm0.36$ &           & & & &     \\
\hline
\multicolumn{9}{c}{\bf intermediate-type galaxies} \\
\hline
$R_\mathrm{c}\le20.5$:  & \\
Pure Schechter          &     &                 &               &  & $-20.43\pm0.17$ & $-0.73\pm0.19$    & $0.01361$ & $0.75$\\
{\bf Gaussian}        & Sb+Sc       & $-19.79\pm0.29$ & $0.88\pm0.17$ & $0.00669$ & & & &     \\
\quad {\bf + Schechter} $^{\mathrm{c}}$ & dSph        &                 &               &  & $-18.85\pm0.33$ & $-1.67\pm0.29$ & $0.00463$ & $0.72$\\
\hline
$R_\mathrm{c}\le21.5$:  & \\
Pure Schechter          &     &                 &               &  & $-20.85\pm0.19$ & $-1.07\pm0.13$    & $0.01361$ & $0.83$\\
Gaussian                & Sb+Sc       & $-20.03\pm0.17$ & $0.92\pm0.13$ & $0.00495$ & & & &     \\
\quad + Schechter $^{\mathrm{c}}$       & dSph        &                 &               &  & $-20.58\pm0.32$ & $-1.49\pm0.32$  & $0.00537$ & $0.78$ \\
{\bf Gaussian}    & Sb+Sc       & $-19.97\pm0.21$ & $0.91\pm0.18$ & $0.00860$ & & & &     \\
\quad {\bf + Schechter} $^{\mathrm{d}}$ & dSph        &                 &               &  & $-18.98\pm0.37$ & $-1.53\pm0.33$  & $0.00934$ & $0.62$ \\
\hline
\multicolumn{9}{c}{\bf late-type galaxies} \\
\hline
$R_\mathrm{c}\le20.5$:  & \\
Pure Schechter          &    &                 &               &  & $-19.84\pm0.24$ & $-1.64\pm0.23$    & $0.00652$ & $0.46$\\
{\bf Gaussian}       & Sc+Sd       & $-18.72\pm0.34$ & $0.86\pm0.14$ & $0.00486$ & & & &   \\
\quad {\bf + Schechter} $^{\mathrm{c}}$       & dI          &                 &               &  & $-17.85\pm0.28$ & $-0.83\pm0.26$ & $0.04394$ & $0.59$  \\
Gaussian                & Sc+Sd       & $-18.72$ F      & $0.86$ F      & $0.00484$ & & & &   \\
\quad + Schechter $^{\mathrm{e}}$ & dI      &                 &               &  & $-17.54\pm0.30$ & $-0.30$ F      & $0.05527$ & $0.41$  \\
Gaussian                & Sd/Sm       & $-17.70       $ & $0.80       $ & $0.00281$ & & & &   \\
\hline
$R_\mathrm{c}\le21.5$:  & \\
Pure Schechter          &     &                 &               &  & $-20.08\pm0.21$ & $-1.48\pm0.16$    & $0.00652$ & $0.51$\\
{\bf Gaussian}        & Sc+Sd       & $-18.86\pm0.29$ & $0.97\pm0.13$ & $0.00440$ & & & &    \\
\quad {\bf + Schechter} $^{\mathrm{c}}$        & dI          &                 &               &  & $-17.50\pm0.26$ & $ 0.39\pm0.21$ & $0.03677$ & $0.61$  \\
Gaussian                & Sc+Sd       & $-18.86$ F      & $0.97$ F      & $0.00392$ & & & &    \\
\quad + Schechter $^{\mathrm{e}}$ & dI       &                 &               &  & $-17.82\pm0.22$ & $-0.30$ F      & $0.04342$ & $0.44$  \\
Gaussian                & Sd/Sm       & $-17.70       $ & $0.80       $ & $0.00287$ & & & &   \\
\hline
\end{tabular}
\smallskip
\\
\end{center}
\begin{list}{}{}
\item[$^{\mathrm{a}}$] For the two-wing Gaussian fits, the parameters
listed in \col labeled $\Sigma$ are $\Sigma_1$ and $\Sigma_2$ respectively.
\item[$^{\mathrm{b}}$] In units of \phiunit.
\item[$^{\mathrm{c}}$] This is the ``iterative fit'' obtained by
iterating over varying values of $\phi_0/\phi^*$ (see text for
details).
\item[$^{\mathrm{d}}$] This fit is obtained with the constraint
$M^*\ge-19.43$ (see text for details).
\item[$^{\mathrm{e}}$] The values of $M_0$ and $\Sigma$ are fixed to
those obtained in the Gaussian+Schechter iterative fits (previous
line), and the value of $\alpha$ is fixed to $-0.30$ (see text for
details).
\end{list}
\end{table*}

For each ESS spectral class, Table \ref{ess_local} recalls the
parameters of the pure Schechter fits listed in Tables \ref{lf_BVR}
and \ref{phistar_BVR}, and then lists the parameters of the composite
fits, denoted ``2-wing Gaussian'' and ``Gaussian+Schechter''.  As for
the pure Schechter fits (see \sct \ref{method}), the composite fits
are obtained using the STY method \citep{sandage79}. The amplitude of
the STY fits plotted in \fgs \ref{lf_local} and \ref{lf_local_2} and
listed in Table \ref{ess_local} are derived by least-square fit
adjustment to the SWML points plotted in \fg \ref{lf_filters}. For the
$R_\mathrm{c}\le21.5$ samples, the same two-step procedure is used as
for the $R_\mathrm{c}\le20.5$ samples: (i) the SWML points are scaled
by least-square adjustment to the pure Schechter STY solution with the
same amplitude $\phi^*$ as for the $R_\mathrm{c}\le20.5$ sample,
listed in Table \ref{phistar_BVR}; (ii) the composite STY fits are
then scaled by least-square adjustment to the scaled SWML points.  We
also list in Table \ref{ess_local} the likelihood ratios for the pure
Schechter fits and the various composite fits.

Note that we only apply the composite fits to the $R_\mathrm{c}$ LFs,
because as shown in \sct \ref{LFfilter}, the LFs in the $B$ and $V$
bands are affected by color incompleteness.  In the following \sctsn,
we justify the choice of the composite functions, and compare the best
fit parameters with those for the local LFs listed in Table
\ref{Vir_Cen_lf}.  We emphasize that the lack of measurement of
intrinsic LFs for \emph{field} galaxies with a statistical quality
comparable to those of \citet{sandage85b} and \citet{jerjen97b} leaves
us with the only option to refer to the group/cluster measurements
listed in Table \ref{Vir_Cen_lf} (we however comment on the sparse
field measurements of \citealt{binggeli90}, see the following \sctsn).

\subsection{The ESS early-type luminosity function \label{LFearly}}

\begin{figure*}
  \resizebox{\hsize}{!}
    {\includegraphics{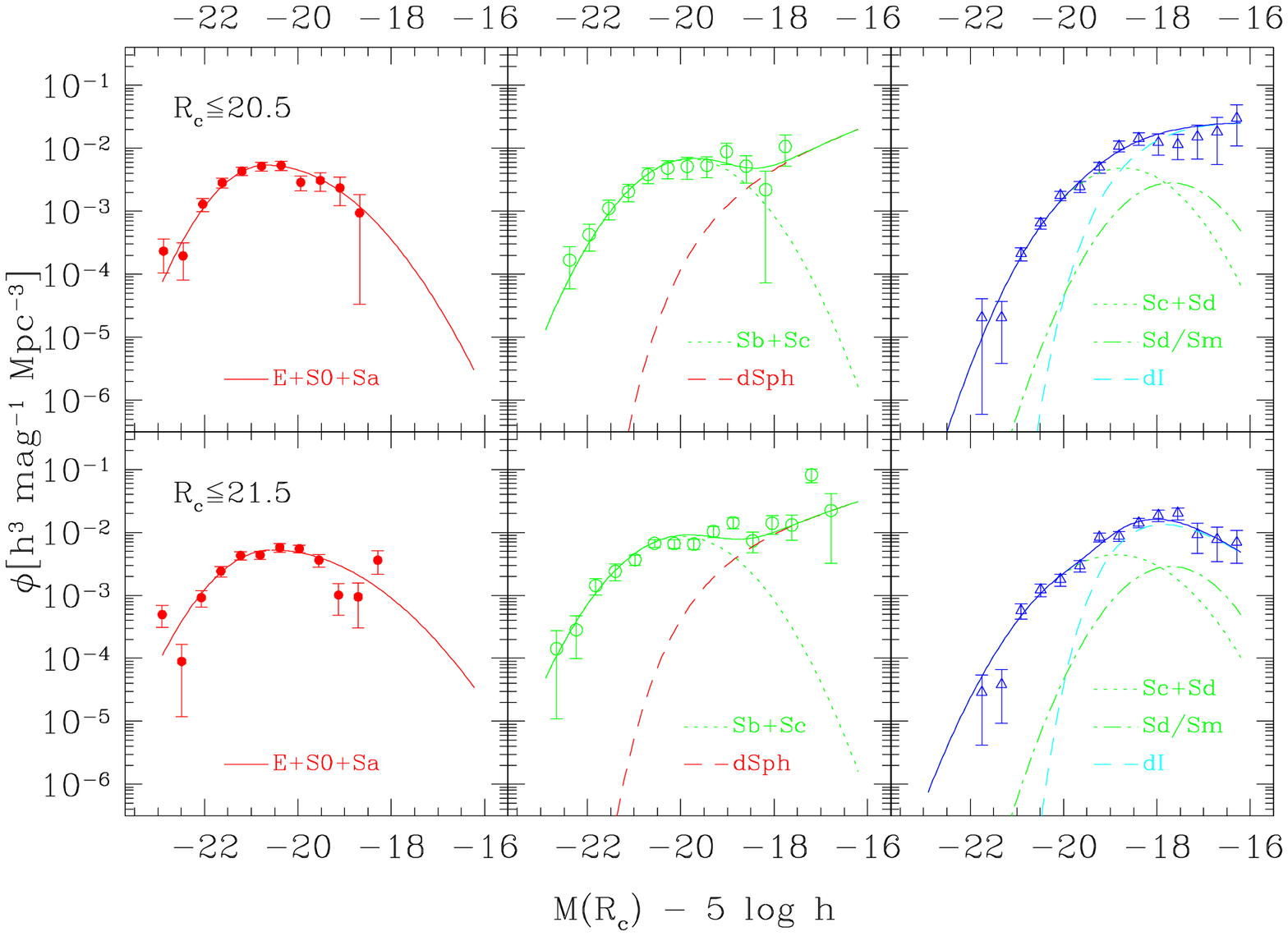}}
\caption{Comparison of the ESO-Sculptor spectral-type luminosity
functions for $R_\mathrm{c}\le20.5$ (top panels) and
$R_\mathrm{c}\le21.5$ (bottom panels) with the composite LFs modeling
the intrinsic LFs (see \protect\citealt{sandage85b} and
\protect\citealt{jerjen97b}).  Left, center, and right panels show the
LFs for the early-type, intermediate-type, and late-type galaxies
respectively. Dotted and dot-dashed lines show the adjusted Spiral
LFs, and dashed lines show the adjusted dwarf galaxy LFs; solid lines
show the composite LFs resulting from the sum of the individual
components, excluding the Sd/Sm component, only shown as indicative
(see text for details). The respective LF parameters are listed in
Table \ref{ess_local} (those for which the ``Type of LF'' is written
in boldface).}
\label{lf_local}
\end{figure*}

As shown in \fg \ref{delta_theta} (\sct \ref{types}), the ESS
early-type spectral class contains predominantly galaxies with E, S0
and Sa/Sab morphological types (see also \sct \ref{spclass}); the
early-type ESS LF can therefore be compared with the sum of the local
LFs for these types. We thus perform the STY fit of a two-wing Gaussian
to the ESS early-type LF, defined as a Gaussian with a different \rms
dispersion ($\Sigma_1$ and $\Sigma_2$) at the faint and bright ends:
\begin{equation}
\label{two_wing_gaussian}
\begin{array}{ll}
\phi(M) dM & = \phi_0 e^{-X} dM \\
{\rm with}& \\
X & = ( M_0 - M ) ^2 / 2 \Sigma_1^2\; {\rm for}\; M\le M_0\\
X & = ( M_0 - M ) ^2 / 2 \Sigma_2^2\; {\rm for}\; M\ge M_0,\\
\end{array}
\end{equation}
For both the $R_\mathrm{c}\le20.5$ and $R_\mathrm{c}\le21.5$ samples,
shown in the top panels of \fg \ref{lf_local}, the two-wing Gaussian
parameterization provides an even better adjustment than with the pure
Schechter function (see \fg \ref{lf_filters} for the
$R_\mathrm{c}\le20.5$ sample), as indicated by the larger likelihood
ratios (see Table \ref{ess_local}). One can also obtain good
adjustments of the ESS early-type LF using the sum of the 2 Gaussian
functions for the S0 and Sa, and a two-wing Gaussian for the E, with
similar peaks but narrower dispersions than for the local LFs in Table
\ref{Vir_Cen_lf}. Such a multi-component parameterization is however
highly degenerate as the relative fractions of E, S0 and Sa in the ESS
are poorly determined from the spectral classification, and because
the 3 Gaussian components have similar peaks and dispersions:
knowledge of the early-type ESS LF is obviously insufficient for
constraining separately the E, S0 and Sa LFs.  We however adopt the
success of a single two-wing Gaussian adjustment, as an indication
that the ESS early-type LF is compatible with a mix of E, S0, and Sa
galaxies having Gaussian LFs.

Note that the local Gaussian LFs for E, S0 and Sa galaxies (shown in
Table \ref{Vir_Cen_lf}) have too large a dispersion to match directly
the ESS early-type LFs, as the two-wing Gaussian (see Table
\ref{ess_local}) cannot be fitted by any combination of the mentioned
local LFs for either the $R_\mathrm{c}\le20.5$ or the
$R_\mathrm{c}\le21.5$ sample.  Although this could partly originate
from evolution and environmental effects, there is a non negligeable
contribution from sampling effects.  At the faint end, the ESS is
limited by its combination of sky coverage and apparent magnitude
limit, which results in a small sampling volume: the local LFs for E,
S0 and Spiral types in Table \ref{Vir_Cen_lf} are defined out to
$M(B_\mathrm{T})\sim-15.5$, that is $M(R_\mathrm{c})\sim-17.0$,
whereas the ESS early-type LF is poorly sampled at $M(R_\mathrm{c})$
fainter than $-19.0$ (see histogram in \fg \ref{lf_filters}).  At the
bright end, the steep exponential decrease of the LF causes an
under-sampling, because of the limited sky coverage of the
survey. Extending the $R_\mathrm{c}\le20.5$ sample to
$R_\mathrm{c}\le21.5$ (which adds 59 early-type, 66 intermediate-type,
and 28 late-type galaxies) is not sufficient to counter-balance this
under-sampling, as the deeper sample is only $\sim 52$\% complete in
redshift measurements (see Table \ref{comptab}). The result of these
combined effects is to skew the ESS early-type LF towards bright
magnitudes. This effect is observed in most magnitude-limited redshift
surveys, and contrasts with the local E LF which is skewed towards
faint magnitudes (see Table \ref{Vir_Cen_lf}). \citet{jerjen97b} also
interpret as incompleteness the early-type LF measured by
\citet{muriel95}, based on the APM survey, which shows a similar
behavior: the low luminosity E are compact and could easily be
misidentified with stars, even on a 2.5-m high resolution Las Campanas
du Pont plate \citep[see][]{jerjen97a}.  Such a bias could also
contribute to a narrow dispersion of the early-type LF in the
ESS. However, there has been so far no detection of a significant
compact population of galaxies which could have been missed in deep
redshift surveys \citep[see for example][]{lilly95}.

\subsection{The ESS intermediate-type luminosity function \label{LFintermediate}}

For the ESS intermediate-type and late-type LFs, the situation is
somewhat different.  The ESS intermediate-type class contains
predominantly Sb and Sc galaxies (see \sct \ref{types} and \fg
\ref{delta_theta}b). \citet{sandage85b} sketch the Sa/Sb and Sc LFs as
2 Gaussian functions with a nearly 1 magnitude brighter peak for the
Sc, and a similar \rms dispersion of $\sim1$ magnitude. Figure
\ref{delta_theta} suggests that in the ESS intermediate-type class,
the Sc are as numerous than the Sb galaxies. Adding to the local Sa/Sb
Gaussian LF (as listed in Table \ref{Vir_Cen_lf}) a contribution from
the Sc local LF would distort the faint end of the Sa/Sb
Gaussian. This would however be insufficient to make the flat
faint-end observed in the ESS intermediate-type LF for both
$R_\mathrm{c}\le20.5$ and $R_\mathrm{c}\le21.5$ (see \fg
\ref{lf_local}). Moreover, examination of \fg 3 of \citet[][based on
\citealt{sandage85b}]{jerjen97b} shows that both the Sb and Sc LFs
decrease to zero galaxies at $M(B_\mathrm{T})\simeq-16$, which
corresponds to $M(R_\mathrm{c})\simeq-17.3$ (using the colors of an
Sbc galaxy listed in Table \ref{conv_BT}), whereas the ESS LF remains
flat out to this limit (see \fg \ref{lf_local}).

Having in mind that there are no dwarf Spiral galaxies in the local
Universe \citep[see][]{sandage85b}, and that dwarf spheroidal galaxies
have bluer colors than giant E galaxies, we propose that the flat
faint-end of the ESS intermediate-type LF is caused by inclusion of
dSph galaxies in this class. Indeed, \citet{caldwell83} suggests that
dE in the Virgo cluster are young and undergo some amount of star
formation indicated by an excess of UV light (the so-called ``UV
upturn phenomenon''): dE with absolute magnitudes $-18\le
M(B_\mathrm{T})\le-16$, that is $-19.7\le M(R_\mathrm{c})\le17.7$ for
the colors of an E galaxy and $-19.3\le M(R_\mathrm{c})\le17.3$ for
the colors of an Sbc galaxy (see Table \ref{conv_BT}), have rest-frame
color $0.6\le U-V\le 1.3$ (see also the similar results of
\citealt{caldwell87} for the Fornax cluster). We thus examine the
colors of the $34$ intermediate-type galaxies in the ESS with
$-19.5\le M(R_\mathrm{c})\le-17.5$. Figure \ref{abs_z} indicates these
galaxies have redshifts in the interval $0.07\le z\le0.25$, with a
median redshift $\sim0.18$. Their apparent colors describe the
interval $0.6\le B-R_\mathrm{c}\le 2.0$, with 73\% of the galaxies in
the interval $0.9\le B-R_\mathrm{c}\le 1.5$.  There is therefore ample
overlap for a population of dE galaxies with rest-frame color $0.6\le
U-V\le 1.3$, as $U-V$ shifts approximately into $B-R_\mathrm{c}$ at
$z\sim0.2$. Independent evidence is brought by the actual spectra of
dE in the Fornax cluster, obtained by \citet{held94}: these spectra
show only a weak or a non-existing break at the location of the H \& K
CaII lines (3933 and 3968 \AA), and display intermediate-color
continua which makes them closely resemblant to Sa and Sb spectra
\citep{kennicutt92}.  If such dwarf Spheroidal galaxies were present
in the ESS, they would be classified as intermediate-type galaxies by
the PCA spectral classification (see \sct \ref{spclass} and
\citealt{galaz98}).

We therefore choose to parameterize the ESS intermediate-type LF
by the sum of a Gaussian LF, modeling the contribution from Sb+Sc
galaxies, and a Schechter component modeling the contribution from
dwarf galaxies. Similarly to the two-wing Gaussian in \eq
\ref{two_wing_gaussian}, the Gaussian LF is defined as
\begin{equation}
\label{gaussian}
\phi(M) dM = \phi_0 e^{-( M_0 - M ) ^2 / 2 \Sigma^2} dM;
\end{equation}
the Schechter LF is defined in \eq \ref{schechter_mag}.  The
Gaussian+Schechter composite LF function has 5 free parameters: the
peak $M_0$ and \rms dispersion $\Sigma$ of the Gaussian, the
parameters $M^*$ and $\alpha$ of the Schechter function, and the ratio
$\phi_0/\phi^*$ of the amplitudes for the 2 functions. 

A general STY fit with all parameters left free is highly unstable and
yields various unrealistic solutions.  We however find that fixing the
value $\phi_0/\phi^*$ is a sufficient constraint for the fit to
converge towards a stable and realistic solution.  We therefore
perform iterative fits in which the ratio $\phi_0/\phi^*$ is fixed to
a series of values separated by some increment; the smallest
increments, used near the maximum of likelihood ratio, are $0.01$. The
best fit is then defined as the STY solution with the largest
likelihood ratio.  In the following, we denote these fits the
``iterative'' STY solutions, or ``iterative fits''.  

The parameters resulting from the iterative fits for the
$R_\mathrm{c}\le20.5$ and $R_\mathrm{c}\le21.5$ samples are listed in
Table \ref{ess_local}, just below the corresponding ``Pure Schechter''
fits.  The iterative composite fits of the ESS intermediate-type LF
provide as good adjustments as the pure Schechter fits: the likelihood
ratios only show a small decrease, from $0.75$ to $0.72$ for the
$R_\mathrm{c}\le20.5$ sample, and from $0.83$ to $0.78$ for the
$R_\mathrm{c}\le21.5$ sample. We have not directly estimated the
uncertainty in the likelihood ratios, but results for fits with
similar LF parameters for the Gaussian and Schechter components
(within 1\%) yield changes in the likelihood ratio by as much a
$0.03$, which provides an underestimate of the true error. The
decrease in the likelihood ratios from the pure Schechter fits to the
iterative fits are therefore within the $\sim1$-$\sigma$ error bars.

In the iterative fit of the intermediate-type LF from the
$R_\mathrm{c}\le21.5$ sample, the value $M^*(R_\mathrm{c})=-20.58$ is
abnormally bright for field dSph galaxies, expected to represent a
significant population in the ESS: as shown by \citet[][see their \fg
10, bottom panel]{binggeli90}, field dSph galaxies might be fainter
than in the Virgo and Centaurus clusters, with
$M(B_\mathrm{T})\ga-17$. We therefore re-run the STY solution for the
$R_\mathrm{c}\le21.5$ sample, with the added constraint that
$M^*(R_\mathrm{c})\ga-19.43$ (the measured value from Virgo, which is
also fainter than for Centaurus, see Table \ref{Vir_Cen_lf}). When
using this constraint on $M^*$, their is no need to perform iterative
fits with varying values of $\phi_0/\phi^*$: leaving all parameters
free yields a stable minimum with $M^*(R_\mathrm{c})=-18.98\pm0.37$
and $\alpha=-1.53\pm0.33$ (other parameters are listed in Table
\ref{ess_local}).  The likelihood ratio decreases to $0.62$, a lower
but still acceptable value. Because the $M^*(R_\mathrm{c})\ga-19.43$
constrained fit to the $R_\mathrm{c}\le21.5$ sample provides shape
parameters for the Gaussian and Schechter components ($M_0$, $\Sigma$,
$M^*$ and $\alpha$) which agree at less than the 1-$\sigma$ level with
those for the iterative fit to the $R_\mathrm{c}\le20.5$ sample (the
uncertainties in 2 measures are added in quadrature in order to
estimate the uncertainty in the difference), we adopt these 2 fits and
plot them in the lower and upper middle panels of \fg \ref{lf_local}
\resp (green dotted lines for the Sb+Sc LF, red dashed line for the
dSph LF); the sum of the Gaussian and Schechter components are plotted
as continuous green lines. The amplitude of each iterative fit is
determined by least-square adjustment to the corresponding SWML
solution (see \sct \ref{LFlocal}).

The values of $M_0$ for the Gaussian component which models the Sb+Sc
contribution to the intermediate-type LF in the $R_\mathrm{c}\le20.5$
and $R_\mathrm{c}\le21.5$ samples, $M_0(R_\mathrm{c})=-19.79\pm0.29$
and $M_0(R_\mathrm{c})=-19.97\pm0.21$ \respn, are both close to that
listed in Table \ref{Vir_Cen_lf} for the Sc galaxies in the
$R_\mathrm{c}$ filter, $M_0(R_\mathrm{c})=-19.8$. Moreover, \fg 3 of
\citet{jerjen97b} shows that the Sb LF may have a similar magnitude
distribution as the Sc LF, in both the Centaurus and Virgo clusters,
whereas the Sa LF has a brighter peak in both clusters. The local
intrinsic LF for Sc galaxies can therefore be used to model the Sb+Sc
LF, thus validating our interpretation of the Gaussian component of
the ESS intermediate-type LF as due to Sb+Sc galaxies. This in turn
suggests that the Spiral galaxies detected in the Centaurus and Virgo
cluster may be representative of those detected in the ESS.

The \rms dispersion $\Sigma$ of the Sb+Sc Gaussian component is
$0.88\pm0.17$ for the $R_\mathrm{c}\le20.5$ sample, and $0.91\pm0.18$
for $R_\mathrm{c}\le21.5$. These 2 values are in good agreement, with
a $0.12$-$\sigma$ difference.  They are however smaller than the
dispersion $\Sigma\sim1.2$ for the local Sc LF (see Table
\ref{Vir_Cen_lf}). As shown in \fg \ref{delta_theta}b, only the Sc
galaxies of earliest spectral type are included in the
intermediate-type class. A narrower dispersion might be expected for
this sub-population. It is also likely that a significant part of the
difference with the ESS dispersion $\Sigma\sim0.9$ results from the
selection effects discussed in \sct \ref{LFearly}, which cause
under-sampling at both the bright and faint ends of the ESS LFs.

The central panels of \fg \ref{lf_local} show that both the
characteristic magnitude $M^*$ and the faint-end slope $\alpha$ of the
dSph Schechter component are poorly constrained by the
intermediate-type LFs, in contrast to the Gaussian component. The
effect on $\alpha$ is more acute for the $R_\mathrm{c}\le20.5$ sample,
as the SWML solution has only few points fainter than the peak of the
Gaussian component.  For the $R_\mathrm{c}\le21.5$, the SWML solution
reaches nearly one magnitude fainter, to $M(R_\mathrm{c})\sim-16.5$,
thus putting tighter constraints on $\alpha$. The differing value of
$M^*$ by $\sim1.6 ^\mathrm{mag}$ obtained for the
$R_\mathrm{c}\le21.5$ sample using the iterative fit and the
$M^*\ga-19.43$ constrained fit \resp (see Table \ref{ess_local})
illustrate the difficulty in constraining $M^*$.

Conversion into the $B_\mathrm{T}$ band of
$M^*(R_\mathrm{c})\simeq-19$, obtained from the iterative fit to the
$R_\mathrm{c}\le20.5$ sample, and from the $M^*\ga-19.43$ constrained
fit to the $R_\mathrm{c}\le21.5$ sample, yields
$M^*(B_\mathrm{T})\simeq-17.3$ (as in Table \ref{Vir_Cen_lf}, we use
the color term for Sab galaxies listed in Table \ref{conv_BT}). We
recall that in the ESS, the LF for the dSph is expected to result from
the combination of the LFs for dSph in groups \emph{and} in the field
(see \sct \ref{ESSlocal}); $M^*(B_\mathrm{T})\simeq-17.3$ is indeed
intermediate between the values for the Virgo and Centaurus clusters
listed in Table \ref{Vir_Cen_lf}, and the fainter value suggested by
field dSph galaxies in the Ursa Major cloud \citep[see \fg 10
of][]{binggeli90}.  We therefore adopt as a likely characterization of
the dSph component included in the ESS intermediate-type LF that
derived from the $R_\mathrm{c}\le21.5$ sample, with
$M^*(R_\mathrm{c})=-18.98\pm0.37$ and $\alpha=-1.53\pm0.33$. Note
however that, even in the $R_\mathrm{c}\le21.5$ sample, the large
uncertainty $\sigma(\alpha)=0.33$ makes the faint-end slope of the
dSph LF component derived from the ESS compatible with those derived
from both the Centaurus and Virgo clusters, at less than the
1-$\sigma$ level.

\subsection{The ESS late-type luminosity function \label{LFlate}}

We propose a similar parameterization for the ESS late-type LF as for
the intermediate-type LF.  Figure \ref{delta_theta} suggests that the
ESS late-type class contains predominantly Sc and Sd/Sm
galaxies. Although the Sc and Sd/Sm populations can be modeled as 2
separate Gaussian LFs with the Sd/Sm LF shifted to fainter magnitudes
(see Table \ref{Vir_Cen_lf} and \citealp{sandage85b}), \fg 3 of
\citet{jerjen97b} shows that the magnitude distribution of the Sd/Sm
galaxies, is included in that for the Sc galaxies.  The contribution
from Sd/Sm galaxies can therefore conveniently be included into the Sc
LF, and we denote Sc+Sd this joint LF.  We then model the ESS
late-type LF as the composite sum of a Gaussian LF for the Sc+Sd
galaxies, and a Schechter function for the Im+BCD galaxies (denoted
dI).  We then show a posteriori that the contribution from Sd/Sm
galaxies to the composite function modeling the ESS late-type LF is
negligeable, as it is dominated at all magnitudes considered by the
contribution from either the Sc or the dI galaxies.

The right panels of \fg \ref{lf_local} show the iterative STY fits of the
Gaussian+Schechter composite LF to the late-type galaxies with
$R_\mathrm{c}\le20.5$ and $R_\mathrm{c}\le21.5$; the Sc+Sd LFs are
shown as dotted lines, and the dI LFs as dashed lines (the
corresponding parameters are listed in Table \ref{ess_local}).  The
increased values of the likelihood ratios ($0.59$ and $0.61$ \respn)
compared with the values for the pure Schechter fits ($0.46$ and $0.51$
\respn) show that the composite fits are better descriptions of the ESS
late-type LFs.  Moreover, the fitted Gaussian peak for the Sc+Sd
component in both the $R_\mathrm{c}\le20.5$ and $R_\mathrm{c}\le21.5$
samples ($-18.72\pm0.34$ and $-18.86\pm0.29$) is remarkably close to
the mean value of $M_0(R_\mathrm{c})$ for the Sc and Sd/Sm local LFs,
$M_0(R_\mathrm{c})=-18.75$ (see Table \ref{Vir_Cen_lf}).  

The measured dispersion of the Sc+Sd Gaussian component is
$\Sigma=0.86\pm0.14$ and $\Sigma=0.97\pm0.13$ for the
$R_\mathrm{c}\le20.5$ and $R_\mathrm{c}\le21.5$ LFs \respn, which
agree at less than $\sim1$-$\sigma$. Values of $1.2\pm0.1$ and
$0.8\pm0.1$ are however listed in Table \ref{Vir_Cen_lf} for the Sc,
and Sd/Sm components respectively. As for the intermediate-type LF
(\sct \ref{LFintermediate}), only part of the Sc galaxies are expected
to be included in the late-type class, those of later spectral-type
(see \fg \ref{delta_theta}a), and this sub-class may have a narrower
dispersion than the full Sc population.  The already mentioned
sampling effects which bias the Gaussian dispersion towards low values
might also affect the ESS late-type LF (see \sct \ref{LFearly}).

The STY composite fits of the late-type LF yield values of
$M^*(R_\mathrm{c})$ for the dI Schechter component which are in
agreement for the $R_\mathrm{c}\le20.5$ and $R_\mathrm{c}\le21.5$
samples ($M^*[R_\mathrm{c}]$ differs by less than $\sim1$-$\sigma$),
with a mean value $M^*(R_\mathrm{c})\simeq-17.7$.  This value is
intermediate between the values for the Virgo and Centaurus cluster
(see Table \ref{Vir_Cen_lf}). Moreover, the faint value
$M^*(R_\mathrm{c})=-16.74$ measured from the Virgo cluster
\citep{jerjen97b} can be excluded: whatever the dispersion of the
Gaussian LF for the Sc+Sd galaxies, and whatever the slope $\alpha$
for the dI component, a faint $M^*(R_\mathrm{c})$ prevents from
adjusting simultaneously the ESS late-type LF in the intervals $-19\le
M(R_\mathrm{c})\le-18$ and $-18\le M(R_\mathrm{c})\le-16$.  Conversion
of $M^*(R_\mathrm{c})\simeq-17.7$ into the $B_\mathrm{T}$ band yields
$M^*(B_\mathrm{T})\simeq-17.1$ (as in Table \ref{Vir_Cen_lf}, we use
the color term for Sm/Im galaxies listed in Table \ref{conv_BT}). This
value appears consistent with that suggested by field dI galaxies in
the Ursa Major cloud \citep[see \fg 10 of][]{binggeli90}.

\begin{figure}
  \resizebox{\hsize}{!}
    {\includegraphics{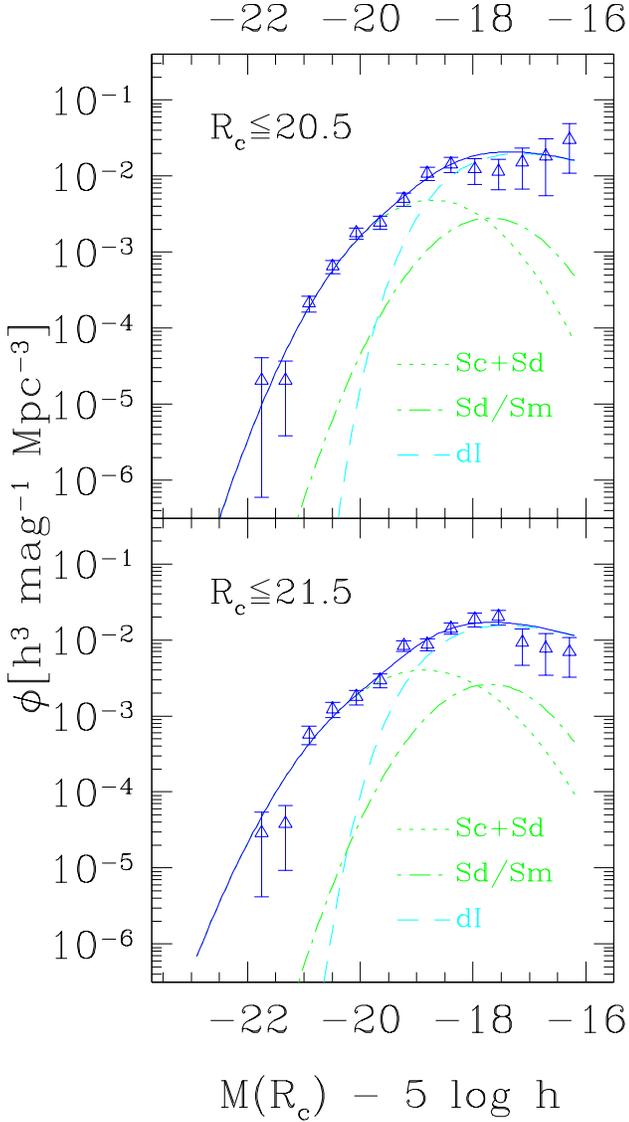}}
\caption{Other composite fits of the ESO-Sculptor late-type
luminosity functions for the $R_\mathrm{c}\le20.5$ and
$R_\mathrm{c}\le21.5$ samples. Dotted and dot-dashed lines show the
Spiral LFs, dashed lines show the dwarf galaxy LFs (dI). Solid lines show
the composite LFs resulting from the sum of the individual components,
except for the Sd/Sm component only shown as indicative (see text for
details).  Here, the Schechter slope $\alpha$ of the dI component is
fixed to $-0.3$, as measured from the Virgo cluster (see Tables
\ref{Vir_Cen_lf} and \ref{ess_local}). The other LF parameters are
listed in Table \ref{ess_local}.}
\label{lf_local_2}
\end{figure} 

In contrast, the values of $\alpha$ for the dI Schechter component of
the ESS late-type LF differ by $3.6$-$\sigma$ for the
$R_\mathrm{c}\le20.5$ and $R_\mathrm{c}\le21.5$ samples: the slopes
are $\alpha=-0.83\pm0.26$ and $\alpha=0.39\pm0.21$ respectively.  The
Centaurus slope $\alpha=-1.35$ (see Table \ref{Vir_Cen_lf}) is too
steep to match the LF of either sample, whereas the slope
$\alpha=-0.31$ measured from the Virgo cluster is acceptable for both
samples: by fixing $M_0(R_\mathrm{c})$ and $\Sigma$ for the Gaussian
component to the best fit values obtained in the iterative fits (Table
\ref{ess_local}), and the Schechter slope $\alpha$ to $-0.30$, the STY
solution yields values of $M^*(R_\mathrm{c})$ which differ by less
than 1-$\sigma$ for the $R_\mathrm{c}\le20.5$ and
$R_\mathrm{c}\le21.5$ samples (and by $\sim1$-$\sigma$ from the
respective values obtained by the iterative STY fits); the
corresponding likelihood ratios are $0.41$ and $0.44$ (the parameters
for these constrained fits are listed in Table \ref{ess_local} just
after the iterative fits, and are plotted in \fg
\ref{lf_local_2}). Similar likelihood ratios are also obtained when
$\alpha$ is fixed to $-0.40$: $0.43$ and $0.42$ for the
$R_\mathrm{c}\le20.5$ and $R_\mathrm{c}\le21.5$ samples
respectively. For $\alpha\le-0.50$ or $\alpha\ge-0.20$, the likelihood
ratios for the 2 samples differ by at least $0.6$. Also, although the
redshift incompleteness is corrected for in the calculation of the
SWML solution (see \sct \ref{method}), the low amplitude of the
faintest 3 points in the $R_\mathrm{c}\le21.5$ LF, which causes the
high value $\alpha=0.39\pm0.21$, could be explained by a differential
bias against late morphological type at this faint limit: beyond the
nominal limit of $R_\mathrm{c}\le20.5$, objects with preferentially
steeper light profile were observed, in order to insure a sufficient
signal-to-noise ratio in the spectra; these objects are likely to have
an earlier morphological type.

Using the faint-end slope $\alpha=-0.83$ obtained for the
$R_\mathrm{c}\le20.5$ sample as the steeper allowed value, and the
common value $\alpha=-0.3$ as an upper limit, we obtain the constraint
that the faint-end slope of the ESS late-type dwarf component lies in
the interval $-0.8\la\alpha\la -0.3$. However, as for the
intermediate-type LF, the ESS weakly constrains the faint-end slope of
the dI component, and we regard these limits on $\alpha$ as tentative.

To evaluate contribution from the Sd/Sm galaxies to the ESS late-type
LF, we also plot in the right panels of \fgs \ref{lf_local} and in \fg
\ref{lf_local_2} the expected Sd/Sm LF with the shape listed in Table
\ref{Vir_Cen_lf} and the amplitude $\phi_0$ defined such as the
integral over the Sd LF is half the integral over the Sc+Sd LF for
$M(R_\mathrm{c})\le-16.6$.  We justify this choice a follows:

\begin{itemize}
\item \fg 3 of \citet{jerjen97b} shows that for
$M(B_\mathrm{T})\le-15.5$, the Sd/Sm galaxies amount to approximately
half the number of Sc galaxies;
\item $M(B_\mathrm{T})\le-15.5$ corresponds to
$M(R_\mathrm{c})\le-16.6$ when using
$M(B_\mathrm{T})-M(R_\mathrm{c})=1.1$ for an Scd galaxy (see Table
\ref{conv_BT});
\item if one assume that about half of the Sc galaxies are included in
each of the ESS intermediate and late-type classes, the Sc galaxies
then contribute in equal amount as the Sd/Sm galaxies to the late-type
LF, and the expected ratio of Sd over Sc+Sd galaxies with
$M(R_\mathrm{c})\le-16.6$ is approximately $0.5$.
\end{itemize}

The resulting amplitudes $\phi_0$ for the Sd/Sm LF in the
$R_\mathrm{c}\le20.5$ and $R_\mathrm{c}\le21.5$ are listed in Table
\ref{ess_local}. Figures \ref{lf_local} or \ref{lf_local_2} show that
in the full magnitude range considered, the Sd/Sm component is a
factor $\sim10$ smaller than the late-type LF. This confirms a
posteriori that the Sd/Sm galaxies have a negligeable contribution to
the ESS late-type LF, and could not be constrained as a separate
component in the composite fits.

\subsection{The ESS peak surface brightness distributions \label{peak_sb}}

We now use the surface brightness (SB) of the ESS galaxies to provide
further evidence for the contribution of dwarf galaxies to both the
intermediate-type and late-type classes. The SExtractor package
\citep{bertin96} was used for image analysis of the ESS photometric
survey \citep{arnouts97}, and among the extracted parameters is the
peak SB of the objects, calculated in the one object pixel with the
highest flux. \citet{galaz02} show that the central SB in the
near-infrared is strongly correlated with fundamental physical
parameters for low-SB galaxies. Extrapolating this result to optical
wavelengths, we use for each galaxy in the ESS its SExtractor peak SB
in the $R_\mathrm{c}$ band (denoted $SB_\mathrm{peak}$) and correct it
for (i) the K-correction of the corresponding galaxy, and (ii) the
dimming due to the expansion of the Universe, which varies with
redshift $z$ as $2.5 \log(1+z)^4$; across the ESS survey, the SB
dimming varies from $0.41 ^\mathrm{mag}$ at $z=0.1$ to $2.04
^\mathrm{mag}$ at $z=0.6$. We obtain a ``rest-frame'' peak SB defined
as
\begin{equation}
\label{sb_cor}
SB_\mathrm{peak/rest} = SB_\mathrm{peak} - 10 \log(1+z) - K(z,\delta^\prime)
\end{equation}
(see \sct \ref{kcor} for the definition of the K-correction
$K[z,\delta^\prime]$).  The resulting $SB_\mathrm{peak/rest}$
describes the interval $18\la \mu_0 \la 22.5$ mag arcsec$^{-2}$ in the
$R_\mathrm{c}$ band for the 617 ESS galaxies with
$R_\mathrm{c}\le20.5$. 

These values of the rest-frame peak SB cannot however be directly
compared among them, because the peak pixel over which they are
calculated corresponds to a \emph{varying} physical aperture at
different redshifts. Moreover, as 2 different telescopes and 4
different CCDs were used over the course of the photometric survey
\citep{arnouts97}, with pixels scales of 0.35 arcsec/pixel, 0.44
arcsec/pixel, and 0.675 arcsec/pixel in the $R_\mathrm{c}$ filter, the
physical transverse size over which the rest-frame peak SB is
calculated can take 3 different values at a given redshift. We thus
calculate for each objet the physical transverse ``radius'' of the
peak pixel, denoted $r_\mathrm{peak}$, and defined as the product of
half the pixel size $A_\mathrm{pix}$ (in radians) by the
angular-distance diameter $d_\mathrm{D}=d_\mathrm{L}/(1+z)^2$, where
$d_\mathrm{L}$ is the luminosity distance given in \eq \ref{d_lum}.
The resulting values of $r_\mathrm{peak}$ vary from $\sim0.15$\hkpc at
$z\simeq0.05$ to $\sim0.5$\hkpcp, $\sim0.6$\hkpcp, $\sim0.95$\hkpc at
$z\simeq0.3$, and to $\sim0.65$\hkpcp, $\sim0.85$\hkpcp,
$\sim1.3$\hkpc at $z\simeq0.6$ (the 3 values correspond to the 3 above
mentioned pixel sizes).

\begin{figure}
  \resizebox{\hsize}{!}
    {\includegraphics{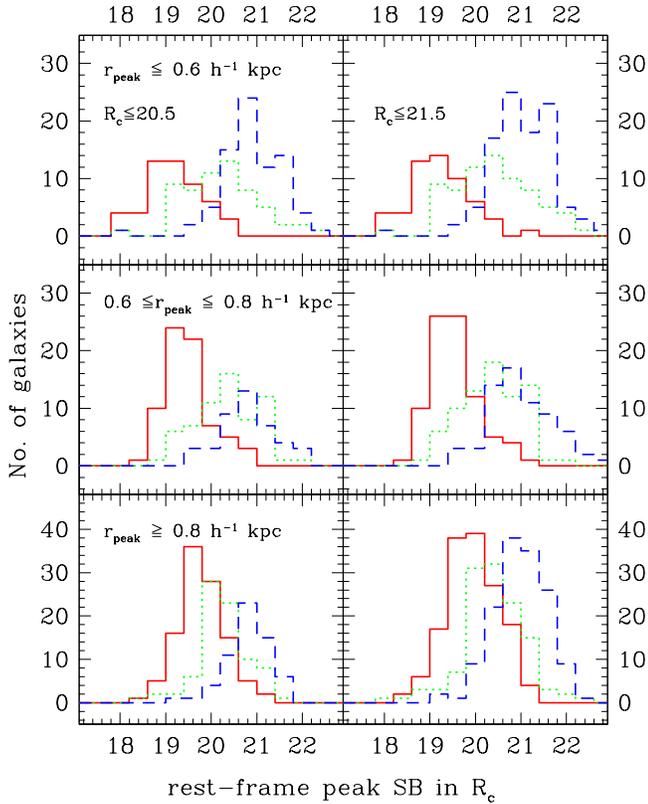}}
\caption{Comparison of the ESO-Sculptor rest-frame peak surface
brightness at $R_\mathrm{c}\le20.5$ (3 left panels) and
$R_\mathrm{c}\le21.5$ (3 right panels), for the early-type (red solid
line), intermediate-type (green dotted line), and late-type (blue
dashed line) galaxies, using redshift increments $\Delta z=0.04$. The
3 spectral classes are separated according to the physical radius
$r_\mathrm{peak}$ of the peak pixel over which is calculated the
surface brightness: top, middle, and bottom panels correspond to
$r_\mathrm{peak}\le0.6$\hkpcp, $0.6< r_\mathrm{peak}\le 0.8$\hkpcp,
and $r_\mathrm{peak}> 0.8$\hkpc respectively.}
\label{sb}
\end{figure}

These variations in $r_\mathrm{peak}$ for the ESS result in
significant variations in the average SB measured within the peak
pixel: for example, as shown by \citet{binggeli91}, the SB profile of
giant and dwarf Elliptical galaxies in the Virgo cluster steeply
decreases outwards, and varies by $\sim3$ to $\sim5$ magnitudes when
the physical radius varies from $\sim0.5$\hkpc to $\sim1.5$\hkpc
\citep[see also][]{binggeli98}.  For comparison of the rest-frame peak
SB among the 3 spectral classes, we therefore separate galaxies within
each spectral class into the following 3 intervals of
$r_\mathrm{peak}$: $r_\mathrm{peak}\le0.6$\hkpcp, $0.6<
r_\mathrm{peak}\le 0.8$\hkpcp, and $r_\mathrm{peak}> 0.8$\hkpcp; these
values are chosen so that there are more than $40$ galaxies in each
sub-sample of each spectral class. Note that the variable seeing
conditions during the course of the survey also affect the measured
peak SB. Seeing is most effective in decreasing the peak SB of objects
with steep profiles, thus decreasing the contrast between objects with
high and low peak SB. The segregation between galaxies with high and
low peak SB detected in \fg \ref{sb} below might thus be intrinsically
larger.

Figure \ref{sb} shows the resulting histograms of rest-frame peak SB for
the 3 intervals of $r_\mathrm{peak}$ within each ESS spectral class,
for $R_\mathrm{c}\le20.5$ and $R_\mathrm{c}\le21.5$.  For
$r_\mathrm{peak}\le0.6$\hkpc and $0.6< r_\mathrm{peak}\le 0.8$\hkpc
(top and middle panels), the intermediate-type and late-type galaxies
with $R_\mathrm{c}\le20.5$ show a low SB tail, which is not present in
the early-type galaxies. For $r_\mathrm{peak}>0.8$\hkpc, the effect is
only visible for the late-type galaxies. For all 3 intervals of
$r_\mathrm{peak}$, the effect persists at $R_\mathrm{c}\le21.5$, with
a larger fraction of galaxies in the low-SB tails.  The absence of
low-SB tail for the intermediate-type galaxies with $r_\mathrm{peak}>
0.8$\hkpc can be explained as follows: the ESS galaxies with
$r_\mathrm{peak}>0.8$\hkpc have $z\ga0.25$, and are therefore brighter
than $M(R_\mathrm{c})\simeq-19.2$, due to their K-corrections (see \fg
\ref{abs_z} in \sct \ref{LFess} above). However, as shown in \fg
\ref{lf_local}, the early-type dwarf contribution to the
intermediate-type LF becomes dominant only at fainter magnitudes than
this limit.  The early-type dwarf galaxies are therefore inherently
excluded from the $r_\mathrm{peak}> 0.8$\hkpc sub-samples, which in
turn explains the absence of the low SB tails for these samples. This
selection effect has a smaller impact on the late-type galaxies, as
these have a smaller K-correction, and a steeply increasing LF at
$M(R_\mathrm{c})\simeq-19.0$: a non-negligeable fraction of the
late-type galaxies are thus included in the $r_\mathrm{peak}>
0.8$\hkpc samples, at both $R_\mathrm{c}\le20.5$ and
$R_\mathrm{c}\le21.5$. 

We also observe a correlation between SB and $M(R_\mathrm{c})$
magnitude for the ESS galaxies in both the
$r_\mathrm{peak}\le0.6$\hkpc and $0.6< r_\mathrm{peak}\le 0.8$\hkpc
sub-samples, with fainter galaxies having fainter SB. The galaxies
with low SB detected in both the intermediate-type and late-type
galaxies are therefore low luminosity objects.  This provides further
evidence that the faint components of the ESS intermediate-type and
late-type LFs are indeed dwarf galaxies, characterized by both low
luminosity and low SB.

For the reasons discussed above, the measured peak SB for the ESS
galaxies cannot be directly compared with the SB measurements derived
from nearby galaxies.  For example, in the Sculptor and Centaurus A
groups, \citet{jerjen00} derive the extrapolated central SB calculated
by adjustment of S\'ersic models to the object profiles: this yields
relatively bright SB. For the Virgo cluster, \citet{binggeli91}
calculate the mean SB within the effective radius defined to contain
half of the total light of the galaxy, which varies by a factor of 3
among the populations of Virgo giant/dwarf Elliptical and Lenticular
galaxies (see their \fg 1). The ESS results can however be compared
with the results of \citet{trentham02a}, who measure the average $B$
band SB of Virgo cluster galaxies within a constant circular aperture
of 6 arc-second radius: at the redshift of Virgo ($z\simeq0.0038$),
this corresponds to $0.33$\hkpcp.  The values of SB measured by
\citet{trentham02a} can thus be compared with those for the ESS
galaxies in the $r_\mathrm{peak}\le0.6$\hkpc sub-sample.  As in the
ESS, \citet{trentham02a} show a tight correlation between SB and
absolute magnitude, with brighter galaxies having brighter SB (see
also \citealt{binggeli91,jerjen00}): the E/S0 and Spiral galaxies
populate the bright part of the Virgo sequence in the SB interval
$\sim 18-23 B$ mag arcsec$^{-2}$ for Virgo, and the early-type and
late-type dwarf galaxies populate the faint part of the sequence, with
$\sim 21-27 B$ mag arcsec$^{-2}$. At a SB of $\ge 22 B$ mag
arcsec$^{-2}$, the dwarf galaxies dominate in numbers over the giant
galaxies. This limit corresponds to $\sim 20.7 R_\mathrm{c}$ mag
arcsec$^{-2}$ for a dE galaxy, assuming the color of an Sab galaxy
(see Table \ref{Vir_Cen_lf}), and to $\sim 21.4 R_\mathrm{c}$ mag
arcsec$^{-2}$ for an Im galaxy (see Table
\ref{conv_BT}). Interestingly, the SB histograms for the ESS
intermediate-type and late-type galaxies with
$r_\mathrm{peak}\le0.6$\hkpc in \fg \ref{sb} both show a local peak
(at faint SB) within less than $0.2 ^\mathrm{mag}$ from these values;
the sharp decrease in objects fainter than these peaks is caused by
incompleteness.  This comparison thus provides evidence that the low
SB tails of the ESS intermediate-type and late-type classes contain
dwarf galaxies similar to those detected in nearby clusters as Virgo.

In \fg \ref{sb}, the variations in the SB distributions as a function
of $r_\mathrm{peak}$ also provide evidence of varying profiles among
the ESS galaxies. When going to larger values of $r_\mathrm{peak}$,
the SB histograms for the intermediate-type and late-type galaxies
maintain a nearly constant median value of SB, whereas the early-type
galaxies show a shift to fainter SB. This can be interpreted as a
signature of the steeper profiles for E galaxies, which according to
\citet{binggeli98} have S\'ersic parameter $n=0.1$ to $0.5$; in
contrast, \citet{binggeli98} show that the SB profile of the
early-type/late-type dwarf and the Spiral galaxies are better fitted by
flatter profiles, with $0.3\la n \la 2.0$ ($n=0.25$ corresponds to the
$r^{1/4}$ law by \citealt{vaucouleurs48}; $n=1$ corresponds to an
exponential profile, as measured by \citealt{freeman70}, for the disk
component of Spiral and S0 galaxies). The effect can be interpreted as
follows: for smaller values of $r_\mathrm{peak}$, steeper parts of the
SB profile of E galaxies are sampled, and brighter values of SB are
derived. The S0 and Sa galaxies also included in the early-type class
might also contribute to the effect, as the bulges have a significant
contribution to the object profile in the central parts of the
galaxies.

\section{Conclusions and prospects \label{concl}}

The present analysis of the ESO-Sculptor Survey (ESS) provides new
measurements of the $B$, $V$, and $R_\mathrm{c}$ luminosity functions
(LF) of galaxies at $z\la0.6$. We use a PCA-based spectral
classification, and a technique providing a parametric estimation of
the K-corrections as a function of redshift and spectral type. From
these, we derive absolute magnitudes accurate to $0.09 ^\mathrm{mag}$
in $R_\mathrm{c}$, $0.13 ^\mathrm{mag}$ in $V$, and $0.16
^\mathrm{mag}$ in $B$ for the nearly complete sample of 617 galaxies
with redshift at $R_\mathrm{c}\le20.5$. The LFs are then calculated
for 3 spectral-type sub-samples with comparable numbers of galaxies,
denoted early-type, intermediate-type, and late-type respectively.
Projection of the \citet{kennicutt92} galaxies onto the ESS spectral
sequence shows that the 3 spectral classes correspond to morphological
types E/S0/Sa, Sb/Sc, and Sc/Sm/Im respectively.

The derived LFs for each spectral type have a similar behavior in the
$B$, $V$ and $R_\mathrm{c}$ bands, which indicates that they measure
physical properties of the underlying galaxy populations. They are
well fitted by Schechter functions, with a dimming of the bright-end and
a steepening of the faint-end when going from early-type to late-type
galaxies. Because the spectroscopic sample was selected in the
$R_\mathrm{c}$ band, the $V$ and $B$ band LFs suffer from
incompleteness in blue galaxies at the faint limit; this bias tends to
weaken the steepening of the faint-end of the LF for late-type
galaxies.

We then compare the ESS spectral-type LFs with the results from the
comparable CNOC2 redshift survey \citep{lin99}, the only other
redshift survey to similar depth and based on a spectral
classification.  The Schechter fits to the ESS LFs in the $B$ and
$R_\mathrm{c}$ bands are in agreement with those from the CNOC2. In
the $V$ band, the ESS provides the \emph{first} estimates of intrinsic
LFs at $z\sim0.3$. Further comparison of the ESS with other redshift
surveys is reported in \citet{lapparent-lc}, in which is performed a
detailed analysis of all the existing measurements of intrinsic LFs in
the $UBVR_\mathrm{c}I_\mathrm{c}$ bands from redshift surveys with
effective depth ranging from $z\simeq0.03$ to $0.6$. By using the
local intrinsic LFs per morphological type as a reference,
\citet{lapparent-lc} shows how the existing redshift surveys may mix
galaxies of different morphological types, thus complicating the
interpretation of their LFs.

The salient results of the present article are obtained by fitting the
3 ESS spectral-type LFs in the $R_\mathrm{c}$ band with composite
functions suggested by the intrinsic LFs measured locally for each
morphological type in the Virgo, Centaurus, and Fornax clusters
\citep{sandage85b,jerjen97b}.  Specifically, we show that the ESS
spectral-type LFs can be modeled as follows:
\begin{itemize} 
\item the early-type LF: by a two-wing Gaussian function representing
the contributions from E, S0, and Sa galaxies;
\item the intermediate-type LF: by the sum of a Gaussian function
representing the Sb+Sc galaxies, and a Schechter function with a steep
slope ($\alpha\la-1.5$) representing the contribution from dwarf
Spheroidal galaxies, which dominates at $M(R_\mathrm{c})\ga -19.0$.
\item the late-type LF: by the sum of a Gaussian function for the Sc+Sd
galaxies, and a Schechter function with a flat or weaker slope
($-0.8\la\alpha\la-0.3$) representing the dwarf Irregular galaxies, which
dominate at $M(R_\mathrm{c})\ga -19.0$.
\end{itemize}

The interesting aspect of the comparison of the ESS spectral-type LFs
with the local intrinsic LFs is that it provides clues on the various
galaxy populations included in the ESS spectral classes. It first
shows that the bright end of the 3 spectral-type LFs is dominated by
giant galaxies (E, S0, Spirals). It also reveals the contribution from
dwarf galaxies to the faint-end of both the ESS intermediate-type and
late-type LFs. These dwarf galaxies lie at $z\la0.2$ and are
characterized by low luminosity ($M[R_\mathrm{c}]\ga-18.5$) and low
surface brightness ($\ga 20.5$ $R_\mathrm{c}$ mag arcsec$^{-2}$
averaged within a physical radius of $0.6-0.8$\hkpcp). This
interpretation of the ESS spectral-type LFs illustrates how a spectral
classification may mix galaxies of different morphological type: the
ESS intermediate-type class may contain both Spiral (Sb+Sc) galaxies
\emph{and} dwarf Spheroidal galaxies.

Comparison of the ESS LF components for the various morphological
types with the local intrinsic LFs by \citet{sandage85b} and
\citet{jerjen97b} suggests that the shape of the LFs for the
individual Hubble types might not vary markedly in the redshift
interval $0\la z\la0.6$: contributions from Gaussian LFs for giant
galaxies (E, S0, Sa, Sb, Sc) with similar peak magnitudes as locally
can be adjusted to the ESS LFs. The systematically narrower dispersion
for the ESS Gaussian components can be explained by selection effects
inherent to magnitude-limited redshift surveys, which cause
under-sampling at both the bright-end and faint-end of the ESS LFs. A
contribution from environmental effects is also expected, such as the
presence of brighter giant galaxies in clusters than in sparse groups
and the field, due to the higher frequency of merging and cannibalism
in dense regions.  The same dimming of the characteristic luminosity
which is observed locally when going to later Spiral type (from Sa, to
Sb, Sc, and Sd/Sm) is observed in the ESS.  Because late-type Spiral
galaxies are brightened in the optical by their higher star formation
rates compared with early-type Spiral galaxies, their dimming in
luminosity is indicative of a systematic decrease in mass.

For the dwarf galaxies, the ESS composite fits suggest a steeper slope
for the early-type dwarf LF ($\alpha\ga-1.5$) than for the late-type
dwarf LF ($\-0.8\la\alpha\la-0.3$), as already detected in several
nearby groups and clusters \citep{sandage85b,jerjen97b,jerjen00}. This
confirms earlier suggestions that the late-type dwarf LF is bounded at
the faint-end \citep{sandage85b,jerjen00}.  Nevertheless, the ESS only
probes the brightest part of the dwarf galaxy LFs, to
$M(R_\mathrm{c})\le-16$, and therefore puts poor constraints on their
actual slope faint-end slope. The characteristic magnitude $M^*$ of
the Schechter LFs for the dwarf galaxies is also poorly constrained by
the composite fits.  We thus emphasize that due to the various
incompleteness effects, the specific ESS composite fits should not be
used as quantitative constraints on the intrinsic LFs at
$z\sim0.3$. These fits should rather be considered as indicative of
the possibilities expected by application to forthcoming larger
samples. These limitations point to the need for field measurements
based solely on dwarf galaxy samples.

Recent results do provide information on the local dwarf galaxy LFs at
faint magnitudes. In their study which combines all available data on
dwarf galaxies in the Sculptor, Centaurus A, and M81 groups, together
with the Local Group, \citet{jerjen00} measure a steep slope
$\alpha=-1.29\pm1.10$ for dwarf galaxies brighter than
$M(B_\mathrm{T})=-14.0$; in these data, late-type dwarf galaxies
dominate over early-type dwarf for $M(B_\mathrm{T})\le-14.0$, and
early-type dwarf galaxies represent an increasing proportion at
fainter magnitudes (out to $M[B_\mathrm{T}]\simeq-9.0$); in these
data, the LF for the late-type dwarf galaxies reaches its maximum in a
``plateau'' located in the interval $-16\le M(B_\mathrm{T})\le -14$
which corresponds to $-16.6\le M(R_\mathrm{c})\le -14.6$ in the ESS
(using Sm/Im colors, see Table \ref{conv_BT}).  Recent observations of
5 nearby clusters and groups (including the Virgo cluster) obtained
with the NAOJ Subaru 8 m telescope on Mauna Kea suggest a similar
faint-end slope for each structure, with an average value
$\alpha=-1.2$ in the interval $-18.0\la M(R)\la-10.0$
\citep{trentham02b}. The fraction of dE over dE+dI galaxies is
estimated to be $83\pm12$\% in the Virgo cluster, the richest of the 5
concentrations, and decreases to $33\pm19$\% in the least dynamically
evolved group, NGC1023. Although the survey by \citet{trentham02b}
does not put constraints on the separate faint-ends of the LFs for the
dwarf Spheroidal and dwarf Irregular galaxies, it suggests a universal
slope $\alpha=-1.2$ for the sum of the two populations. The mean LF
measured by \citet{trentham02b} is also dominated by the Gaussian
component for giant galaxies at $M(R)\le-19.5$, and is separated from
the power-law behavior at faint magnitudes by a transition region in
the interval $-19.5\la M(R)\la-18.0$, characterized by a knee. Note
that in the pure Schechter fits, the faint-end slope is actually
determined by the LF in this very magnitude interval. This yields the
steep slope $\alpha=-1.64\pm0.23$ for the ESS late-type LF, whereas a
flatter slope $-0.8\la\alpha\la-0.3$ is derived when the LF is
decomposed into its intrinsic components.  This casts further doubt on
the adjustment of the LFs from redshift surveys by pure Schechter
functions, and emphasizes the usefulness of the composite fits such as
performed here.

Because giant and dwarf galaxies show marked differences in both their
LFs and their spatial distributions, we expect that their detailed
description produce crucial constraints for the N-body models, thus
yielding clues on the mechanisms for galaxy formation
\citep[see][]{mathis02a,mathis02b}.  Note that an evolutive sequence
among the dwarf galaxies, which could be closely linked to galaxy
interactions and merging, is suggested both by observations
\citep{sung98} and models of galaxy formation
\citep{valageas99,okazaki00}. Measuring the intrinsic LFs for each
class of dwarf galaxies, in various environments, could help in
constraining these evolution scenarii.

Most importantly, the present analysis of the ESS LFs, and their
comparison with the local intrinsic LFs per morphological type points
to the importance of separating the galaxy populations which have
different intrinsic LFs.  The ESS spectral-type LFs also illustrate
the limits of measuring intrinsic LFs from redshift surveys in which
galaxies are \emph{solely} classified from their spectra, as spectral
classification is insufficient to separate giant and dwarf galaxies.
The best approach for measuring intrinsic LFs is to use a
morphological classification. Several schemes for quantitative galaxy
classification have been proposed so far
\citep{bershady00,abraham00,refregier02,odewahn02}. The present
analysis of the ESS LFs suggests that a useful morphological
classification for measuring intrinsic LFs could also include the
surface brightness profile of the galaxies, as it provides key
information for separating the giant and dwarf galaxies
\citep{binggeli91,ferguson94,binggeli98,marleau98}.

Knowledge of surface brightness also allows measurement of the
bi-variate brightness distribution, defined as the variations of the
galaxy LF with absolute magnitude and surface brightness. As shown by
\citet{andreon01c}, the ``general'' bi-variate brightness distribution
in the Coma cluster steepens and shifts to fainter luminosities at
lower surface brightnesses, in agreement with the steep LF measured
for dwarf galaxies in nearby groups and clusters of galaxies
\citep{jerjen00,trentham02b}. \citet{cross01} show that accounting for
the distribution in surface brightness provides unbiased estimates of
the ``general'' luminosity density \citep[see also][]{cross02}.
\citet{barazza01} also detect a higher means surface brightness in
field and group late-type dwarf galaxies than in cluster late-type
dwarf, which in turn suggest different histories in the various
environments. Significant improvements over the existing analyses
could be brought by including morphological classification into the
analyses of the bi-variate brightness distribution, as surface
brightness alone is not sufficient to discriminate among the different
morphological types present at a given surface brightness
level. Colors and spectral classification might provide part of this
additional information, as they may enable one to differentiate among
the giant galaxies (E, S0 and Spiral) on one hand, and among the dwarf
galaxies (dE, dS0 and dI) on the other hand.  Measurement of the
bi-variate brightness distribution for each morphological type appears
as the ultimate goal to aim at.

The ESS sample analyzed here is not large enough for measuring either
the intrinsic LFs per morphological type or the bi-variate brightness
distribution. Such detailed analyses require large redshift samples,
with at least $\sim 10^5$ galaxies.  Samples that large are being
obtained at $z\la0.2$ by 2 dedicated surveys, the Sloan Digital Sky
Survey (see http://www.sdss.org/; \citealt{blanton02}), and the 2dF
Galaxy Redshift Survey (see http://www.mso.anu.edu.au/2dFGRS/). As
shown by \citet{lapparent-lc}, the scheme used so far for galaxy
classification in the 2dF survey (based on PCA spectral
classification, \citealp{madgwick02a}, and interpreted in terms of
star formation history, \citealp{madgwick02b}) appears insufficient
for measuring the intrinsic LFs, whereas the SDSS estimates based on
colors \citep{blanton01} seem more successful.  Useful results on the
intrinsic LFs at $z\sim1$ should be obtained from the DEEP2
\citep{davis02} and VIRMOS \citep{lefevre01} surveys, thanks to
efficient multi-slit spectrographs on the Keck
\citep{cowley97,james98} and ESO-VLT \citep{lefevre01} telescopes,
respectively. The present analysis of the ESS emphasizes the need that
these various surveys use objective algorithm for galaxy
classification which are able to separate the giant and dwarf galaxy
populations along with the different morphological types within these
2 populations.

Another forthcoming survey is also expected to provide significant
contributions to the measurement of the intrinsic LFs to $z\la1$: the
Large-Zenith-Telescope project \citep[][see also
http://www.astro.ubc.ca/LMT/lzt.html]{cabanac02,hickson98b}, which
will provide ``photometric redshifts'' with an accuracy
$\sigma(z)\le0.05$ at $z\la1$, using 40 medium-band filters. As shown
by \citet{lapparent-lc}, the CADIS \citep{fried01} and COMBO-17
\citep{wolf03} surveys with similar redshift accuracy
($\sigma(z)\le0.03$) succeed in measuring spectral-type LFs in good
agreement with the ESS and CNOC2 (for which $\sigma(z)\la0.0005$).  We
therefore expect that the Large-Zenith-Telescope provides detailed
measurements of the intrinsic LFs and bi-variate brightness
distributions to $z\la1$, thanks to its expected $10^6$ galaxies.

If complemented by detailed and quantitative morphological
information, the mentioned next-generation surveys will allow one to
study whether and how the intrinsic LFs vary with redshift and local
density.  Most redshift surveys to $z\ga0.5$ have detected significant
evolution in several of the intrinsic LFs
\citep{lilly95,heyl97,fried01}. A marked evolution is also detected in
the ESS, and is reported and compared with the previous measurements
in \citet{lapparent03b}. In contrast, the existing detection of a
variation of the LF with local density \citep{bromley98} is poorly
conclusive.  The mentioned next-generation surveys with $\sim10^5$ to
$10^6$ galaxies should address these issues in further details and
with improved statistics.

\begin{acknowledgements}
V. de L. is grateful to Harry Van der Laan, ex-director general of
ESO, for the advent of the key-programme concept, allowing the
observations for long-term projects such as presented here to reach
completion. V. de L. thank Helmut Jerjen for his prompt and kind reply
to many questions on dwarf galaxies and local luminosity
functions. The authors are also grateful to an anonymous referee, an
anonymous reader, and Stefano Andreon, whose comments on the first
submitted version of this article helped in improving its content.

Christian Oberto is thanked for one full year of assistance with the
spectroscopic data-reduction, which lead to completion of the
spectroscopic data-reduction phase. Many thoughts also go to
Christ\`ele Bellanger, for her involvement and dedicated work as a
Ph. D. student at the early and thankless stages of the survey. V. de
L.  gratefully acknowledges the support of Laurence Courriol-Nicod
which helped in the management of the project and in keeping alive the
confidence in its eventual success.
\end{acknowledgements}

\bibliography{lapparent}
\bibliographystyle{aa}

\end{document}